\documentclass{article}

\usepackage{ifthen,amssymb}

\usepackage{mik-makro}

\usepackage{amsmath,xspace,graphicx,color,proof,array}
\usepackage[all]{xypic}

  \newcommand{\deriv}{\mathsf{der}}
\newcommand{\promonmul}[1]{\bar \mu_{#1}}
\newcommand{\promonun}[1]{\bar \eta_{#1}}
\newcommand{\monmul}[1]{\mu_{#1}}
\newcommand{\monun}[1]{\eta_{#1}}
\newcommand{\stone}[1]{\mathsf{Stone}#1}

\newcommand{\deselect}{\mathsf{deselect}}
\newcommand{\filter}{\mathsf{filter}}

\newcommand{\subleft}[1]{#1_{\mathrm{left}}}
\newcommand{\subright}[1]{#1_{\mathrm{right}}}

\newcommand{\eqexplain}[1]{\mbox{\small (#1)}}
\newcommand{\closealg}[1]{\overline{#1}}
\newcommand{\restrictalg}[1]{#1_\monad}
\newcommand{\distance}{\mathrm{dist}}
\newcommand{\promonad}{{\overline \monad}}

\newcommand{\sem}[1]{[\![#1]\!]}

\newcommand{\upol}[1]{\pol_{#1}}

\newcommand{\upeval}[3]{[\![#2]\!](#3)}
\newcommand{\ident}[1]{\mathrm{id}_{#1}}

\newcommand{\pol}{\mathsf{pol}}

\newcommand{\synt}[1]{\mathsf{synt}#1}
\newcommand{\unarymonad}{\mathsf {Point}}
\newcommand{\forest}{\mathsf F}
\newcommand{\mult}{\mathrm{mul}}
\newcommand{\unit}{\mathrm{unit}}
\newcommand{\mso}{{\sc mso}\xspace}
\newcommand{\msoinf}{{\sc mso+}inf\xspace}
\newcommand{\msol}[1]{{\sc mso}$_\monad$($#1$)}

\newcommand{\shuffle}{\mathrm{shuffle}}
\newcommand{\langvar}{\mathbb L}
\newcommand{\algvar}{\mathbb A}

\newcommand{\monad}{\mathsf T}

\newcommand{\clo}{\mathrm{clo}}

\newcommand{\makealg}[1]{\mathsf{Alg}\ \!#1}

\newcommand{\makelang}[1]{\mathsf{Lan}\ \!#1}

		\newcommand{\eqdef}{\quad \stackrel  {\mbox{\tiny{def}}} = \quad}
		\newcommand{\unitt}[2]{
		\promonun {#2}(#1)}
 \newcommand{\alg}{{\bf A}}
  \newcommand{\balg}{{\bf B}}
    \newcommand{\calg}{{\bf C}}
	  \newcommand{\salg}{{\bf S}}
	    \newcommand{\malg}{{\bf M}}
  \newcommand{\powerset}[1]{\mathsf{P}#1}
  \newcommand{\repr}{\monad_0}
\newcommand{\reco}{\mathsf{rec}}

  \newcommand{\syntb}{\synt_\balg f}
  \newcommand{\syntim}{A_f}
  \newcommand{\syntimalg}{\alg_f}

  	\newcommand{\lncshide}[1]{}
	
	\newtheorem{theorem}{Theorem}[section]
	
	\newtheorem{lemma}[theorem]{Lemma}
	
	\newtheorem{corollary}[theorem]{Corollary}
	\newtheorem{fact}[theorem]{Fact}
	\newtheorem{claim}{Claim}[theorem]

	\newcommand{\mypart}[2]{\pagebreak \part{#1} #2 \pagebreak}

\begin{document}

	\newcounter{monadcounter}
	\setcounter{monadcounter}{-1}
	
	\title{Recognisable Languages over Monads}
	\author{Miko{\l}aj Boja\'nczyk}
	\lncshide{\institute{University of Warsaw}}
	\pagestyle{plain}

	\maketitle

	 \tableofcontents
\mypart{Introduction}{In this part, we introduce monads and their algebras. Section~\ref{sec:monads-and-their-algebras} contains the basic definitions: first illustrated on examples of finite words and $\infty$-words, and then formally defined. Sections~\ref{sec:myhill-nerode}-\ref{sec:mso} show how some results about languages can be stated and proved on the level of monads, including: the  Myhill-Nerode theorem  (Section~\ref{sec:myhill-nerode}),  Eilenberg's pseudovariety theorem (Section~\ref{sec:pseudovariety}), and some parts of the connection between regular languages and \mso (Sections~\ref{sec:representing-an-algebra} and~\ref{sec:mso}). }
\section{Introduction}
The  principle behind algebraic language theory for various kinds of structures, such as  words or   trees, is to use a compositional function from the structures into a finite set. To talk about compositionality, one needs some way of composing structures into bigger structures. It so happens that category theory has an abstract concept for this, namely a monad. The goal of this paper is to propose monads as a unifying framework for discussing existing algebras and designing new algebras.
To introduce monads and their algebras, we begin with two examples, which use a monad style to present algebras for finite and infinite words.

\begin{myexample}\label{ex:finite-words} 
	 Consider the following non-standard definition of a semigroup. Define a $+$-algebra $\alg$ to be a set $A$ called its \emph{universe}, together with a \emph{multiplication operation}  $\mult_\alg : A^+ \to A$,  which is the identity on single letters, and which is associative in the sense that the 
	following diagram commutes.
		\begin{align*}
			\vcenter{\xymatrix @R=1pc { (A^+)^+  \ar[rr]^{\monmul A} \ar[d]_{{(\mult_\alg)}^+} && A^+ \ar[d]^{\mult_\alg}  \\
			A^+ \ar[rr]_{\mult_\alg}& & A
			}},
		\end{align*}
In the diagram, $(\mult_\alg)^+$ is the function that applies $\mult_\alg$ to each label of a word where the  alphabet is $A^+$, and 
		 $\monmul A$ is the function which flattens a word of words into a  word, e.g.
		 \begin{align*}
 (abc)(aa)(acaa) \qquad \mapsto \qquad abcaaacaa.
		 \end{align*}
Restricting the multiplication operation in a $+$-algebra to words of length  two (the semigroup binary operation) is easily seen to be a one-to-one correspondence between $+$-algebras and semigroups. \end{myexample}

The second example will be running example in the paper.
\begin{running}\label{ex:infty-words} Let us define an algebra for infinite words in the spirit of the previous example.	 Define $A^\infty$ to be the $\infty$-words over $A$, i.e.~$A^+ \cup A^\omega$. Define an $\infty$-algebra $\alg$ to be a set $A$, called its \emph{universe}, together with a \emph{multiplication operation} $\mult_\alg : A^\infty \to A$, which is the identity on single letters, and which is associative in the sense that the 
	following diagram commutes.
		\begin{align*}
			\vcenter{\xymatrix @R=1pc { (A^\infty)^\infty  \ar[rr]^{\monmul A} \ar[d]_{{(\mult_\alg)}^\infty} && A^\infty \ar[d]^{\mult_\alg}  \\
			A^\infty \ar[rr]_{\mult_\alg}& & A
			}}
		\end{align*}
In the diagram, $(\mult_\alg)^\infty$ is the function that applies $\mult_\alg$ to the label of every position in a $\infty$-words where the alphabet is  $A^\infty$, and  $\monmul A$ flattens an $\infty$-word of $\infty$-words into an $\infty$-word.  If the argument of $\monmul A$  contains an infinite word on some position, then all subsequent positions are ignored.
		
		 An $\infty$-algebra is essentially the same thing as an $\omega$-semigroup, see~\cite{perrin_pin_words}, with the difference that $\omega$-semigroups have separate sorts for finite and infinite words. There is also a close connection with Wilke semigroups~\cite{DBLP:conf/icalp/Wilke91}, which will be described as the running example develops.
\end{running}

The similarities in the examples suggest that the should be an abstract notion of algebra, which would cover the examples and possibly other settings, e.g.~trees. A closer look at the examples  reveals that  concepts  of algebraic language theory such as  ``algebra'', ``morphism'', ``language'', ``recognisable language'' can be  defined only in terms of the following  four basic concepts (written below in the notation appropriate to $+$-algebras):
\begin{enumerate}
	\item  how a set $A$ is transformed into a set $A^+$;
	\item how a function $f : A \to B$ is lifted to a function $f^+ : A^+ \to B^+$;
	\item  a flattening operation from $(A^+)^+ \to A^+$;
	\item how to represent an element of  $A$ as an element of $A^+$.
\end{enumerate}
These four concepts, subject to certain axioms, are what constitutes a monad, a fundamental concept in category theory (and  recently, programming languages).

The point of this paper is that, based on a monad one can also define things like:   ``syntactic algebra'', ``pseudovariety'',  ``\mso logic'',  ``profinite object'', and even prove some theorems about them. Furthermore, monads as an abstraction cover practically every setting where algebraic language theory has been applied so far, including labelled scattered orderings~\cite{DBLP:journals/jcss/BedonR12}, labelled countable total orders~\cite{DBLP:conf/icalp/CartonCP11}, ranked trees~\cite{DBLP:books/el/treeauto1992/Steinby92}, unranked trees~\cite{DBLP:conf/birthday/BojanczykW08}, preclones~\cite{esik2003logically}. 

The paper has three parts.

Part I of this paper shows  that several results of formal language theory can be stated and proved on the abstract level of monads, including: the Myhill-Nerode theorem on syntactic algebras (Section~\ref{sec:myhill-nerode}), the Eilenberg pseudovariety theorem (Section~\ref{sec:pseudovariety}), or the Reiterman theorem (Section~\ref{sec:topology}) on profinite identities defining pseudovarieties. Another example is  decidability of \mso (Section~\ref{sec:mso}), although here monads only take care of the symbol-pushing part, leaving out the combinatorial part that is specific to individual monads, like applying the Ramsey theorem in the case of infinite words. When proving such generalisations of classical theorems, one is naturally  forced to have a closer look at notions such as ``derivative of a language'', or ``finite algebra'', which are used in the assumptions of the theorems.

Part II includes shows how existing algebraic settings can be seen as a special case of monads.  Part II also contains  some new settings, illustrating how new kinds of algebras  can be easily produced using monads. Specifically,  Section~\ref{sec:unary} describes a monad for words with a distinguished position, where standard theorems and definitions come for free by virtue of being a monad.

Part III, is devoted to profinite constructions. It is shown that every monad has a corresponding profinite monad, which, like any monad, has its own notion of recognisability, which does not reduce to recognisability in the original monad.  For example, the monad for finite words has a corresponding monad of profinite words, and recognisable languages of profinite words turn out to be a generalisation of languages of infinite words definable in the logic {\sc mso+u}.

\paragraph*{Thanks.} I would like to  thank Bartek Klin (who told me what a monad is), Szymon Toru\'nczyk, Joost Winter and Marek Zawadowski for discussions on the subject.

\section{Monads and their algebras}
\label{sec:monads-and-their-algebras}
This paper uses only the most rudimentary notions of category theory:  the definitions of a category (objects and composable morphisms between them), and of a functor (something that maps objects to objects and morphisms to morphisms in a way that is consistent with composition).  Almost all examples in this paper use the category of sets, where objects are sets and morphisms are functions; or possibly the category of sorted sets, where objects are sorted sets for some fixed set of sort names, and morphisms are sort-preserving functions.

A \emph{monad} over a category  is defined to be a functor $\monad$ from the  category to itself, and for every object $X$ in the category, two  morphisms
	\begin{align*}
		\monun X : X \to \monad X \qquad \mbox{and} \qquad
		\monmul X  : \monad \monad X \to \monad X,
	\end{align*}
	which are called the unit and multiplication operations. The monad must satisfy the  axioms given in Figure~\ref{fig:monad-axioms}. In the language of sets, an intuition appropriate for this paper is that a monad inputs a set $X$, and produces the set of all ``structures'' whose ``nodes'' are labelled by elements of $X$. Depending on the monad, the structures could be words, or trees, or graphs, etc. The function $\monun X$ inputs a label and  produces a one-node structure that uses this label; while the function $\monmul X$, which is the essence of the monad, flattens a structure of structures into a single structure. Basing on this intuition,  we will use the name \emph{$\monad$-structures over $X$} for elements of $\monad X$.

	\begin{figure}[htbp]
		\centering
	\begin{align*}
		\vcenter{\xymatrix @R=2pc { X  \ar[r]^{f} \ar[d]_{\monun X} & Y \ar[d]^{\monun Y}  \\
		\monad X \ar[r]_{\monad f}& \monad Y
		}} \qquad 		\vcenter{\xymatrix @R=2pc { \monad \monad X  \ar[r]^{\monad \monad f} \ar[d]_{\monmul X} & \monad \monad Y \ar[d]^{\monmul Y}  \\
		\monad X \ar[r]_{\monad f}& \monad Y
		}}.
	\end{align*}
	\begin{align*}
		\vcenter{\xymatrix @R=2pc @C=3pc {\monad \monad \monad X  \ar[r]^{\monmul{ \monad X}} \ar[d]_{\monad{ \monmul X}} & \monad \monad X \ar[d]^{\monmul X}  \\
		\monad \monad X \ar[r]_{ \monmul X}& \monad X
		}} 
		\qquad
		 		\vcenter{\xymatrix @R=2pc { \monad X \ar[dr]^{\mathrm{id}_X}  \ar[r]^{\monun {\monad X}} \ar[d]_{\monad \monun X} & \monad \monad X \ar[d]^{\monmul X}  \\
		\monad \monad X \ar[r]_{ \monmul X }& \monad X
		}}
	\end{align*}
		\caption{The axioms of a monad are that these four diagrams  commute for every object $X$ in the category and every morphism $f : X \to Y$. The upper diagrams say that the unit and multiplication are natural. The lower left diagram says that  multiplication is associative, and the lower right says that the unit is consistent with multiplication.} 
		\label{fig:monad-axioms}
	\end{figure}
	
We already saw two monads in Example~\ref{ex:finite-words} and in the running example.
	
	For this paper, the most important thing about monads is that they have a natural corresponding notion of algebra. An \emph{Eilenberg-Moore algebra in a monad $\monad$}, or simply $\monad$-algebra, is a pair $\alg$ consisting of a \emph{universe $A$}, which is an object in the category underlining the monad, together with a multiplication morphism
	\begin{align*}
		\mult_\alg : \monad A \to A,
	\end{align*}
	such that the $\mult_\alg \circ \monun A$ is the identity, and which is associative in the sense that the following diagram commutes.
	\begin{align*}
		\vcenter{\xymatrix @R=1pc @C=3pc { \monad \monad A  \ar[r]^{\monmul A} \ar[d]_{\monad \mult_\alg} & \monad A \ar[d]^{\mult_\alg}  \\
		\monad A \ar[r]_{\mult_\alg} & A
		}}
	\end{align*}
	Observe that this associativity is similar to the lower left axiom in Figure~\ref{fig:monad-axioms}. In fact, the lower left axiom in Figure~\ref{fig:monad-axioms} and the upper half of the lower right axiom say  that $\monmul X$ induces a  $\monad$-algebra with universe  $\monad X$.
	
	We use the convention that an algebra is denoted by a boldface letter, while its universe is written without boldface. 	A \emph{$\monad$-morphism} between two $\monad$-algebras $\alg$ and $\balg$ is defined to be a function $h$ between their universes which respects their multiplication operations in the sense that the following diagram commutes.
	\begin{align*}
		\vcenter{\xymatrix @R=1pc @C=3pc { \monad  A  \ar[r]^{\monad h} \ar[d]_{ \mult_\alg} & B \ar[d]^{\mult_\balg}  \\
		A \ar[r]_{h} & B
		}}
	\end{align*}

This completes the definition of monads and their algebras. 
\paragraph*{Languages and colourings.} To develop the basic definitions of recognisable  languages over a monad, we require the following parameters, which we call the \emph{setting}: the underlying category, the monad,  a notion of finite alphabet, and a notion of finite $\monad$-algebra. So far, we do not place any restrictions on the notions of finiteness, e.g.~when considering sets with infinitely many sorts, reasonable settings will often have finite algebras whose universe is not be finite in the same sense as a finite algebra. Actually, for  some monads, it is not clear what a finite algebra should be, e.g.~this is the case for infinite trees, and this paper sheds little new light on the question.
 Fix a setting, with the  monad being $\monad$, for the following definitions.


 A \emph{colouring} of a $\monad$-algebra is defined to be a morphism from its universe to some object in the underlying category. For example, when the category is sets, then a coloring is like a multivalued language, i.e.~instead of saying only ``yes'' or ``no'' to each input, a colouring can have multiple values. A coloring is said to be recognised by a $\monad$-morphism if it factors through it. A coloring is called \emph{$\monad$-recognisable} if it is recognised by some $\monad$-morphism with a finite target, according to the notion of finite $\monad$-algebra given in the setting.

Almost all examples in this paper are in sets, or in sorted sets. When the category is sets or sorted sets, we will  focus mainly on the special case of colourings, namely languages,  where colourings  have two possible values on every sort.
Consider a finite alphabet, according to the notion of finite alphabet given in the setting.  In all of the examples of this paper where the category is sorted sets, a finite alphabet will be a possibly sorted set with finitely many elements. In particular, if there are infinitely many sorts, then a finite alphabet will use only finitely many.
   A $\monad$-language over a finite alphabet $\Sigma$  is defined to be  any subset $L \subseteq \monad \Sigma$. Notions of recognisability are inherited from colourings, using the characteristic function of a language.   Colourings are a mild generalisation of languages, for example, when the category is sets, then a colouring with finitely many colours is $\monad$-recognisable if and only if for every color, its inverse image is a recognisable language.

When the monad $\monad$ is clear from the context, we will sometimes skip the prefix $\monad$-, and simply write language, algebra, morphism, structure.

\paragraph*{Beyond recognisable languages.} The recognisable languages will play the role of regular languages in the monad. One could go beyond regular languages. For instance, there is a natural monad version of context-free grammars, where the production rules have right hand sides in the monad applied to the terminals and nonterminals, and one can prove some theorems, like closure  of context-free languages  under intersection with recognisable languages. Context-free languages are beyond the scope of this paper.

\section{Syntactic morphisms}
\label{sec:myhill-nerode}
This section presents
 a monad generalisation of the Myhill-Nerode theorem, which gives a sufficient condition for colourings, and therefore also languages, to have  a syntactic (i.e.~minimal) morphism. The generalisation is proved only in the setting of sorted sets, and therefore also in the setting of normal sets\footnote{Bartek Klin has an alternative proof, which  works in arbitrary categories, but requires some additional assumptions.}.
Fix a category of sorted sets, for some choice of, possibly infinitely many, sort names. A finite sorted set is one which has finitely many elements, in particular it can use only finitely many sorts.

\paragraph*{Finitary algebras.}  If $\monad$ is a monad, then  a $\monad$-algebra $\alg$ is called \emph{finitary} if for every $w \in \monad A$, there is some finite $A_0 \subseteq A$ such that $w \in \monad A_0$. Sometimes, a monad is such that every $\monad$-algebra is finitary, e.g.~this is the case for the monad of finite words $A^+$. 

\begin{theorem}\label{thm:syntactic-morphism}[Syntactic Morphism Theorem] Consider a monad $\monad$ in a category of  sorted sets. Let $f$ be a colouring of an algebra $\alg$, which is recognised by a $\monad$-morphism $h$ into some finitary $\monad$-algebra.  There exists a   surjective $\monad$-morphism into a  $\monad$-algebra
	\begin{align*}
		\synt f : \alg \to \alg_f,
	\end{align*}
	 called the \emph{syntactic morphism of $f$}, 
	 which  recognises $f$ and which  factors through  every surjective $\monad$-morphism recognising~$f$. Furthermore, $\synt f$ is unique up to isomorphisms on  $\alg_f$.
\end{theorem}

Note that if $\alg$ itself is finitary, then $f$ is recognised by the identity $\monad$-morphism on $\alg$.  Therefore, if a monad $\monad$ is such that every $\monad$-algebra is finitary, then  every colouring of a $\monad$-algebra has a syntactic morphism.  This implies that every colouring has a syntactic morphism in monads such as  the monad of finite words that corresponds to monoids, the monad of nonempty finite words that corresponds to semigroups, and several monads for describing finite trees that will be described later in the paper. 
Before proving the theorem, we give an example which  shows how that a syntactic morphism might not exist in general. 

\begin{running}
	Consider the monad of $\infty$-words  and the language 
	\begin{align*}
		L = \set{ a^{n_1} b a^{n_2} b \cdots : \mbox{the sequence $n_i$ is unbounded, i.e.~$\limsup {n_i} = \infty$.}}
	\end{align*}
We will prove that $L$ does not have a syntactic morphism. 
%
%
Consider an equivalence relation $\sim$ on natural numbers such that every equivalence class is finite. For example, $\sim$ could identify all numbers that are between two consecutive powers of two. Define a function 
	\begin{align*}
		h_\sim : \set{a,b}^\infty \to \underbrace{\Nat \cup (\Nat^2 \times \Nat/_\sim) \cup \set{\bot,\top}}_A
	\end{align*}
as follows. If the input is infinite, then $h_\sim$ returns $\bot$ or $\top$ depending on whether the input belongs to $L$. If the input has no $b$'s, then $h_\sim$ returns the length.  Finally, if the input contains at least one $b$, then $h$ returns the triple consisting of:  the number of $a$'s before the first $b$; the number of $a$'s after the last $b$; the equivalence class of the largest $n$ such that the input has an infix $ba^nb$  (or the equivalence class of $0$ if there is no such $n$). One can show that the kernel of $h_\sim$ is a congruence in the natural sense, and therefore $A$ can be equipped with the structure of an $\infty$-algebra which makes $h$ an $\infty$-morphism recognising $L$.
	
	Consider  two equivalence relations $\sim_1$ and $\sim_2$ on natural numbers, such that their transitive closure has infinite equivalence classes, e.g.~$\sim_1$ identifies even numbers with their successors, while $\sim_2$ identifies even numbers with their predecessors. If there were a syntactic morphism $h$, then it would need to factor through both $h_{\sim_1}$ and $h_{\sim_2}$, and therefore it would need to assign the same value to all words in $ba^*b$. By associativity, $h$ would assign the same value to all $\infty$-words with infinitely many $b$'s, and therefore it would not recognise $L$.
\end{running}

The rest of Section~\ref{sec:myhill-nerode} is devoted to proving the Syntactic Morphism Theorem.  

\subsection{Proof of the Syntactic Morphism Theorem}
We are working in a category of sorted sets; fix therefore a set of sort names, and a monad $\monad$.
We first show that  the syntactic morphism, if it exists, is unique up to ismorphisms on the target algebra. This is a consequence of the following lemma.  In the lemma, the crucial distinction is between a function between universes of two $\monad$-algebras, and   such a function which is a $\monad$-morphism, i.e.~one that is consistent with the multiplication in the two algebras.
\begin{lemma}\label{lem:triangular-morphism}
	Let $\monad$ be a monad, 	let $\alg,\balg,\calg$ be $\monad$-algebras,  let
	\begin{align*}
		f : \alg \to \balg \qquad \mbox{and}\qquad g :  \alg \to \calg
	\end{align*}
be  $\monad$-morphisms, with $f$ being surjective, and let $h : B \to C$ be a function  such that, as functions on universes, the following diagram commutes.
	\begin{align*}
		\xymatrix{ A \ar[r]^f \ar[dr]_g & B \ar[d]^h \\ & C}
	\end{align*}
Then $h$ is a $\monad$-morphism.
\end{lemma}
\begin{pr}
	This might be a standard lemma on Eilenberg-Moore algebras, although this proof uses right inverses, and it will therefore not work in every category.
	Consider the following diagram.
	\begin{align*}
		\xymatrix
		{
		& \monad B \ar[dr]^{\mult_\balg}\ar@/_7pc/[dd]_{\monad h}\\
		\monad A \ar[ur]^{\monad f} \ar[dr]_{\monad g} \ar[r]^{\mult_{\alg}} & A \ar[dr]_g \ar[r]^{f} & B\ar[d]^h  \\  & \monad C \ar[r]_{\mult_\calg}   & C}
	\end{align*}
	The right  triangular face (involving $A,B,C$) commutes by assumption of the lemma, and the left triangular face (involving $\monad A$, $\monad B$, $\monad C$) commutes by $\monad$ applied to the assumption of the lemma. The two quadrangular faces commute because $f$ and $g$ are $\monad$-morphisms. Therefore, the entire diagram commutes.
	Because $f$ is surjective, and we are in the category of sorted sets, $f$  has a right inverse, i.e.~a function $f^{-1} : B \to A$  such that $f \circ f^{-1}$ is the identity on $B$. By the previous commuting diagram, the two paths from $\monad A$ to $C$ in the following diagram describe the same function.
   	\begin{align*}
   		\xymatrix{\monad B\ar[r]^{\monad f^{-1}} & \monad A \ar[r]^{\monad f} & \monad B \ar[d]_{\monad h}\ar[r]^{\mult_{\monad B}} & B \ar[d]^h \\ & &  \monad C \ar[r]^{\mult_{\monad C}} & C}
   	\end{align*}
	   Because $\monad$ is a functor, it follows that the path connecting the two copies of $\monad B$ in the above diagram is actually the identity on $\monad B$, and therefore the square face above commutes, which proves that $h$ is a $\monad$-morphism.\end{pr}

\paragraph*{Congruences.} Define a \emph{congruence} in an $\monad$-algebra $\alg$ to be a  surjective function $g : A \to B$ from the universe of $\alg$ to some set such that $g \circ \mult_\alg$ factors through $\monad g$.

\begin{lemma}\label{lem:congruence}
	If $\alg$ is a $\monad$-algebra and $g : A \to B$ is a congruence, then there is a multiplication operation on $B$ which makes $g$ into a $\monad$-morphism.
\end{lemma}
\begin{pr}
	The assumption that $g$ is a congruence says that there is a function, call it $\mult_\balg$, which makes the following diagram commute.
	\begin{align*}
		\xymatrix{
		\monad A \ar[d]_{\mult_\alg} \ar[r]^{\monad g}& \monad B \ar[d]^{\mult_\balg}\\
		A \ar[r]_g& B
		}
	\end{align*}
	To prove the lemma, we need to show that $\mult_\balg$ is associative, which is explained in the following diagram.
	\begin{align*}
		\xymatrix{
		\monad \monad B \ar[ddd]_{\monad \mult_\balg} \ar[rrr]^{\monmul B}  & & & \monad B\ar[ddd]^{\mult_\balg}\\	
		& \monad \monad A \ar[ul]_{\monad \monad g} \ar[d]_{\monad \mult_\alg}\ar[r]^{\monmul A} & \monad A \ar[d]^{\mult_\alg} \ar[ur]^{\monad g}\\
		& \monad A \ar[r]_{\mult_\alg} \ar[dl]^{\monad g}& A \ar[dr]^g\\
		\monad B \ar[rrr]_{\mult_\balg}& & & B
		}
	\end{align*}
	The upper face commutes because $\monad g$ is a $\monad$-morphism between the free algebras $\monad A$ and $\monad B$. The right and lower faces commute by the assumption on $\mult_\balg$, while the left face commutes by $\monad$ applied to this assumption. It follows that all paths that begin in $\monad \monad A$ and end in $B$ denote the same function. Since $g$ is surjective, we can use the same argument as in the end of Lemma~\ref{lem:triangular-morphism} to show that the perimeter of the diagram commutes.
\end{pr}

 Therefore, a congruence is simply a $\monad$-morphism with the algebraic structure on the target being ommitted.
	%
	%

\newcommand{\extpol}[1]{\mathsf{upol}_{#1}}

\paragraph*{Polynomials.} In universal algebra, a polynomial is a term with some constants from the algebra. We  generalise this notion to monads. For a set $X$,  define the set of \emph{polynomials over $\alg$ with variables $X$} to be 
\begin{align*}
	\pol_X \alg \eqdef \monad (A \sqcup X).
\end{align*}
For a valuation $v : X \to A$, we consider the evaluation function
\begin{align*}
	\upeval X {\_} v : \pol_X \alg \to A 
\end{align*}
which first replaces the variables in the argument polynomial by the valuation $v$, and then applies the multiplication in $\alg$. The notion of polynomials makes sense in arbitrary categories, not just those in sorted sets,  but the following definition is specific to the category of sorted sets. Suppose that $p$ is a polynomial over $\alg$ with variables $X$. Define 
\begin{align*}
	\sem p : \alg^X \to \alg 
\end{align*}
to be the function $v \mapsto \sem p (v)$. A problem is that although $\sem p$ is a well-defined function,  it is not a morphism in the category, because it is not necessarily sort preserving. For example, if $X$ has just one variable $x$,  then the function $\sem p$ is sort preserving only when the (output) sort of $p$ is the same as the sort of the variable $x$.   In the language of category theory, this problem is that monads in sorted sets need not be strong. The problem goes away when there is only one sort.

If $h : \alg \to \balg$ is a $\monad$-morphism, and $p \in \pol_X \alg$, then $h(p) \in \pol_X \balg$ is defined by applying $h$ to the constants in $p$ and leaving the variables alone. Formally speaking~$h(p)$ is obtained by applying  $\monad h'$ to $p$, where $h'$ is the disjoint union of $h$ and the identity on the variables.  From the definition of $\monad$-morphism, it follows  that the $\monad$-morphisms commute with polynomials in the sense that the following diagram commutes:
\begin{align}\label{eq:commute-polynomials}
	\xymatrix @C=4pc{
	\alg^X \ar[d]_{h^X} \ar[r]^{\sem p} & \alg \ar[d]^h\\
	\balg^X \ar[r]_{\sem {h(p)}} & \balg
	}
\end{align}
Note that the diagram is not in the category of sorted sets, because the horizontal arrows are not necessarily sort preserving.

\paragraph*{Unary polynomials.} In our proof of the Syntactic Morphism Theorem, special attention is devoted to unary polynomials. In the setting of (unsorted) sets, which covers the well-known versions of the Syntactic Morphism Theorem for monoids or finite automata, the classical construction is to identify elements that cannot be distinguished by unary polynomials. To define unary polynomials in the setting of sorted sets, one needs a little care with the sorts.
%
%
In the following, we assume that the name of each sort is also an element of its own sort.
 For sort names $\tau$ and $\sigma$, a \emph{unary polynomial with input sort $\tau$ and output sort $\sigma$ over $\alg$} is defined to be a polynomial over $\alg$, which has sort $\sigma$, and which  uses just one variable, namely the sort name $\tau$. By abuse of notation, if $\tau$ is a sort name then we write $\upol \tau \alg$ and  $\upeval \tau p a$, respectively, instead of the formally correct $\pol_{\set{\tau}} \alg$ and $\upeval \tau p {\tau \mapsto a}$. 
 In the setting of (unsorted) sets, there is only one sort and unary polynomials can be composed forming a monoid. In the setting of sorted sets, to compose unary polynomials one needs to take care that the output sort of one unary polynomial matches the input sort of the other.


\newcommand{\syna}[1]{f_{#1}}
\paragraph*{Definition of the syntactic morphism.}
Consider a colouring 
\begin{align*}
	f : \alg \to C,
\end{align*}
as in the assumptions of the Syntactic Morphism Theorem.  
\label{page:syntactic-equivalence}
Define an equivalence relation $\sim$ on the universe $\alg$ which identifies $a,b \in A$ if they have the same sort $\tau$ 
and 
\begin{align*}
	f(\upeval \tau p a) = 	f(\upeval \tau p b) \qquad \mbox{for every $p \in \upol \tau \alg$.}
\end{align*}
Define $\syntim$ to be the equivalence classes of $\sim$, and define the syntactic morphism 
\begin{align*}
	\synt f : A \to \syntim
\end{align*}
 to be the function which maps $a$ to its equivalence class under $\sim$.  We will  show that  $\synt f$
is a congruence, and therefore by Lemma~\ref{lem:congruence} there is a multiplication operation on $\syntim$ image which makes $\synt f$ into a surjective $\monad$-morphism. Let us begin by showing that $\sim$ is a congruence with respect to polynomials with finitely many variables, as expressed in the following lemma.

\begin{lemma}\label{lem:n-ary-congruence}
	Let $X$ be a finite set of variables, and let $p \in \pol_X \alg$. If $v_1,v_2 : X \to A$ are valuations then
	\begin{align*}
		\bigwedge_{x \in X} v_1(x) \sim v_2(x) \qquad \mbox{implies} \qquad \sem p (v_1) \sim \sem p (v_2).
	\end{align*}
\end{lemma}
\begin{pr}
	The idea is that the definition of $\sim$ guarantees the lemma for unary polynomials, and then induction extends the result to polynomials of higher finite aritites.
	
	Consider first the case when $X$ has exactly one variable, i.e.~$p$ is a unary polynomial.
	Let $a_1,a_2$ be the values of the valuations $v_1,v_2$ on the unique variable. We need to show that $a_1 \sim a_2$ implies
	\begin{align*}
 \sem p (a_1) \sim \sem p (a_2).
	\end{align*}
	Unraveling the definition of $\sim$, we need to show that 
   	\begin{align*}
  (f \circ \sem q  \circ \sem p) (a_1) = (f \circ \sem q \circ \sem p) (a_2)
   	\end{align*}
	holds for every unary polynomial $q$ whose input sort is the output sort of $p$.  
		Composing the polynomials $q$ and $p$ yields a unary polynomial $r$  such that that $\sem q \circ \sem p$ and $\sem {r}$ describe the same function. 
			By assumption that $a_1,a_2$ are $\sim$-equivalent, they have the same values under $f \circ \sem r$, which proves the above equality, and   completes the proof of the special case of the lemma when $X$ has one variable.
			
			The case when $X$ has more than one variable is proved by a straightforward induction on the size of $X$ as follows. Let then $v_1,v_2 : X \to A$ be as in the assumption of the lemma.  Choose some parition $X = X_1 \cup X_2$ with both $X_i$ being nonempty, and  define $p_i$  for $i \in \set{1,2}$ to be the polynomial obtained from $p$ by substituting the variables from $X_i$ with their values under $v_i$. Then
			\begin{align*}
				\sem p (v_1) = \sem {p_1} (v_1 |_{X_2}) \sim \sem {p_1} (v_2 |_{X_2}) =
				\sem {p_2} (v_2 |_{X_1}) \sim \sem {p_2} (v_2 |_{X_1}) = \sem p (v_2).
			\end{align*}
\end{pr}

Let us restate a special case of the above lemma in terms of commuting diagrams.
\begin{corollary}\label{cor:commuting-polynomials}
	If $X$ is a finite set then
	\begin{align*}
	\begin{tabular}{ m{3cm} m{1cm} m{3cm} }
	$	\xymatrix{
		X \ar[d]_{ v_2} \ar[r]^{{v_1}} &  A \ar[d]^{\synt f} \\
		A \ar[r]_{\synt f} & \syntim
		}
	$ &  implies &
	$
		\xymatrix{
		\monad X \ar[d]_{\monad v_2} \ar[r]^{\monad {v_1}} & \monad A \ar[r]^{\mult_\alg}  & A \ar[dd]^{\synt f} \\ \monad A \ar[d]_{\mult_\alg} \\
		A \ar[rr]_{\synt f} & &\syntim
		}
	$
	 \\
		\end{tabular}
	\end{align*}
\end{corollary}
\begin{pr}
	This is a special case of Lemma~\ref{lem:n-ary-congruence} where the polynomials have no constants in them, i.e.~they are built entirely out of variables.
\end{pr}

\begin{lemma}\label{lem:syntf-factors-through-recognisers}
	If $h : \alg \to \balg$ is a $\monad$-morphism that recognises $f$, then $\synt f$ factors through $h$.
\end{lemma}
\begin{pr}
The value of $\synt f$ for an element of $\alg$ is determined by the values of $f \circ \sem p$ on the element, ranging over all unary polynomials $p$ of appropriate input sort. Therefore, to prove the lemma it suffices to show that $f \circ \sem p$ factors through $h$ for every  unary polynomial $p$.  This is the content of diagram~\eqref{eq:commute-polynomials} and the assumption that $h$ recognises $f$. %
%
\end{pr}

We now resume the proof of the Syntactic Morphism theorem.
Recall the assumption that the coloring $f$ is recognised by a morphism
\begin{align*}
	h : \alg \to \balg
\end{align*}
into a finitary $\monad$-algebra. By  Lemma~\ref{lem:syntf-factors-through-recognisers}, the syntactic morphism factors through $h$, and therefore there is a function $\syntb$ which makes the following diagram commute.
\begin{align*}
	\xymatrix{ \alg \ar[r]^h \ar[dr]_{\synt f} & \balg \ar[d]^{\syntb} \\ & \syntim}
\end{align*}
We will show in the following lemma that $\syntb$  is a congruence on $\balg$. The lemma will complete the proof of the Syntactic Morphism Theorem, because by Lemma~\ref{lem:congruence}, there is a multiplication operation on $\syntim$ which makes it into an algebra $\syntimalg$ such that    $\syntb$ is a $\monad$-morphism.  Therefore, $\synt f$ is a $\monad$-morphism from $\alg$ to $\syntimalg$, as the composition of  $\monad$-morphisms $\syntb$ and $h$.

\begin{lemma}\label{lem:}
$\syntb$ is a congruence in $\balg$.
\end{lemma}
\begin{pr}
	By the assumption that we are in a category of sorted sets, and the assumption that $\syntb$ is surjective, there is a right inverse
	\begin{align*}
		\syntb^{-1} : \syntim \to B,
	\end{align*}
i.e.~a function such that $\syntb \circ \syntb^{-1}$ is the identity on $\syntim$. Define 
\begin{align*}
	i \eqdef \syntb^{-1} \circ \syntb.
\end{align*}
Similarly, let  $h^{-1} : B \to A$ be a right inverse of $h$, i.e.~a function such that $h \circ h^{-1}$ is the identity on $B$. From the definitions of $h^{-1}$ and $i$ we see that the following diagram commutes.
\begin{align}\label{eq:tri-triangular}
	\xymatrix@C=4pc{
	&& B \ar[dd]^{\syntb} \ar[dr]^{h^{-1}}\\
	& B \ar[dr]^{\syntb}\ar[ur]^i & & A\ar[dl]^{\synt f}\\
	A\ar[ur]^{h}\ar[rr]_{\synt f}&& \syntim
	}
\end{align}
Later in the proof, we will use the above commuting diagram to show that that the assumptions of  Corollary~\ref{cor:commuting-polynomials} are satisfied, when the mappings $v_1,v_2$ from the Corollary are the identity and $h^{-1} \circ i \circ h$, restricted to a finite subset of $A$.

	We will prove that $\monad i$ does not affect the value under $		\syntb \circ \mult_\balg$, i.e.~that the following diagram commutes
	\begin{align}\label{eq:right-inverse-pentagon}
		\xymatrix { \monad B\ar[d]_{\mult_\balg} \ar[r]^{\monad i} & \monad B\ar[d]^{\mult_\balg}  \\
		   B\ar[dr]_{\syntb} & B\ar[d]^{\syntb} \\ & \syntim}
	\end{align}
	Before proving that the diagram above commutes, we show how it implies the statement of the lemma. The statement is that $\syntb$ is a congruence, which means that if
	$w,w' \in \monad B$
	have the same image under $\monad \syntb$, then they have the same image under $\syntb \circ \mult_\balg$. By definition $i$ factors through $\syntb$, and therefore if $w,w'$ have the same image under $\monad \syntb$, then they have the same image under $\monad i$, and therefore they have the same image under $\syntb \circ \mult_\balg$ thanks to~\eqref{eq:right-inverse-pentagon}.
	
	To prove~\eqref{eq:right-inverse-pentagon}, we use the assumption that $\balg$	 is finitary, i.e.~every structure in $\monad B$ already belongs to $\monad B_0$ for some finite subset $B_0 \subseteq B$. Therefore, to prove that the above diagram commutes, it suffices to prove that it commutes when the upper left $\monad B$ is replaced by $\monad B_0$ for some finite $B_0$. Let then  $B_0$ be a finite subset of $B$. Define $A_0$ to be the image of $B_0$ under $h^{-1}$, and  define $j : A \to A$ to be the restriction of $h^{-1} \circ i \circ h$ to $A_0$.
	\begin{align*}
		\xymatrix@C=4pc{
		& \monad B_0\ar[d]^{\monad h^{-1}}\\
		& \monad A_0 \ar[ddl]_{\mult_\alg} \ar[rr]^{\monad j} \ar[d]^{\monad h}&  & \monad A \ar[d]^{\monad h} \ar[ddr]^{\mult_\alg}\\
		& \monad B_0 \ar[d]^{\mult_\balg} \ar[r]^{\monad i}& \monad B \ar[ur]^{\monad h^{-1}} \ar[dr]^{\mult_\balg}  & \monad B \ar[d]^{\mult_\balg}\\
		A \ar[r]^h \ar[drr]_{\synt f}& B\ar[dr]^{\syntb} &  & B\ar[dl]_{\syntb} &   A \ar[l]^h \ar[dll]^{\synt f}\\
		& & \syntim
		}
	\end{align*}
	 We claim that all paths that begin in the upper $\monad B_0$ and end in $\syntim$ denote the same function. Thanks to~\eqref{eq:tri-triangular}, the  functions $j$ and the identity on $A_0$ satisfy the assumptions  in Corollary~\ref{cor:commuting-polynomials}. By the Corollary,  the  two paths on the permieter that  begin in $\monad B_0$ and end in $\syntim$ describe the same function.  The two upper quadrangular face commutes by definition of $j$. The face which uses $\monad B$ twice commutes because $h \circ h^{-1}$ is the identity on $B$. The two faces which use $\mult_\alg$ commute because $h$ is a $\monad$-morphism. The two triangular faces commute by definition of $\syntb$. 
	 
	Since $\monad h \circ \monad h^{-1}$ is the identity on $\monad B_0$, we have proved that the diagram~\eqref{eq:right-inverse-pentagon} commutes assuming that the top left corner is replaced by $\monad B_0$; and therefore by the assumption on $\balg$ being finitary we  have proved that the diagram~\eqref{eq:right-inverse-pentagon} commutes in general.%
\end{pr}

\section{Pseudovarieties}
\label{sec:pseudovariety}
This section is dedicated to a monad version of 
Eilenberg's pseudovariety theorem. Eilenberg's theorem says that, in the case of semigroups, language pseudovarieties and algebra pseudovarieties, which will be defined below,  are in bijective correspondence. The theorem implies that if $\langvar$ is a language pseudovariety, then the membership problem $L \in \langvar$ can be decided only by looking at the syntactic semigroup of $L$, and one need not look at the accepting set, nor at the information about which letters are mapped to which elements of the semigroup.  A typical application of the pseudovariety theorem is that definability in  first-order logic, or various fragments thereof, can be determined based only on the syntactic monoid. The theorem does not give an algorithm to determine this, the algorithm needs to be  found in a case-by-case way.

In this section we prove that the pseudovariety theorem works in general for monads when the category is (possibly sorted) sets, with the same proof as in the case of monoids. Surely Eilenberg must have known this, since he invented both the pseudovariety theorem and algebras in abstract monads, but I have not found this result in his book~\cite{0317.94045}. 
Our generalised pseudovariety theorem   subsumes pseudovariety theorems for: finite words in both monoid and semigroup variants~\cite{0317.94045}, $\infty$-words~\cite{DBLP:conf/icalp/Wilke91}, scattered linear orderings~\cite{DBLP:journals/jcss/BedonR12}, finite trees~\cite{DBLP:books/el/treeauto1992/Steinby92}; it also gives pseudovariety theorems for other known settings which have not had their pseudovariety theorems yet, such as forest algebra.


\paragraph*{Algebra pseudovarieties.} The definition of an algebra pseudovariety is a straightforward generalisation of the definition given by Eilenberg for   semigroups or monoids. It is  a class of finite algebras, according to the notion of finiteness given in the setting, which is closed under products, morphic images and subalgebras, as defined below in more detail.

\begin{itemize}
	\item {\bf Products.} A class of $\monad$-algebras is called \emph{closed under  products} if whenever $\alg,\balg$ are in the class, then so is $\alg \times \balg$.
	\item {\bf Morphic images.} A class of $\monad$-algebras is called \emph{closed under morphic images} if whenever  $h : \alg \to \balg$ is a surjective $\monad$-morphism and $\alg$ is in the class,  then so is $\balg$.
	\item {\bf Subalgebras.} A class of $\monad$-algebras is called  \emph{closed under subalgebras} if whenever  $\alg$ is in the class, then every subalgebra of $\alg$ is in the class. A subalgebra of $\alg$ is obtained by restricting the universe to a subset $B$ such that $\mult_\alg$ maps elements of $ \monad B$ to $B$.
	\item{\bf Algebra pseudovariety.} A class of finite $\monad$-algebras is called an \emph{algebra pseudovariety} if it has all three closure properties defined above.
\end{itemize}

\begin{running}\label{run:definite-alg}  Call an $\infty$-algebra $\alg$   \emph{definite} if the multiplication operation
	\begin{align*}
		\mult_\alg : A^\infty \to A
	\end{align*}
	is such that the value of the multiplication depends only on the first $n$ letters of the argument, for some $n$ depending only on the algebra.  	Definite $\infty$-algebras are easily seen to form an algebra pseudovariety.
\end{running}

\paragraph*{Language pseudovarieties.}  Unlike for algebras, the notion of language pseudovariety requires  some discussion. For intuition, let us recall the original notion of language pseudovariety for semigroups that was introduced by Eilenberg. Eilenberg defines a language pseudovariety for semigroups to be a class of recognisable languages of finite words which is closed under Boolean combinations,  inverse images of semigroup morphisms $h : \Sigma^+ \to \Gamma^+$, and derivatives. Here a derivative of a language $L \subseteq \Sigma^+$ is defined to be any  language of the form 
\begin{align*}
	w^{-1}  L  v^{-1} \eqdef \set{ u \in \Sigma^+ : wuv \in L}.
\end{align*}
for some $w,v \in \Sigma^*$. 

It is not immediately obvious how to generalise the notion of derivative to abstract monads. We propose two solutions: one using unary polynomials, which we call a polynomial derivative, and one using syntactic morphisms, which we call a syntactic derivative. The advantage of polynomial derivatives is that they are closer to the derivatives used by Eilenberg, while the advantage of syntactic derivatives is that they make sense in settings without a clear notion of unary polynomials (recall that unary polynomials have only been defined for sorted sets).  The two notions of derivative  lead to different   notions of language pseudovariety, which happen to coincide in settings that use sorted sets. The precise definitions are given below.

\begin{itemize}
	\item {\bf Boolean combinations.} A class of $\monad$-languages is called \emph{closed under Boolean combinations}, if whenever it contains languages $L \subseteq \monad \Sigma$ and $K \subseteq \monad \Gamma$, then it also contains
	\begin{align*}
		L \cap K \qquad L \cup K \qquad \monad \Sigma - L
	\end{align*}
	Of course, in the presence of complementation, only one of $\cup,\cap$  is needed.
	\item {\bf Morphic preimages.}  A class of $\monad$-languages is called \emph{closed under morphic preimages} if whenever the  class contains a language $L \subseteq \monad \Sigma$ and  $h : \monad \Gamma \to \monad \Sigma$ is a $\monad$-morphism with $\Gamma$ being a finite alphabet, then the class also contains $h^{-1}(L)$.
	
	\item {\bf  Polynomial derivatives.}  (This definition assumes that the setting has a notion of unary polynomial, which have only been defined for sorted sets in this paper.) A class of $\monad$-languages is called \emph{closed under polynomial derivatives} if whenever the class contains a language $L \subseteq \monad \Sigma$ and $p$ is a unary polynomial in the $\monad$-algebra $\monad \Sigma$, then the class also contains the language
	\begin{align*}
		p^{-1}L \eqdef \set{w \in \monad \Sigma : \sem{p}(w) \in L}.
	\end{align*} 
	
	\item {\bf Syntactic derivatives.}  A class of $\monad$-languages is called \emph{closed under polynomial derivatives} if whenever the class contains a language, then it also contains all other languages recognised by its syntactic algebra.
	
	\item{\bf Polynomial language pseudovariety.} A \emph{polynomial language pseudovariety} is a class of recognisable $\monad$-languages that is closed under Boolean combinations, morphic preimages, and polynomial derivatives.
	\item{\bf Syntactic language pseudovariety.} A \emph{polynomial language pseudovariety} is a class of recognisable $\monad$-languages that is closed under Boolean combinations, morphic preimages, and syntactic derivatives. (Since complementation is covered by syntactic derivatives, it suffices to have only closure under union and not all Boolean combinations.)
	
\end{itemize}

As usual for pseudovarieties,  a $\monad$-language in the above definitions is formally treated as its characteristic function, which means that a language comes with a description of its domain. The reason for this is that it is sometimes important to know the input alphabet of a language. Before continuing, let us observe the following simple fact.

\begin{fact}\label{fact:pol-is-synt}
	Polynomial derivatives are a special case of syntactic derivatives.
\end{fact}
\begin{pr}	
Thanks to~\eqref{eq:commute-polynomials},	any $\monad$-morphism, not necessarily the syntactic morphism, which recognises $L$ will also recognise every polynomial derivative $p^{-1}L$.
\end{pr}


\begin{running}\label{run:definite-language} Call an $\infty$-language \emph{definite} if there is some $n \in \Nat$ such that membership in the language depends only on the first $n$ letters. Examples of definite $\infty$-languages include: ``words that begin with $a$'', or ``words of length at least two''. Clearly definite $\infty$-languages are closed under Boolean combinations. They are also closed under inverse images of $\infty$-morphisms, because if 
	\begin{align*}
		h : \Sigma^\infty \to \Gamma^\infty
	\end{align*}
	is an $\infty$-morphism, then the first $n$ letters of $h(w)$ are uniquely determined by the first $n$ letters (or less) of $w$. 	 Here  it is important that the monad of $\infty$-words does not allow the empty word; there is a natural variant of the monad which does have the empty word, and in this variant the definite $\infty$-languages do not form a pseudovariety. 
	
  The same argument as for $\infty$-morphisms applies to functions $\Sigma^\infty \to \Sigma^\infty$ defined by unary polynomials.  Therefore, definite $\infty$-languages are closed under polynomial derivatives as well. Summing up, definite $\infty$-languages form a polynomial language pseudovariety. Definite $\infty$-languages also form a syntactic language pseudovariety, but this takes a little more effort to check, and will follow from Corollary~\ref{thm:synt-is-pol}.
\end{running}


\subsection{The Syntactic Pseudovariety Theorem.}
\label{sec:synt-pseudovar}
We have defined two versions of language pseudovarieties, syntactic and polynomial, and therefore there will be two version of Pseudovariety Theorem. In this section we present the version which talks about syntactic varieties. The proof is essentially a monad version of half of Eilenberg's proof, because the definition of syntactic derivative eliminates the other half. A closer similarity with Eilenberg's full theorem is the polynomial version, which is presented in the next section, but which  comes at the cost of restricting to settings that use sorted sets.

For a class $\langvar$ of recognisable $\monad$-languages, define $\makealg \langvar$ to be the  class of finite $\monad$-algebras  which only recognise $\monad$-languages from $\langvar$. For a class $\algvar$ of finite $\monad$-algebras, define $\makelang \algvar$ to be the $\monad$-languages recognised by $\monad$-algebras from~$\algvar$. The Pseudovariety Theorem says that these mappings are mutual bijections when restricted to pseudovarieties.

 \begin{theorem}\label{thm:eilenberg-unsorted}[Syntactic Pseudovariety Theorem] Consider a setting  with the following properties.
 	\begin{itemize}
 		\item Every recognisable language has a syntactic morphism;
 		\item Every finite  algebra is finitely generated, i.e.~its universe has a finite subset $G$ such that multiplication is surjective when restricted to $\monad G$;
 		\item Every  finite algebra $\alg$ has a finite subset of its universe $A_0$ with the following property. If $h : \alg \to \balg$ is a surjective $\monad$-morphism which is injective when restricted to $A_0$, then $h$ is an isomorphism.
 	\end{itemize}
  Then the mapping $\makelang$ is a bijection between algebra pseudovarieties and syntactic language pseudovarieties, and its inverse is $\makealg$.\end{theorem}

The rest of Section~\ref{sec:synt-pseudovar} is devoted to proving the Syntactic Pseudovariety Theorem.
We begin by showing that $\makealg$ and $\makelang$ produce pseudovarieties  when given pseudovarieties (of appropriate types, respectively); actually not all closure properties are needed for this part.
If   $\langvar$  is a syntactic language pseudovariety, then  $\makealg \langvar$ is easily seen to be an algebra pseudovariety. Actually, to prove this, we only need to assumption that $\langvar$ is closed under Boolean combinations. This is because every $\monad$-language recognised by $\alg \times \balg$ is a Boolean combination of $\monad$-languages recognised by $\alg$ and $\balg$. If $\algvar$ is any class of finite $\monad$-algebras, in particular an algebra pseudovariety,  then $\makelang \algvar$ is easily to be a syntactic language pseudovariety.

To finish the  proof of the Syntactic Pseudovariety Theorem, it remains to  show that if $\langvar$ and $\algvar$ are  pseudovarieties of  $\monad$-languages and $\monad$-algebras respectively, then
\begin{align*}
 \makealg \makelang \algvar = \algvar \qquad \mbox{and} \qquad \makelang \makealg \langvar = \langvar.
\end{align*}
By definition, the class $\makelang \makealg \langvar$   consists of $\monad$-languages that are recognised by some finite $\monad$-algebra which only recognises $\monad$-languages from $\langvar$.  Therefore 
\begin{align*}
	\makelang \makealg \langvar \subseteq \langvar.
\end{align*}
For the converse inclusion, consider a language $L \in \langvar$. By assumption on the setting, $L$ has a syntactic algebra, and by definition of language pseudovarieties, every language recognised by this syntactic algebra belongs to $\langvar$. Therefore, $L$ is recognised by some algebra which only recognises languages from $\langvar$. Here we have profited from the definition of syntactic derivatives; with polynomial derivatives this part of the proof will need to be more involved.

More effort is required for the equality
\begin{align*}
	\makealg \makelang \algvar = \algvar.
\end{align*}
By definition, the class $\makealg \makelang \algvar$ consists  of finite $\monad$-algebras $\alg$ such that every finitely sorted $\monad$-language  recognised by $\alg$ is recognised by some $\monad$-algebra from $\algvar$. This gives the right-to-left inclusion. The converse inclusion is proved in the following lemma.
 
\begin{lemma}\label{lem:}
	Let $\alg$  be a finite $\monad$-algebra such that every $\monad$-language recognised by $\alg$ is recognised by some $\monad$-algebra from $\algvar$. Then $\alg \in \algvar$.
\end{lemma}
\begin{pr}
 Let $G$ be a finite generating subset of the universe of $\alg$, i.e.~a subset such that $\mult_\alg$ is surjective when restricted to $\monad G$. By the assumptions on the setting, there is a finite subset $A_0$  of the universe of $\alg$ such that if a surjective $\monad$-morphism $f : \alg \to \balg$ is injective on $A_0$ then it is an isomorphism.  For $a \in A_0$ define 
	 \begin{align*}
	L_a = \set{ w \in \monad G : \mult_\alg (w) = a}.
	 \end{align*}
Let  the syntactic morphism of $L_a$ be
	 \begin{align*}
	 	h_a : \monad G \to \balg_{a}.
	 \end{align*}
The syntactic morphism exists because $L_a$ is recognised by a finite algebra, namely $\alg$, and therefore the Syntactic Morphism Theorem can be applied. Furthermore, by the assumption of the lemma, $L_a$, like any language recognised by $\alg$, is also recognised by some algebra from $\algvar$. Therefore, the  syntactic algebra $\balg_a$ is  an image of some algebra in $\algvar$, and therefore itself belongs to $\algvar$ by closure of algebra pseudovarieties under morphic images. Using the definition of syntactic morphism again, the syntactic morphism $h_a$ must factor through $\mult_\alg$. Summing up, $\balg_a \in \algvar$ and there is a surjective morphism $f_a$ which makes the  following diagram commute.
	 \begin{align*}
	 	\xymatrix{\monad{G} \ar[r]^{\mult_\alg} \ar[dr]_{h_a}  & \alg \ar[d]^{f_a}  \\  & \balg_a }
	 \end{align*}
Define $h$ to be  the product of the morphisms $h_a$ ranging over $a \in A_0$, which is surjective onto its image
	 \begin{align*}
	 	h:  \monad G \to \balg \subseteq  \prod_{a \in A_0} \balg_{a}.
	 \end{align*}
	 The algebra $\balg$ belongs to $\algvar$, by closure of $\algvar$ under finite products and subalgebras.  Defining $f$ to be the product of all $f_a$, we see that the following diagram commutes.
	 \begin{align*}
	 	\xymatrix{\monad{G} \ar[r]^{\mult_\alg} \ar[dr]_{h}  & \alg \ar[d]^{f}  \\  & \balg }
	 \end{align*}
To prove that $f$ is actually an isomoprhism, it suffices to show that $f$ is inective when restricted to $A_0$. Because $G$ are generators, every element  $ a \in A_0$ is the image under $\mult_\alg$ of some $w_a \in \monad G$. Furthermore, if $a \neq b$, then $h(w_a) \neq h(w_b)$, because only one of $w_a,w_b$ belongs to the language $L_a$ that is recognised by $h$.
\end{pr}
 
 \subsection{The Polynomial Pseudovariety Theorem}
 \label{sec:poly-pseudovariety}
In this section, we prove that if the setting uses sorted sets with finitely many sorts, then both the syntactic and polynomial versions of language pseudovariety coincide. The key result is Lemma~\ref{lem:pol-is-synt} below, which says that syntactic derivatives can be represented as inverse morphic images of Boolean combinations of polynomial derivatives. This lemma is essentially the other half of Eilenberg's proof, which was not used in the syntactic version of the pseudovariety theorem.  
We state Lemma~\ref{lem:pol-is-synt} in the more general setting with possibly infinitely many sorts. In such a setting, call a language finitely sorted if on all but finitely many sorts it is full or empty. When there are finitely many sorts, then all languages are finitely sorted.

\begin{lemma}\label{lem:pol-is-synt} Consider a setting where  the category  is sorted sets, with possibly infinitely many sorts, and that the notion of finite algebra is such that finiteness of an algebra implies that the universe is finite on every sort.
		Then for every  recognisable $\monad$-language $L$,  every finitely sorted  syntactic derivative of $L$ is an inverse image, under some $\monad$-morphism, of a Boolean combination of polynomial derivatives of $L$.
\end{lemma}

Before proving the above lemma, let us note two corollaries, and an example of a setting where the conclusion of the lemma is violated.

\begin{corollary}\label{thm:synt-is-pol}
	Consider a setting where the category is sorted sets with finitely many sorts, finite alphabets are finite sorted sets, and  the notion of finite algebra is such that finiteness of an algebra implies that the universe is  finite.   Then syntactic language pseudovarieties are the same thing as polynomial language pseudovarieties.
\end{corollary}
\begin{pr}
	Lemma~\ref{lem:pol-is-synt} implies that every polynomial language pseudovariety is closed under syntactic derivatives, and is therefore a syntactic language pseudovariety.  The converse is Fact~\ref{fact:pol-is-synt}, which says that closure under syntactic derivatives implies closure under polynomial derivatives.
\end{pr}

Actually, the above corollary would also be true with infinitely many sorts, with a modified definition of language pseudovariety where only finitely sorted languages are allowed.
By combining the above corollary with the Syntactic Pseudovariety Theorem, we get the Polynomial Pseudovariety Theorem stated below.
\begin{corollary}\label{thm:eilenberg-unsorted}[Polynomial Pseudovariety Theorem] Under  assumptions  on the setting as in Corollary~\ref{thm:synt-is-pol},  $\makelang$ is a bijection between $\monad$-algebra pseudovarieties and polynomial $\monad$-language pseudovarieties, and its inverse is $\makealg$.
\end{corollary}

An advantage of the polynomial version of the pseudovariety theorem is that it is sometimes easier to check if a class is closed under polynomial derivatives, as compared to syntactic derivatives. This was the case for definite $\infty$-languages discussed previously in the running example.
\begin{running}\label{run:definite-pseudovariety}
	As an illustration  of the Polynomial Pseudovariety Theorem, it is easy to see that   $\makealg$ takes the class of definite $\infty$-languages to the class of definite $\infty$-algebras, and the mapping $\makelang$ goes the other way. An $\infty$-language is definite if and only if its syntactic $\infty$-algebra is definite.
\end{running}

Here is  an example of a setting which violates the conclusions of the Polynomial Pseudovariety Theorem, and therefore also the conclusions of Lemma~\ref{lem:pol-is-synt}.

 \begin{myexample}
	 In the proof of the Lemma~\ref{lem:pol-is-synt}, we will use the following property of the category of sorted sets: if $g : X \to Y$ is surjective, then there is an inverse $g^{-1} : Y \to X$ such that $g \circ g^{-1}$ is the identity on $Y$. An example of a category where this assumption fails is nominal sets. In nominal sets,   the Polynomial Pseudovariety Theorem  also fails, as we show in this example.  The example assumes familiarity with nominal sets, and orbit-finite sets.
	
	Consider the category where objects are finitely supported  nominal sets  and morphisms are finitely supported  functions. Consider the monad of finite words in this category, where algebras are finitely supported semigroups.  To complete the definition of the setting,  define finite alphabets to be finitely supported sets which are orbit-finite, and define finite algebras to be finitely supported orbit-finite semigroups. This  setting was studied in~\cite{DBLP:journals/mst/Bojanczyk13}, although not using the monad terminology. The Syntactic Morphism Theorem holds in this setting, as was shown in Lemmas 3.3 and 3.4 of~\cite{DBLP:journals/mst/Bojanczyk13}. We will show that the Pseudovariety Theorem fails in this setting.
	
\newcommand{\twopower}[1]{\mathsf P_2 #1}
\newcommand{\atoms}{\mathbb A}
Here is the  property of finitely supported functions that  will make the Polynomial Pseudovariety Theorem fail. Let $\atoms$ denote the atoms underlying the  nominal sets, let $\twopower \atoms$ be size two sets of atoms, i.e.~unordered pairs of atoms. One can show that  if 
\begin{align*}
	f : \twopower \atoms \to \atoms^+
\end{align*}
is a  function supported by a finite set $S$ of atoms,
	\begin{align}\label{eq:not-in-support}
		f(\set{a,b}) = f(\set{c,d}) \qquad \mbox{for }a,b,c,d \not \in S.
	\end{align}

Define $\langvar$ to be the polynomial  language pseudovariety generated by all recognisable languages over the alphabet $\atoms$. It is not difficult to see that  a language $L \subseteq \Sigma^+$ belongs to $\langvar$ if and only if there is a finitely supported monoid morphism
\begin{align*}
	h : \Sigma^+ \to \atoms^+ 
\end{align*}
such that $L$ is an inverse image under $h$ of some recongisable subset of $\atoms^+$.  We will show that $\makealg \langvar$ is not an algebra pseudovariety, because it is not closed under morphic images.

Consider the following two languages.
\begin{enumerate}
	\item The alphabet is  ordered pairs of atoms, i.e.~ $\atoms^2$. The language  consists of two letter words  over this alphabet such that the two atoms which appear in the first letter are pairwise distinct from the two atoms that appear in the second letter. In other words, this language is
	\begin{align*}
		L_1 = \set{(a,b)(c,d) : \set{a,b} \cap \set{c,d} = \emptyset} \subseteq (\atoms^2)^+
	\end{align*}
	This language is recognised by a semigroup, call it $\salg_1$, whose  universe is
		\begin{align*}
			\atoms^2 \cup \set{\top,\bot},
		\end{align*}
	 with elements of $\atoms^2$ describing one letter words, with  $\top$ describing words in the language, and with  $\bot$  describing  words of length at least two that are outside the language. Although $\salg_1$ recognises the language $L_1$, it is not its syntactic semigroup. To get the syntactic semigroup, one needs to identify ordered pairs that correspond to the same set, i.e.~the syntactic semigroup, call it $\salg_2$, has universe
	\begin{align*}
		\twopower \atoms \cup \set{\top,\bot}.
	\end{align*}
Clearly $\salg_2$ is an image of $\salg_1$ under a finitely supported semigroup morphism, namely the function which forgets the order in pairs. Therefore, any algebra pseudovariety with $\salg_1$ will also contain $\salg_2$.
	\item Here is a language that is recognised by $\salg_2$. The alphabet is unordered pairs  of atoms, i.e.~$\twopower \atoms$. The language  consists of two letter words  over this alphabet such that the set in the first letter is disjoint with the set in the second letter. In other words, this language is
	\begin{align*}
		L_2 = \set{\set{a,b} \set{c,d} : \set{a,b} \cap \set{c,d} = \emptyset} \subseteq (\twopower \atoms)^+
	\end{align*}
	
\end{enumerate}

 We claim that  $\makealg \langvar$ contains $\salg_1$ but not $\salg_2$, and is therefore not an algebra pseudovariety.  It is not difficult to show that $\salg_1$ recognises only languages from $\langvar$, and therefore it belongs to $\makealg \langvar$. We only show that $L_2$ is not in $\langvar$, and therefore $\salg_2$  is not in $\makealg \langvar$.  To this end, we need to show that there is no finitely supported semigroup morphism
	\begin{align*}
		h : (\twopower \atoms)^+ \to \atoms^+
	\end{align*}
	such that $L_2$ is an inverse image of some recognisable subset of $\atoms^+$. Indeed, by~\eqref{eq:not-in-support}, the function $h$ would need to assign the same value to two different letters in $\twopower \atoms$, and therefore it could not recognise $L_2$.
 \end{myexample}

The rest of this section is devoted to proving Lemma~\ref{lem:pol-is-synt}.
  
  \begin{lemma}\label{lem:derivative-kinds}
Assume the assumptions of Lemma~\ref{lem:pol-is-synt}.
  	Let  $L$ be a recognisable  $\monad$-language. Every  finitely sorted language recognised by the syntactic morphism of $L$ is  is a Boolean combination of polynomial derivatives of $L$.
  \end{lemma}
  \begin{pr}
  		Let  $L \subseteq \monad \Sigma$ be a recognisable $\monad$-language.  
  %
  %
  %
  %
   Let  $K$ be   a finitely sorted language recognised by the syntactic morphism of $L$, in particular the $K$ is also a subset of $\monad \Sigma$. We want to show that $K$ is a Boolean combination of derivatives of $L$. A finitely sorted language is a finite union of single-sorted languages, and therefore without loss of generality, we can assume that $K$ entirely included in a single sort, call it $\tau$. 	
  %
  %
  %
  %
  		 \begin{claim}
  		 	There is a finite set $P \subseteq \upol \tau {\monad \Sigma}$ such that structures $w,w' \in \monad \Sigma$ of sort $\tau$ have the same image under the syntactic morphism of $L$ if and only if 
  			\begin{align}\label{eq:mn-poly}
   p(w) \in L \quad \mbox{iff}\quad p(w') \in L \qquad \mbox{for every $p \in P$.}
  			\end{align}
  		 \end{claim}
  		 \begin{pr}
  			 By construction of the syntactic morphism in the proof of the Syntactic Morphism Theorem, structures $w,w'$ have the same image under the syntactic morphism  if and only if~\eqref{eq:mn-poly} holds for every polynomial $p \in \upol \tau {\monad \Sigma}$, not necessarily from some finite set $P$.   In other words, one can choose for every $w,w'$ a polynomial $p_{w,w'}$ such that $w$ and $w'$ have the same image under the syntactic morphism if and only if 
			 \begin{align*}
			 \sem{p_{w,w'}}(w) \in L \qquad \mbox{iff} \qquad \sem{p_{w,w'}}(w') \in L.
			 \end{align*}
			 Furthermore, because the syntactic morphism recognises $L$, the choice of  $p_{w,w'}$ need  need only depend on the  images of $w$ and $w'$ under the syntactic morphism, for which there are finitely many possibilities.
  	\end{pr}
	
  Stated differently, the claim says that structures   in  sort $\tau$ have the same image under the syntactic morphism if and only if they belong to the same polynomial  derivatives $p^{-1}L$ for $p$ belonging to the finite set $P$ in the statement of the claim. This means that $K$, being a subset of sort $\tau$ that is recognised by the syntactic morphism, is a Boolean combination of finitely many  derivatives.
  \end{pr}
  
 \begin{lemma}\label{lem:derivative-preimage}
	 Let $\Gamma$ be a set and let $f : \alg \to \balg$  and $h : \monad \Sigma \to \balg$  be $\monad$-morphisms. If $f$ is surjective, then there is 
 some $\monad$-morphism $g$ which makes the following diagram commute
	 	\begin{align*}
	 		\xymatrix{ \monad \Sigma \ar[dr]_h \ar[r]^g & \alg \ar[d]^{f} \\ & \balg}
	 	\end{align*}
 \end{lemma}
 \begin{pr}
	 Because $f$ is surjective, there is a function $g'$ which makes the following diagram commute.
 	 \begin{align*}
 		 \xymatrix@C=4pc{
 		\alg \ar[dr]_{f} & \Sigma \ar[l]_{g'} \ar[r]^{\monun \Sigma} & \monad \Sigma \ar[dl]^h \\
 		& \balg
 		}
 	 \end{align*}
 	 Consider the following diagram.
 	 \begin{align*}
 		 \xymatrix@C=4pc{
		 & \monad A  \ar[ddl]_{\mult_\alg}& \\ 
		 & A \ar[u]^{\monun A} &
		 \\
 		\alg \ar[dr]_{f} & \Sigma \ar[u]^{g'} \ar[l]_{g'} \ar[r]^{\monun \Sigma} & \monad \Sigma \ar[dl]^h \ar[uul]_{\monad g'} \\ 
 		& \balg
 		}
 	 \end{align*}
 	 By definition of $g'$, the lower face commutes. The upper left face commutes because multiplication in an algebra must maps units to themselves, while the upper right face comes from the assumption that the unit in a monad is a natural transformation. All arrows on the perimeter of the diamond-shaped diagram describe $\monad$-morphisms.  Therefore, both paths which begin with the edge  $\monun \Sigma$ and end in $\balg$ describe the same function.  Because $\monad \Sigma$ is generated by the units of $\Sigma$, and both paths from $\monad \Sigma$ to $\balg$ are  (compositions of) $\monad$-morphisms, it follows that both paths from $\monad \Sigma$ to $\balg$ describe the same $\monad$-morphism.  Therefore, $g$ in the statement of the lemma can be taken to be $\mult_\alg \circ \monad g'$.
  \end{pr}

\begin{pr}[of Lemma~\ref{lem:pol-is-synt}] The lemma says that  if $L \subseteq \monad \Gamma$ is a recognisable $\monad$-language, then every finitely sorted  syntactic derivative of $L$ is an inverse image, under some $\monad$-morphism, of a Boolean combination of polynomial derivatives of $L$. Let the syntactic morphism of $L$ be
	\begin{align*}
		f : \monad \Gamma \to \balg.
	\end{align*}
Suppose that $K \subseteq \monad \Sigma$ is  a finitely sorted syntactic derivative of $L$, i.e.~it is recognised by some $\monad$-morphism 
\begin{align*}
	 h : \monad \Sigma \to \balg.
\end{align*}
By Lemma~\ref{lem:derivative-preimage}, there is a $\monad$-morphism $g$ which makes the following diagram commute.
	 	\begin{align*}
	 		\xymatrix{ \monad \Sigma \ar[dr]_h \ar[r]^g & \monad \Gamma \ar[d]^{f} \\ & \balg}
	 	\end{align*}
In other words, $K$ is an inverse image, under $g$, of some language $M$ recognised by the syntactic morphism $f$. We can assume without loss of generality that $M$ is empty (respectively, full) on sorts where $K$ is empty (respectively, full), and therefore $M$ is also finitely sorted. By Lemma~\ref{lem:derivative-kinds}, $M$   is a Boolean combination of polynomial derivatives of $L$.
\end{pr}

This completes the proof of the Polynomial Pseudovariety Theorem.

\section{Representing an algebra}
\label{sec:representing-an-algebra}
In all interesting cases, the monad $\monad$ produces infinite sets, even on finite arguments. Therefore, the finiteness of the universe of a $\monad$-algebra $\alg$ does not, on its own, imply that the algebra itself has  a finite representation, because one needs some way of representing the algebra's multiplication operation 
\begin{align*}
	\mult_\alg : \monad A \to A.
\end{align*}
 In this section, we present one such way. We assume that the monad is in the category of sets, or sorted sets. The idea is to find a function  $\repr$,  which chooses for every finite set $A$ a finite subset $\repr A \subseteq \monad A$ such that:
\begin{enumerate}
	\item for every finite $\monad$-algebra $\alg$ with universe $A$, the multiplication operation is uniquely determined by its values on $\repr A$;
	\item the function $A \mapsto \repr A$ can be computed, modulo some representation of elements in $\repr A \subseteq \monad A$.\end{enumerate}
For instance, in the monad of finite words,  the function $\repr$ maps a set $A$ to word over $A$ of length two, because  a semigroup is uniquely determined by its neutral element and its binary multiplication. In the example of $\infty$-algebras, the function $\repr$ maps $A$ to words over $A$ of length  two and to infinite words of the form $a^\omega$ for some $a \in A$. We now describe these notions in more detail.

\paragraph*{Subfunctors.}  Because the monad is in the category of sets, or sorted sets,  the notion of subset can be used.  Define a \emph{subfunctor} of a monad $\monad$  to be a mapping which takes every  set $X$ to a subset $\repr X \subseteq \monad X$. A subfunctor on its own is not a monad (as defined here it is not even a functor), however it can be used to generate a monad as follows.
 For an ordinal number $\alpha$, define $\repr^\alpha X \subseteq \monad X$ as follows by transfinite induction: $\repr^0 X$ is the units of $X$, while for $\alpha > 0$ we have
\begin{eqnarray*}
	\repr^{\alpha} X  \eqdef \bigcup_{\beta < \alpha}\mult_{\monad X} \repr \repr^{\beta} X.
\end{eqnarray*}
By monotonicity, this sequence  must stabilise at some value, which is denoted by $\repr^* X$.
If the monad is finitary, i.e.~every element $w \in \monad X$ belongs to $w \in \monad Y$ for some finite $Y \subseteq X$, then the sequence stabilises at $\omega$, i.e.~induction only on natural numbers is needed.
 It is not difficult to show that  $\repr^*$ is a submonad of $\monad$, i.e.~a subfunctor with the  monad structure inherited from $\monad$. A subfunctor $\repr$ is said to \emph{span} an algebra~$\alg$ 
if \begin{align*}
	\mult_\alg \repr^* X = \mult_\alg \monad X 
\end{align*}
holds for every  subset $X$ of the universe. A subfunctor is called \emph{complete} if it spans every $\monad$-algebra, and  \emph{finitely complete} if it spans every finite $\monad$-algebra; note how this depends on the notion of finite $\monad$-algebra.

\begin{running}\label{ex:omega-monad-represented}
	Consider the monad $\infty$ for infinite words. Define 
	\begin{align*}
		\repr X \eqdef \set{  xy, x^\omega : x,y \in X}.
	\end{align*} 
It is not difficult to check that the submonad $\repr^*$ maps $X$ to  the finite and ultimately periodic words over alphabet $X$. Using the Ramsey Theorem, in the same way as it is used explicitly by Wilke in~\cite{DBLP:conf/icalp/Wilke91}, and implicitly by B\"uchi in~\cite{buchi_decision}, we show that $\repr$ is finitely complete. Indeed, let $X$ be a subset of the universe in some finite $\infty$-algebra $\alg$. To show that $\repr$ spans $\alg$, we need to show that if $w \in X^\infty$, then there is some ultimately periodic word $v$ over $X$ such that
	\begin{align*}
		\mult_\alg(w) = \mult_\alg(v).
	\end{align*}
If $w$ is finite, then it already is ultimately periodic. Otherwise, using the Ramsey Theorem, one can decompose $w$ as 
\begin{align*}
	w = w_0 w_1 w_2 \cdots \qquad \mbox{with }w_0,w_1,\ldots \in X^+
\end{align*}
such that $\mult_\alg$ gives the same result for all the finite words $w_1,w_2,\ldots$. Let $a_i$ be the image of $w_i$ under $\mult_\alg$. By assumption that all $a_i$ are the same for $i \ge 1$ and by associativity, we have
\begin{align*}
	\mult_\alg (w) = \mult_\alg(a_0 a_1 a_1 \cdots)= \mult_\alg(w_0 (w_1)^\omega),
\end{align*}
and the latter uses an ultimately periodic word. As we shall see, the argument made in this example is the only part of the proof of decidability of \mso on $\infty$-words that needs to be proved by hand; the remainder of the proof will follow from abstract principles stated in Theorem~\ref{thm:decide-mso}.
\end{running}

\paragraph*{Reducts.}  Consider a subfunctor $\repr$ that is finitely complete for a monad $\monad$.   For a finite $\monad$-algebra $\alg$, define its \emph{$\repr$-reduct}  to be the pair consisting of the universe $A$ of $\alg$, and the restriction of the multiplication operation from $\alg$ to the subfunctor:
 \begin{align*}
 \mult_{\alg}|_{\repr A} :	\repr A  \to A
 \end{align*}
 The $\repr$-reduct is a special case of what category theorists call an \emph{algebra over signature $\repr$.}  Straight from the definition it follows that if   $\repr$ spans $\alg$, then $\alg$ is uniquely determined by its $\repr$-reduct. In particular, if $\repr$ is complete, then every  algebra over signature $\repr$ extends to at most one  $\monad$-algebra. Note the ``at most one'' in the previous sentence; some algebras over signature $\repr$ might not extend to $\monad$-algebras, e.g.~not every binary operation extends to a semigroup operation, because for this associativity is needed. The same holds for finite completeness and finite algebras.  The point of using $\repr$-reducts is that sometimes $\repr$ can be chosen so that it preserves finiteness, and therefore $\repr$-reducts can be manipulated by algorithms, at least as long as finite objects and functions between them can be manipulated by algorithms.
 
 \begin{running}\label{example:wilke-semigroup} The $\repr$-reduct of a finite $\infty$-algebra consists of a finite universe $A$ together with two operations, of aritites two and one:
	\begin{align*}
 	\_ \cdot \_ : A \times A \to A \qquad \_^\omega : A \to A.
	\end{align*}
	This is essentially the same thing as a Wilke semigroup.
	Not every choice of finite universe and three operations above will yield an $\repr$-representation of some finite $\infty$-algebra; this requires the operations to satisfy certain axioms, e.g.~Wilke gives such axioms in Definition 3 of~\cite{wilke_algebraic}.
 \end{running}

\paragraph*{Computing the syntactic $\monad$-morphism.} The point of $\repr$-reducts is to have a finite representation of $\monad$-algebras so that they can be manipulated by algorithms. We give one example of such an algorithm, namely the Moore\footnote{This is not the same Moore as in Eilenberg-Moore algebras.}  algorithm. This algorithm computes the syntactic morphisms in polynomial time. To state this result, we need to explain how morphisms are represented.  Consider a subfunctor $\repr$. We assume that it is \emph{effective}, in  the sense that $\repr \Sigma$ can be computed up to isomorphism based on $\Sigma$ for finite $\Sigma$, in particular $\repr$ preserves finiteness. The $\repr$-representation of a finite $\monad$-algebra is simply the finite  multiplication table that gives the values of $\mult_\alg$ for arguments from $\repr A$. The $\repr$-representation of a $\monad$-morphism $h : \monad \Sigma \to \alg$ consists of the $\repr$-representation of the algebra, as well as the values of $h$ for units.  If $\repr$ has polynomial size increase, as is the case in the examples of monoids or  $\infty$-algebras discussed in Examples~\ref{ex:omega-monad-represented} and~\ref{example:wilke-semigroup}, then the $\repr$-representation of an algebra will be of size polynomial with respect to the size of the universe. However, there will be examples where $\repr$ has  exponential size increase, e.g.~in Section~\ref{sec:chain} in the case of countable chains. 
\begin{lemma}\label{lem:moore}
	Let $\monad$ be a monad in a category of sorted sets, with finitely many sorts, and let $\repr$ be a subfunctor that is complete for finite algebras. Then syntactic $\monad$-morphisms can be computed for $\monad$-recognisable languages, in polynomial time with respect to $\repr$-representation.
\end{lemma}
\begin{pr}
	Using the Moore algorithm.	%
	%
\end{pr}

\section{Monadic second-order logic}
\label{sec:mso}
An important part of the theory of regular languages is the connection between recognisability and definability in monadic second-order logic \mso.  This connection says that languages recognised by finite recognisers are the same thing as \mso definable languages. Examples where this connection holds include: finite words (as proved independently by B\"uchi, Elgot and Trakhtenbrot), infinite words (as proved by B\"uchi), finite trees (as proved by Thatcher and Wright), infinite trees (as proved by Rabin), etc.
There are common ingredients in all of the proofs, and there are parts that are specific to each domain. In this section, we show that the common ingredients  can be stated and proved on the abstract level  of monads. This takes care of much of the symbol pushing in the proofs, and leaves only the combinatorial parts to be proved in each specific case, e.g.~nothing is left to be proved for finite words or trees, or only the Ramsey theorem needs to be applied in the case of $\infty$-words.


\subsection{Language theoretic definition of  \mso}
\label{sec:abstract-mso}

To establish the connection between \mso and recognisability, consider the following lemma, see~\cite{thomas_languages}, which characterises \mso in a way that does not talk about  ``positions'' or ``sets of positions'' of a structure, but is defined in purely language theoretic terms. 
\begin{lemma}\label{lem:closure-properties}
	A language $L \subseteq \Sigma^*$ is definable in \mso if and only if it belongs to the least class of languages that is closed under Boolean combinations, images and inverse images of morphisms $h : \Sigma^* \to \Gamma^*$, and which contains the languages
	\begin{align*}
		0^* \subseteq \set{0,1}^* \qquad \mbox{and} \qquad 0^*1^* \subseteq \set{0,1}^*.
	\end{align*}
\end{lemma}

A similar lemma holds for infinite words (instead of $0^*1^*$ one uses $0^*1^\infty$), and also for finite and infinite trees, etc.
Motivated by the above, we define an abstract notion of \mso in a monad~$\monad$. In the abstract version, predicates are modelled by  languages.  For a set $\Ll$ of  $\monad$-languages, define \msol{\Ll} to be the smallest class of $\monad$-languages which contains $\Ll$, is closed under Booolean operations, images and inverse images of $\monad$-morphisms.


The following lemma is in the category of sets, or more generally, in categories which have a powerset functor that preserves finiteness. A non-example is the category of nominal sets with orbit-finite sets, where powerset does not preserve orbit-finiteness, and also \mso contains non-recognisable languages, see~\cite{DBLP:journals/mst/Bojanczyk13}.
\begin{lemma}\label{lem:mso-recognisable}
	If $\Ll$ contains only $\monad$-recognisable $\monad$-languages, then so does \msol{\Ll}.
\end{lemma}
\begin{pr}
	To prove the lemma, one needs to show that 	$\monad$-recognisable languages are closed under Boolean operations, images of $\monad$-morphisms, inverse images of $\monad$-morphisms.
	For Boolean operations we use products, for inverse images the property is immediate. The only nontrivial part is the images, where we use the powerset construction, defined as follows. We write $\powerset X$ for the powerset of $X$. If $X$ is a set, then we say that $w \in \monad X$ belongs pointwise to $v \in \monad \powerset X$ if there is some element of 
	\begin{align*}
 \monad \set{(a \in X,b \in \powerset X) : a \in b}
	\end{align*} which projects to $w$ and $v$ respectively on the first and second coordinates.
For a $\monad$-algebra $\alg$, define its powerset to be the $\monad$-algebra
\begin{align*}
	\powerset \alg : \monad \powerset A \to \powerset A
\end{align*}
 whose multiplication operation maps $w \in \monad \powerset A$ to the set 
\begin{align*}
	\set{ \mult_\alg (v) : \mbox{$v \in \monad A$ belongs pointwise to $w$}}.
\end{align*}
It is not difficult to check that this is indeed a $\monad$-algebra, for the distrustful see Johnstone~\cite{}. 
\end{pr}

\subsection{Deciding satisfiability of \mso}
For a monad $\monad$, we define   \emph{\mso satisfiability over $\monad$} to be the following decision problem. An instance is what one can see as an \mso formula, which is formalised as an expression that uses the constructors of \mso formulas, with the predicates being represented by $\monad$-morphisms recognising them. The question is whether the language corresponding to the instance is nonempty.

In this section we give a sufficient criterion for the decidability of \mso satisfiability.
We assume that the monad is in the setting of finitely sorted sets.

\paragraph*{Strongly effective subfunctor.} Recall the notion of an effective subfunctor $\repr$  from Section~\ref{sec:representing-an-algebra}, which said that if $\Sigma$ is finite then  $\repr \Sigma$ is also finite and can be computed based on $\Sigma$.  As discussed in Section~\ref{sec:representing-an-algebra}, if a monad $\monad$ has an effective subfunctor $\repr$ that is finitely complete, then a finite $\monad$-algebra can be represented by its multiplication table restricted to $\repr$, while a $\monad$-morphism
\begin{align*}
	h : \monad \Sigma \to \alg
\end{align*} 
where $\Sigma$ is a finite alphabet and $\alg$ is finite can be  represented by its values on generators, i.e.~units of $\Sigma$. For the results on \mso of this section, we will need a stronger assumption, which says that algebras recognising singleton sets can be computed. A subfunctor $\repr$ is called \emph{strongly effective} if for every finite set $\Sigma$ and every    $w \in \repr \Sigma$, one can compute a representation of a $\monad$-morphism 
	\begin{align*}
		h : \monad \Sigma \to \alg
	\end{align*}
into a finite $\monad$-algebra	 that recognises $\set w$.

\begin{myexample}\label{ex:wilke-is-effective}
	Consider the monad $\infty$ of infinite words, and the  subfunctor 
	\begin{align*}
		\repr X \eqdef \set{\epsilon, xy, x^\omega : x,y \in X}.
	\end{align*} 
which was	 considered in Examples~\ref{ex:omega-monad-represented} and~\ref{example:wilke-semigroup}, and proved to be finitely complete. We claim that $\repr$ is strongly effective. Clearly $\repr$ preserves finiteness and can be computed, as $\repr X$  is isomoprhic to $1 \sqcup X^2 \sqcup X$. For a finite alphabet $\Sigma$ and  $a,b \in \Sigma$ it is not difficult to compute $\repr$-reducts of  $\infty$-algebras that recognise the languages $\set{ab}$ and $\set{a^\omega}$. Let us do the case of $\set{a^\omega}$. The $\infty$-algebra  has four elements in its universe, representing the empty word, finite words in $a^+$,  the unique infinite word $a^\omega$, and finally words that use some letter other than $a$. \end{myexample}

The following theorem shows that a sufficient crieterion for decidable \mso satisfiability is having a subfunctor that is finitely complete and strongly effective. 

\begin{theorem}\label{thm:decide-mso}   Let $\monad$ be a monad in the setting of finitely sorted sets. If there is a subfunctor $\repr$  that is strongly effective and finitely complete, then \mso satisfiability  is decidable.
\end{theorem}

As mentioned at the beginning of this section, 
Theorem~\ref{thm:decide-mso} is abstract nonsense in the sense  that it does not resolve the  actual combinatorics necessary to prove satisfiability of \mso.  This can be seen in the series of Examples~\ref{ex:omega-monad-represented},~\ref{example:wilke-semigroup} and~\ref{ex:wilke-is-effective}, which show that the monad of infinite words has a subfunctor that is finitely complete and effective, and therefore Theorem~\ref{thm:decide-mso} can be invoked to show that satisfiability of \mso is decidable over infinite words. The decidability proof that comes from  these examples has the same structure as the    original proof of B\"uchi~\cite{buchi_decision}, or its algebraic version in~\cite{wilke_algebraic}.  What the examples show is that a large part of the proof is sufficiently generic to be stated on the abstract level of monads; and the only  challenge is finding a subfunctor that is finitely complete and strongly effective, with finite completeness being essential part.

Theorem~\ref{thm:decide-mso} follows immediately from the following lemma.
\begin{lemma}\label{lem:} 
From multiplication tables of $\repr$-reducts of a finite  $\monad$-algebras $\alg,\balg$, one can compute multiplication tables of the $\repr$-reducts of $\powerset \alg$ and $\alg \times \balg$.
\end{lemma}
\begin{pr}
The Cartesian product is immediate, the interesting case is the powerset $\powerset \alg$. For $ w \in \repr (\powerset A)$, we need to compute $\mult_{\powerset \alg} (w).$ By strong effectivity of $\repr$, we can compute a $\monad$-morphism
	\begin{align*}
		h : \monad (\powerset A) \to \balg
	\end{align*}
	that recognises the singleton $\set w$.
Define $\Sigma$ to be the finite set of pairs $(a,A_0)$ such that $a \in A_0 \subseteq A$ and consider the $\monad$-morphism
	\begin{align*}
		g: \monad \Sigma \to \alg \times \balg
	\end{align*}
	which works like $\mult_\alg$ on the first coordinate, and like $h$ on the second coordinate.
	By definition of the powerset algebra, 
	\begin{align*}
		\mult_{\powerset \alg} (w) = \set{ a : \mbox{ some $v \in \monad \Sigma$ satisfies $g(v)=(a,h(w))$}}.
	\end{align*}
	Therefore, to compute the above, it suffices to be able to compute the image 
	\begin{align*}
		g(\monad \Sigma) \subseteq \alg \times \balg.
	\end{align*}
	Because $\repr$ spans every finite $\monad$-algebra, the above image is the same thing as the smallest subset of $\alg \times \balg$ that contains images of single letters from $\Sigma$, and which is closed under $g$ restricted to $\repr$. This subset can be computed.
\end{pr}

\mypart{Example Monads}{In this part, we give examples of how monads can be used  to describe algebraic approaches to the languages for labelled chains (Section~\ref{sec:chain}), unary queries over finite words (Section~\ref{sec:unary}) and various kinds of trees (Section~\ref{sec:trees}). These examples illustrate the general theorems from the first part, i.e.~the Syntactic Morphism Theorem, the Eilenberg Pseudovariety Theorem, and the results on \mso.
}
\section{Monads for chains}
\label{sec:chain}
\stepcounter{monadcounter}
In this section, we show monads for representing chains, which are a generalisation of infinite words, where the set of positions can be any total order, e.g.~the rational or even real numbers.
A \emph{chain} over an alphabet $\Sigma$ is defined to be a nonempty totally ordered set of \emph{positions}, together with a labelling of these positions  by $\Sigma$. Chains form a monad, modulo the issue that all chains over a given alphabet do not form a set. The unit of this monad interprets an element $a \in \Sigma$ as a chain with a single position labelled by $a$.  The multiplication of a  chain of chains $w$ is defined by taking positions to be pairs  $(i,j)$ such that $i$ is a position in $w$, and $j$ is a position in the label of position $i$, ordered lexicographically. 

Shelah showed in~\cite{shelah_composition}  that it is undecidable if a sentence of  \mso is true in ordered real numbers $(\mathbb R, \le)$, which can be seen as an unlabelled chain, or equivalently, a chain over a one-letter alphabet. This implies  that satisfiability of \mso is undecidable on arbitary chains, or even on chains of cardinality continuum, i.e.~one cannot decide, given an \mso formula with a binary predicate for the order, whether or not the formula is true in some chain. The binary predicate for the order can be seen as the language of chains  over the alphabet $\set{0,1}$ where all zeros are before all ones. It follows that the assumptions of Theorem~\ref{thm:decide-mso} cannot be met, even for chains of cardinality at most continuum. These problems go away if one considers countable chains.

\paragraph*{Countable chains.}
\stepcounter{monadcounter}
A \emph{countable} chain is one where the   set of positions is countable. A countable chain is called \emph{scattered} if its indexing set is scattered, i.e.~its positions do not embed an isomorphic copy of the rational numbers. A special case of a scattered chain is a countable well-chain, i.e.~one where the positions are well-ordered. These three kinds of chains are submonads of the monad of chains, i.e.~they form monads when equipped with the unit and multiplication inherited from the monad of all chains.

The following theorem shows that in all three cases, the algebras admit finitely complete subfunctors, as defined in Section~\ref{sec:representing-an-algebra}, which are also strongly effective as defined in Section~\ref{sec:mso}.  The cases of countable well-founded and countable scattered chains are  simple enough to warrant a self-contained proof,  modulo the Hausdorff theorem on scattered chains. The case of arbitrary countable chains is more involved and follows from~\cite{shelah_composition}, see also~\cite{DBLP:conf/icalp/CartonCP11}.

\begin{theorem}\label{thm:countable-chains} \ \\
	\begin{enumerate}
		\item	Every finite algebra in the monad of countable well-chains is spanned by
	\begin{align*}
		X \mapsto \set{x \cdot y, x^\omega : x,y \in X}
	\end{align*}
		\item	Every  finite algebra in the monad of countable  scattered chains is spanned by
	\begin{align*}
		X \mapsto \set{x \cdot y, x^\omega , x^{-\omega} : x,y \in X}
	\end{align*}		
	\item	Every finite algebra in the monad of countable   chains is spanned by
	\begin{align*}
		X \mapsto \set{x \cdot y, x^\omega , x^{-\omega},\shuffle Y : x,y \in X, Y \subseteq X}
	\end{align*}
where $\shuffle Y$ is the chain where the positions are rational numbers and where every $y \in Y$ labels a dense subset (such a chain is unique up to isomorphism).
	\end{enumerate}
\end{theorem}
\begin{pr}[of the first two cases]
The  Hausdorff theorem on scattered chains says that scattered chains are the smallest class of chains that contains the finite chains, chains indexed by $\omega$ and $-\omega$, and is closed under substitution. For well-founded countable chains, the same holds, but $-\omega$ is not allowed. The result then follows, using the Ramsey theorem in the same way as in the case of $\infty$-algebras. \end{pr}

\begin{corollary}\label{cor:}
	Satisfiability for \mso is decidable on: all countable chains, scattered chains, and well-ordered countable chains.
\end{corollary}
\begin{pr}
	It is easy to see that the subfunctors given in Theorem~\ref{thm:countable-chains} are strongly effective. Therefore, the result follows from  Theorem~\ref{thm:decide-mso}.
\end{pr}

In particular, for the well-chains and the scattered chains, we get a simple self-contained proof of decidability for \mso. This proof is no different from the known ones, but the advantage of using monads is that they clearly identify which part of the argument is specific to the monad being used.

%
%

\section{Pointed words}
\label{sec:unary}
\stepcounter{monadcounter}
This section presents a monad which generates a new kind of algebra, which, although simple, has not appeared in the literature up to the author's best knowledge. 
The monad, call it $\unarymonad$,  is defined by
\begin{align*}
	\unarymonad A \eqdef A^* \underline A A^*,
\end{align*}
where $\underline A$ is a disjoint copy of the set $A$. Elements of  $\unarymonad A$ are called \emph{pointed words}\footnote{Similar ideas would work for pointed chains, pointed trees, etc.}. The idea is that a pointed word represents a nonempty word over $A$ where the underlined position is selected, and therefore a pointed word can be used as an input to a unary query that tests properties of positions in a word. Therefore we will use the term unary query for a set of pointed words. The unit operation is $a \mapsto \underline a$, while the monad multiplication operation is the same as in the monad of finite words, except that the underlined position is the underlined position in the underlined word.

A pointed word can be viewed in two ways: as a nonempty word over alphabet $\Sigma$  with a distinguished position, or as a special case of a non-pointed word over an extended alphabet $\Sigma \cup \underline \Sigma$. 
 In logical terms,  the first view proposes that  sets of pointed words are defined by unary queries (i.e.~formulas with one free individual variable) over the alphabet $\Sigma$, and the second view proposes that sets of pointed words are defined by Boolean queries (i.e.~with no free variables) over the extended alphabet $\Sigma \cup \underline \Sigma$. For some logics, the two views are essentially the same.  For instance a set of pointed words is \mso definable in the first view  if and only if it is \mso-definable in the second view.  The same is true for first-order logic with the order predicate.  Therefore, for some logics such as \mso or first-order logic, characterising unary queries reduces to characterising Boolean queries. However, for some logics this is not the case.
 
 In Section~\ref{sec:unary-fotwo}, we will show that for two-variable first-order logic, characterising unary queries is not easily reducible to characterising Boolean queries over extended alphabets. We also show how finite $\unarymonad$-algebras are useful in characterising unary queries. Along the way, we use much of the machinery developed in Part I of this paper, in particular the Syntactic Morphism theorem, the Pseudovariety Theorem, and the results on representation. All of these would be relatively straightforward to prove by hand in the special case of the monad $\unarymonad$, but deducing them from abstract nonsense allows us to focus on the more specific and  combinatorial parts of the proof.

Much of the material in Section~\ref{sec:unary-fotwo} is specific to unary queries definable in two-variable first-order logic, and the reader who is more interested in the general principles of monads is advised to skip it.

 \subsection{Unary queries definable in two-variable first-order logic}
\label{sec:unary-fotwo}
To illustrate the monad  of pointed words $\unarymonad$,   consider the fragment of first-order logic that uses only two variables, but which is allowed to reuse them by requantifying.  The logic has access to predicates for the labels and the order, but not for the successor, although similar results are true for other choices of predicates.  We say that a set of pointed words, i.e.~a unary query, is two-variable definable if it can be defined by a formula of two-variable first-order logic that has one free variable, say $x$,  and which uses the predicates described above. In the semantics of the formula, the free variable binds the selected position, but once the free variable of the query is requantified, the selected position is forgotten.  For example the unary query ``the distinguished position is followed by at least two  positions with label $a$''   can be defined in two-variable first-order logic, although only thanks to using requantification:
\begin{align*}
\varphi(x) = 	\exists y  \ (x < y \land a(y) \land \exists x \ (y < x \land a(x))).
\end{align*}
It is not immediately clear how to define the unary query ``the successor of the distinguished position has label $a$'', because the natural formula would use three variables to define successor in terms of order:
\begin{align*}
\psi(x) = \exists y \ (x<y \land a(y) \land \forall z (z \le x \lor y \le z)).
\end{align*}
In fact, the unary query $\psi(x)$ cannot be defined using two variables, as long as the  vocabulary has predicates just for the order and labels, which is our chosen setting in this section. This example illustrates that with only two variables, the choice of vocabulary is more important than in first-order logic with arbitrarily many variables. 

The two-variable fragment  of first-order logic is a well-studied logic for non-pointed words, i.e.~for Boolean queries on words, see e.g.~\cite{DBLP:conf/stoc/TherienW98}, but  it also makes sense for unary queries, as it corresponds to unary queries definable in XPath with only the transitive axis $/\!/$ and its inverse\footnote{To be fair, the XPath motivation would be best justified by studying the tree variant of the logic. Preliminary research indicates that the results from this section can be generalised to trees.}.

 We will show that two-variable definable languages form a pseudovariety, and therefore by the Pseudovariety Theorem, definability of a language in two-variable logic depends only the syntactic $\unarymonad$-algebra of the language. The Pseudvariety Theorem alone does not give an algorithm to decide this definability, but such an algorithm  is given  in Theorem~\ref{thm:fotwo}.


\paragraph*{The transformation monoids.} In every $\unarymonad$-algebra there is a hidden monoid, actually two monoids.
Consider a $\unarymonad$-algebra $\alg$. For $a \in A$, define its \emph{left transformation} to be the function $A \to A$ defined by
\begin{align*}
	b \mapsto  \mult_\alg(a \underline b).
\end{align*}
Likewise we define the right transformation. Left transformations form a   monoid, equipped with function composition, call it the \emph{left monoid}. If $A$ is finite then so is the left monoid. Likewise one can define right transformations and the right monoid. It is not difficult to see that a $\unarymonad$-algebra is uniquely specified by its universe $A$ and the the left and right transformations for each $a \in A$.  In other words, using the terminology of Section~\ref{sec:representing-an-algebra}, the  subfunctor
\begin{align*}
   A  \mapsto \set{ a \underline b, \underline a b : a,b \in A}
\end{align*}
 is complete for all $\unarymonad$-algebras. It is also strongly effective as defined in Section~\ref{sec:mso}. It follows that a finite $\unarymonad$-algebra can be represented in space polynomial in the size of its universe; and that syntactic algebras can be computed in polynomial time (by Lemma~\ref{lem:moore}).
 
The following example shows that just looking at the left and right monoids of a unary query is not sufficient to decide if it is two-variable definable. Stated in the language of temporal logic, the example shows that two-variable logic does not have the separation property.

\begin{myexample} Let us revisit the successor query discussed at the beginning of this section. 
	Let the alphabet be $\set{a,b}$, and consider the unary query ``the successor of the selected position has label $a$'', i.e.
	\begin{align*}
		\set{w \underline \sigma a v :  \mbox{$w,v \in \set{a,b}^*, \sigma \in \set{a,b}$} } \subseteq \unarymonad \set{a,b}.
	\end{align*}
	When seen as a language over an extended alphabet, the above is definable by a formula of two-variable logic without free variables. The formula says that there exists a position with label $a$, such that one can go one step to the left and find the underlined position, but one cannot go two steps to the left and find the underlined position. 
	When seen as a unary query over the alphabet $\Sigma$, the above is not  two-variable definable, which will follow from Theorem~\ref{thm:fotwo}.
	
Also, one can observe that just looking at the left and right monoids is not sufficient to understand the query. 	In this case, the left monoid is trivial, i.e.~contains only the identity transformation, while the right monoid is the syntactic monoid of the language ``words beginning with $a$''. Both monoids have the property that they recognise only languages definable in two-variable first-order logic.
\end{myexample}

The above example shows that characterising unary queries definable in two-variable logic does not simply reduce to characterising languages (i.e.~Boolean queries) definable in two-variable logic over an extended alphabet.

\paragraph*{An Ehrenfeucht-Fra\"iss\'e game.} We now show that two-variable definable unary queries form a pseudovariety of $\unarymonad$-languages. Therefore, by the Pseudovariety Theorem, the syntactic $\unarymonad$-algebra of a unary query has sufficient information (unlike the left and right monoids) to decide if the query is two-variable definable. 

We do this using  Ehrenfeucht-Fra\"iss\'e games in a standard way.
Consider two pointed words $w_0,w_1$.  For $n \in \Nat$, define the following game, which is played by players Spoiler and Duplicator. At the beginning of the game, the labels of the  selected positions in the two pointed words are checked; if they are different then Spoiler wins immediately and the game is terminated. If the selected positions have the same labels, then $n$ rounds of the game are played as follows. At the beginning of each round Spoiler chooses  $i \in \set{0,1}$ and a  direction, which is one of ``left'', ``stay'' or ``right''. Then Spoiler changes the selected position in the pointed word $w_i$ according to the direction, i.e.~if the direction is ``left'' then the selected position is moved somewhere to the left, if it is ``stay'' than it is not changed, and if it is ``right'' then it is moved to the right.  Duplicator responds by choosing a choosing a new selected position in the other pointed word $w_{1-i}$, according to the direction chosen by Spoiler, and such that the new selected positions have the same labels. If Duplicator cannot do this, then Spoiler wins immediately and the game is terminated. Otherwise, another round is played with the new selected positions; and if all $n$ rounds are played without Spoiler winning, then Duplicator wins.

We write $w_0 \sim_n w_1$ if Duplicator has a winning strategy in the $n$-round game. 
It is not difficult to show that $w_0 \sim_n w_1$ holds  if and only if  $w_0,w_1$ satisfy the same unary queries of two-variable logic of quantifier depth $n$. The following lemma, which is proved by composing winning strategies for Duplicator in an obvious way,  	says that equivalence under $\sim_n$ is preserved under unary polynomials and $\unarymonad$-morphisms.

\begin{lemma}\label{lem:sim-n-preserved}
If pointed words satisfy $w_0 \sim_n w_1$ then
\begin{align*}
	\begin{array}{rcrl}
			\sem p(w_0) &\sim_n& \sem p(w_1)  & \mbox{for every unary polynomial in $\unarymonad \Sigma$}	\\
		h(w_0) &\sim_n& h(w_1) &  \mbox{for every $\unarymonad$-morphism $f : \unarymonad \Sigma \to \unarymonad \Gamma$}	
	\end{array}
\end{align*}
\end{lemma}

A corollary of the above lemma is that unary queries that are two-variable definable form a pseudovariety of $\unarymonad$-languages. Closure under Boolean combinations is immediate, while for closures under derivatives and inverse images under $\unarymonad$-morphisms, one uses Lemma~\ref{lem:sim-n-preserved} and the fact that a $\unarymonad$-language is two-variable definable if and only if it is a finite union of equivalence classes of $\sim_n$ for some $n$.

\paragraph*{An effective characterization}
As stated above, two-variable unary queries form a pseudovariety of $\unarymonad$-languages, and therefore the Pseudovariety Theorem can be invoked to show that whether or not a unary query is two-variable depends only on its syntactic $\unarymonad$-algebra. Is this dependency effective? In the case of Boolean queries, i.e.~languages of non-pointed words, this problem was solved in~\cite{DBLP:conf/stoc/TherienW98}, where it was shown that a language $L \subseteq \Sigma^*$ is two-variable definable if and only if its syntactic monoid belongs to a class of monoids called {\sc da}. The class {\sc da} is a pseudovariety of monoids that can be defined by two identities, and therefore membership in it is decidable.   In the following theorem, we  extend the result of~\cite{DBLP:conf/stoc/TherienW98}  from Boolean queries to unary queries, i.e.~from non-pointed words to pointed words.
\begin{theorem}\label{thm:fotwo}
	Let  $q \subseteq \unarymonad \Sigma$ be a unary query recognisable by a finite $\unarymonad$-algebra. Then $q$ is two-variable definable if and only if its syntactic $\unarymonad$-algebra $\alg$ satisfies:
	\begin{itemize}
		\item the left and right monoids of $\alg$ belong to {\sc da};
		\item for every  $w \in A^+$ which all letters in a set $B \subseteq A$, every $b \in B$ and every $v \in \unarymonad B$,  the following equality holds for all but finitely many $n \in \Nat$:
	\begin{align*}
		\mult_\alg (w^n b v  w^n)  = \mult_\alg(w^n v w^n) = \mult_\alg (w^n v b  w^n).
	\end{align*}
	\end{itemize}
\end{theorem}

The rest of Section~\ref{sec:unary-fotwo} is devoted to proving the above theorem. We begin with a corollary of the theorem, which says that definability of unary queries in two-variable logic can be decided in polynomial time. When talking about polynomial time, we refer to representation of $\unarymonad$-algebras with respect to 
\begin{align*}
   A  \mapsto \set{ a \underline b, \underline a b : a,b \in A}.
\end{align*}
When $a,b$ are in the universe of a $\unarymonad$-algebra $\alg$, we will treat $a \underline b$ as an element of $\alg$, although the more formally correct notation would  be $\mult_\alg(a \underline b)$.

\begin{corollary}\label{cor:fotwo-decidable}
	Whether or not a recognisable $q \subseteq \unarymonad \Sigma$ is two-variable definable can be decided in polynomial time with respect to the recognising morphism.
\end{corollary}
\begin{pr}
	By Lemma~\ref{lem:moore}, the syntactic morphism can be computed in polynomial time based on any recognising morphism into a finite algebra. Therefore, it suffices to show that the  conditions in Theorem~\ref{thm:fotwo} can be checked in polynomial time, when given on input an $\unarymonad$-algebra $\alg$.
	
	 For the first condition, one computes the left and right monoids. These monoids are quotients of $\alg$ under an equivalence relation that can be checked in polynomial time, and therefore they can be computed in polynomial time. Then one checks in polynomial time if the left and right monoids satisfy the identities for {\sc da}. 
	
	Let us show how to check the  second condition.  A naive algorithm would check all possible subsets $B \subseteq A$, which would take exponential time. To overcome this, define an ordering $\preceq$ on the universe of $\alg$, such that  $a \preceq b$ holds if there exist pointed words $w,v \in \unarymonad A$ such that 
\begin{align*}
	a = \mult_\alg(w) \qquad b=\mult_\alg(v) 
\end{align*}
and every letter that appears in $w$ also appears in $v$, ignoring the underlining. The relation $\preceq$ is not necessarily transitive, due to taking the image under $\mult_\alg$.  In terms of $\preceq$, the second condition in the statement of the theorem says that  for all but finitely many $n$, 
	\begin{align*}
		\mult_\alg (a^n b \underline c   a^n)  = \mult_\alg(a^n \underline c a^n) =  \mult_\alg(a^n \underline c b a^n) \qquad \mbox{for every $a \succeq b,c$}
	\end{align*}
It is not difficult to show that the above need only be checked for $n$ which are linear in the size of left and right monoids of $\alg$, and therefore the only remaining thing to do is compute $\preceq$.

It is not difficult to show that $\preceq$  is the smallest relation which contains every pair $a \preceq a$ and which satisfies the following implications for every $a,b,c,d$ in $\alg$.
\begin{eqnarray*}
a \preceq b &\mbox{implies}& a \preceq b \underline c\\
a \preceq b &\mbox{implies}& a \preceq \underline b  c\\
a \preceq b \mbox{ and } c \preceq d &\mbox{implies}& a \underline c \preceq b  \underline d\\
a \preceq b \mbox{ and } c \preceq d &\mbox{implies}& \underline a c \preceq  \underline b  d.
\end{eqnarray*}
In particular, $\preceq$ can be computed in polynomial time using a fixpoint algorithm.
\end{pr}

The rest of Section~\ref{sec:unary-fotwo} is devoted to proving Theorem~\ref{thm:fotwo}. We begin with the easier implication. 
\begin{lemma}\label{lem:necessary-fotwo}
	If a unary query is two-variable definable, then its syntactic algebra satisfies the conditions in Theorem~\ref{thm:fotwo}.
\end{lemma}
\begin{pr}
	Let $q \subseteq \unarymonad \Sigma$ be a unary query definable in two-variable first-order logic, and let $n$ be the quantifier depth of the defining formula.
	By Lemma~\ref{lem:sim-n-preserved}, the equivalence relation $\sim_n$ is preserved under unary polynomials, and therefore by the Syntactic Morphism theorem it is a congruence, i.e.~the set of equivalence classes can be equipped 
	 a  multiplication operation which makes it into a finite $\unarymonad$-algebra, call it $\alg_n$. Since $q$ is a finite union of equivalence classes under $\sim_n$, it is recognised by $\alg_n$, and therefore the syntactic algebra of $q$ is an image of $\alg_n$ under a $\unarymonad$-morphism. The conditions in Theorem~\ref{thm:fotwo}  are easily seen to be closed under images of $\unarymonad$-morphisms, and therefore it suffices to show that these conditions are satisfied by $\alg_n$. We only sketch the proof for the second condition: this boils down to showing that if $w \in \Sigma^*$ is a word which uses all letters in a set $B \subseteq \Sigma$, then
	 \begin{align*}
	 	w^m v_1 \underline b v_2 w^m  \sim_n w^m \underline b  w^m \qquad \mbox{for every $m \ge n$, $v_1,v_2 \in B^*$ and $b \in B$.}
	 \end{align*}
This is proved by induction on $n$. Here is one of the cases that needs to be considered: if in the first round, Spoiler moves the selected position of the first pointed word to some position in $v_1$, the Duplicator responds by moving the selected position in the second pointed word to a position in the last copy of $w$ before $\underline b$ which has the same label, such a position exists by assumption on $w$ using all letters from $B$.
\end{pr}

The rest of Section~\ref{sec:unary-fotwo} is devoted to showing the converse implication in Theorem~\ref{thm:fotwo}. 
	A possibly partial function $f : \unarymonad \Sigma \to X$ with $X$ finite  is  called two-variable definable if the  inverse image of every $x \in X$ is two-variable definable.   We will prove that if $\alg$ is a finite $\unarymonad$-algebra that  satisfies the conditions in the theorem, then every $\unarymonad$-morphism $h : \unarymonad \Sigma \to \alg$ is definable in two-variable logic.
	The proof is by induction on the size of the alphabet $\Sigma$.  We begin by introducing some auxiliary results.

 \paragraph*{Filtering.} For a unary query $q \subseteq \unarymonad \Sigma$, define 
 \begin{align*}
 	\filter q : (\Sigma^* \cup \unarymonad \Sigma) \to (\Sigma^* \cup \unarymonad \Sigma)
 \end{align*}
 to be the function which inputs a pointed or non-pointed word, and only keeps positions that are selected by $q$. The output is a pointed word if the input was a pointed word and $q$ selected the selected position, otherwise the the output is a non-pointed word.  For example, if $q \subseteq \unarymonad \set{a,b}$ is the set of pointed words where the distinguished position has label $a$, then 
 \begin{align*}
 	(\filter q) (a\underline b a ) = aa \in \Sigma^* \qquad  	(\filter q) (a\underline a a ) = a\underline a a \in \unarymonad \Sigma.
 \end{align*}

The following simple fact is proved by   relativising formulas in the obvious way.
\begin{fact}\label{fact:filter}
	If  $f : \unarymonad \Sigma \to X$ and $q \subseteq \unarymonad \Sigma$  are definable in two-variable logic, then so is the partial function $f \circ \filter q$. (The function is partial because it is undefined when $\filter q$ removes the selected position.)
\end{fact}
 
 \newcommand{\tranmon}[1]{\mathsf{mon}#1}
 
 \paragraph*{A monoid.}
 Let $\tranmon \alg$ be the product of the left and right transformation monoids of $\alg$.  It is not difficult to see that the left and right transformations of $h(w)$ for $w \in \unarymonad \Sigma$ do not depend on the selected position. In other words, there is a function $\tranmon h$ which makes the following diagram commute
 \begin{align}\label{eq:hath}
 	\xymatrix@C=4pc{
	\unarymonad \Sigma \ar[dr]_{\mathsf{mon}} \ar[r]^{\deselect} & \Sigma^* \ar[d]^{\tranmon h} \\ & \tranmon A}
 \end{align}
 where $\deselect$
 is the function that ignores the selected position and $\mathsf{mon}$ is the function that computes the left and right transformation.
  It is not difficult to show that $\tranmon h$ is a monoid morphism. By the assumption of the theorem, the monoid $\tranmon \alg$ is in {\sc da}, as a product of two monoids in {\sc da}.
 
\paragraph*{Known results about {\sc da}.} We now recall some results about sets of words (not pointed words) definable in two-variable logic.  Let $\malg$ be a monoid. Define the right ideal generated by $m \in \malg$ to be the set
\begin{align*}
	\set{m n : n \in \malg} \subseteq \malg.
\end{align*}
 We say $m, n \in \malg$ are $\Rr$-equivalent, denoted by $m \sim_\Rr n$, if they generate the same right ideals. 
 There is a symmetric notion of $\Ll$-equivalence that uses left ideals.  The theorem below summarises some results from~\cite{DBLP:conf/stoc/TherienW98}. It only mentions $\Rr$-classes, i.e.~equivalence classes under $\Rr$-equivalence, but the symmetric results also hold for $\Ll$-classes.

\begin{theorem}\label{thm:da}
For a  monoid morphism  $g : \Sigma^* \to \malg$ into a monoid in {\sc da}:
	\begin{enumerate}
		\item every language recognised by $g$ is a  two-variable definable Boolean query, i.e.~it is definable by a  formula of two-variable first-order logic without free variables;
		\item for every $\Rr$-class $R$  of $\malg$ there is a  two-variable definable unary query  which selects a position if and only if the prefix up to and including  that position is mapped by $g$ to $R$.
\item for every $m \in \malg$ the  set $	\set{ n : mn \sim_\Rr m}$ is a submonoid of $\malg$.
	\end{enumerate}
\end{theorem}

Apply the above theorem to the morphism $\tranmon h$ defined in diagram~\eqref{eq:hath}. By the second item of the theorem and its symmetric variant for $\Ll$-classes, for every $\Rr$-class $R$ and every $\Ll$-class $L$ of the monoid $\tranmon \alg$, there are two-variable definable unary queries, call them  $q_R$ and $q_L$, such that
\begin{align*}
	u\underline av \in q_R \qquad \mbox{iff} \qquad \tranmon h (ua) \in R\\
	u\underline av \in q_L \qquad \mbox{iff} \qquad \tranmon h (av) \in L
\end{align*}
holds for every $u,v \in \Sigma^*$ and $a \in \Sigma$.

The more difficult implication from Theorem~\ref{thm:fotwo} will follow from the following lemma. 
\begin{lemma}\label{lem:}
	Let $R$ be an $\Rr$-class in the monoid $\tranmon \alg$, and let $L$ be an $\Ll$-class in the monoid $\tranmon \alg$. Then the partial function obtained from $h$ by restricting its domain to $q_R \cap q_L$ is two-variable definable.
\end{lemma}

Before proving the lemma, observe that it implies that $h$ is two-variable definable. This is because every pointed word belongs to $q_R \cap q_L$ for some choice of $R$ and $L$.

\medskip
\begin{pr}
Call an $\Rr$-class \emph{minimal} if the corresponding right ideal is  minimal  with respect to inclusion. Likewise we define a minimal $\Ll$-class. Consider three cases: when $R$ is not minimal, when $L$ is not minimal, and when both $R$ and $L$ are minimal. The first two cases are not disjoint.

		\paragraph*{The $\Rr$-class $R$ is not minimal.} We  prove a stronger result, namely the restriction of $h$ to $q_R$  is two-variable definable.
		Let us decompose $q_R$ into a disjoint union of two unary queries, both of which are two-variable definable: $q^0_R$ selects the leftmost position that satisfies $q_R$, and $q^+_R$ § selects the remaining positions. 

	\begin{itemize}
		\item 	Let us first show that $h$ is two-variable definable when  its domain is restricted to $q^0_R$. 	Define $\subleft q$  be the unary query which selects positions that are strictly to the left  of some position that satisfies $q_R^0$. Likewise define $\subright q$ to be the positions that are strictly to the right  of some position that satisfies $q_R^0$.
		 Because $q_R^0$ is two-variable definable, then the queries $\subleft q$ and $\subright q$ are also two-variable definable. Consider some  $w \in q_R^0$.  The word $\deselect w$ underlying $w$ splits into three consecutive intervals: 
		 \begin{enumerate}
		 	\item  first come the positions that satisfy $\subleft q$;
			\item then comes the single position that satisfies $q_R^0$;
			\item finally come the positions that satisfy $\subright q$.
		 \end{enumerate}
Because the intervals are consecutive, when restricted to arguments from $q_R^0$, the  function  $h$ factors through the following three functions.
\begin{align*}
	\begin{array}{rclcl}
		\tranmon h \circ \filter \subleft q &:& \unarymonad \Sigma &\to& \tranmon \alg\\
				 h \circ \filter q_R^0 &:& \unarymonad \Sigma &\to&  \alg\\
		\tranmon h \circ \filter \subright q &:& \unarymonad \Sigma &\to& \tranmon \alg\\		
	\end{array}
\end{align*} The first and third functions are  two-variable definable by Theorem~\ref{thm:da} and  Fact~\ref{fact:filter}. The middle function depends only on the label of the selected position, and is therefore also two-variable definable. Therefore, $h$ is two-variable definable when restricted to arguments from $q_R^0$.
	\item Let us now show that $h$ is two-variable definable when restricted to $q^+_R$. The proof is the same as above, the only difference is in the proof that 
	\begin{align}\label{eq:hfilter}
		h \circ \filter q^+_R : \unarymonad \Sigma \to \alg
	\end{align}
	is two-variable definable. To prove this, let $m$ be an element of the $\Rr$-class $R$, and consider the set
	\begin{align*}
 M_m \eqdef \set{ n : mn \sim_\Rr m}
	\end{align*}
	which is a submonoid of $\tranmon \alg$ by item 3 of Theorem~\ref{thm:da}. Because $M_m$ is a submonoid, it is not difficult to show that it does not depend on the choice of $m \in R$, i.e.~it only depends on the $\Rr$-class $R$.  Furthermore, this submonoid cannot be all of $\tranmon \alg$, since otherwise $R$ would be a minimal $\Rr$-class. Therefore, there is some  $a \in \Sigma$ such that $\tranmon h(a)$ does not belong to the submonoid, which means that
	\begin{align*}
		m  \tranmon h (a) \not \sim_\Rr m \qquad \mbox{for every $m \in R$}
	\end{align*}
	 It follows that positions selected by $q^+_R$ cannot have label $a$. Therefore the function in~\eqref{eq:hfilter} is two-variable definable by induction assumption on a smaller alphabet.
	\end{itemize}
	
	\paragraph*{The $\Ll$-class $R$ is not minimal.} This case is symmetric to the previous one.
	\paragraph*{Both $L$ and $R$ are minimal.} We are left with the case where $L$ is a minimal $\Ll$-class of $\tranmon \alg$, and $R$ is a minimal $\Rr$-class of $\tranmon \alg$. We will prove that, when restricted to  pointed words in $q_R \cap q_L$, the value of $h$ depends only on the  label of the selected position. In other words, we claim that if
		\begin{align*}
			w = \subleft w  \underline a  \subright w \qquad 		v = \subleft v  \underline a  \subright v
		\end{align*}
		are pointed words in  $q_R \cap q_L$ with the same label $a \in \Sigma$ of the distinguished position, then both have the same value under $h$.
		Let $u \in \Sigma^*$ be a non-pointed word which uses all letters in $\Sigma$, whose $\Rr$-class is $R$ and whose $\Ll$-class is $L$. Such a word exists by minimality, for example, $u$ can be obtained from either of the pointed words $w$ or $v$ by replacing the selected position by some non-pointed word which uses all letters from $\Sigma$. Let $n$ be a sufficiently large number. We will show that
		\begin{align*}
			h(w)= h(u^n  \underline a  u^n) = h(v).
		\end{align*}


By symmetry, we only prove the left equality above.
	Because  $\tranmon h(\subleft w)$ is in the same $\Rr$-class as $\tranmon h(\subleft w  u^n)$, we have
	\begin{align*}
		\tranmon h(\subleft w  u^n)  m =  \tranmon h (\subleft w)  \qquad \mbox{for some $m \in \tranmon \alg$.}
	\end{align*}
	Therefore, there must exist some $\subleft x \in \Sigma^*$ such that $\subleft w  u^n  \subleft x$ induces the same left transformation in $\alg$ as $\subleft w$.
	Using a symmetric reasoning for $\Ll$-classes, we obtain some  $\subright x \in \Sigma^*$ such that $h(w)$ is equal to
	\begin{align*}
 h(\subleft w  u^n  \subleft x  \underline a  \subright x  u^n  \subright w) 
	\end{align*}
	The elements $\tranmon h(\subleft w  u^n)$ and $\tranmon h (u^n)$ are in the $\Rr$-class $R$, because both have prefixes in this $\Rr$-class and the class is minimal; and they are both in the $\Ll$-class $L$ because both have suffixes in this $\Ll$-class.  (Here we use minimality.) In a monoid from {\sc da}, or more generally in an aperiodic monoid,  and element is uniquely determined by its $\Ll$-class and $\Rr$-class. Therefore, $\subleft w  u^n$ and $u^n$ induce the same left transformation in $\alg$. By this observation, and a symmetric one for $\Ll$-classes, it follows that $h(w)$ is equal to
	\begin{align*}
	h(u^n  \subleft x  \underline a  \subright x  u^n ).
	\end{align*}
	From the assumption on $\alg$ in the theorem, the above is equal to
	\begin{align*}
	h(u^n  \underline a   u^n ).
	\end{align*}
	By the same reasoning, $h(v)$ is also equal to the above, which completes the proof of the lemma.
\end{pr}

\section{Monads for trees}
\label{sec:trees}
In this section, we present a series of monads for modelling trees.
\subsection{Ranked trees over a fixed alphabet}
\stepcounter{monadcounter}
We begin with a monad that represents finite trees over a fixed ranked alphabet.
Algebras in this monad will be deterministic bottom up tree automata over the ranked alphabet.

 Consider a ranked alphabet $\Sigma$, i.e.~a finite set where each element has an associated rank, which is a natural number. A ranked tree over such an alphabet is a finite tree labelled by $\Sigma$, where a node has as many children as the rank of its alphabet, and these children are ordered. In other words, this is a ground term over $\Sigma$ seen as a signature.  We  define a monad $\monad_\Sigma$, which is parametrised by $\Sigma$, and which will model ranked trees over $\Sigma$. Although the alphabet $\Sigma$ is ranked, the monad $\monad_\Sigma$ itself is in the category of sets, i.e.~sets without any  arity structure imposed. 
 
 Define $\monad_\Sigma$ to be the monad which maps a set $\Gamma$ to the set of terms over the signature $\Sigma$ extended by variables from $\Gamma$ (i.e.~trees where labels from $\Gamma$ can occur in the leaves).  The multiplication operation 
 \begin{align*}
 \mult_{\monad_\Sigma \Gamma} :	\monad_\Sigma \monad_\Sigma \Gamma \to \monad_\Sigma \Gamma
 \end{align*}
  is term substitution, while the unit maps a variable $a \in \Gamma$ to the term that consists only of this variable. In the language of category theory, this is the monad generated by $\Sigma$ interpreted as a polynomial functor.

If $\Gamma$ is a finite alphabet, then a $\monad_\Sigma$-language over $\Gamma$ is a set of trees over the ranked alphabet $\Sigma$, extended by rank zero symbols for letters from~$\Gamma$. In the special case of $\Gamma = \emptyset$, a $\monad_\Sigma$-language over the empty alphabet is a set of   ranked trees over the alphabet $\Sigma$.

\begin{myexample}\label{ex:dfa} In this example, we show that when $\Sigma$ contains only letters of rank one, then the monad $\monad_\Sigma$ can be seen as modelling deterministic word automata with input alphabet $\Sigma$.
	Consider a ranked set $\Sigma$, which has only letters of rank one. If $Q$ is a set, then elements of $\monad_\Sigma Q$ are trees with unary branching where inner nodes are from $\Sigma$ and the unique leaf is from $Q$. For example an  element of  $\monad_{\set{a,b}} \set q$ can look like this:
	\begin{center}
		\includegraphics[scale=0.5]{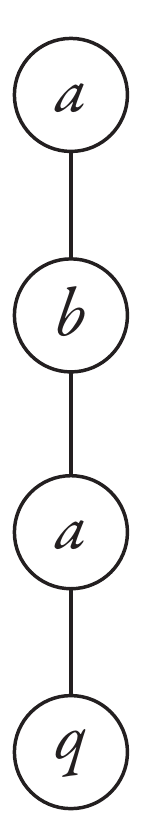} 
	\end{center}
When written bottom-up, such a tree can be seen as a word in $Q \cdot \Sigma^*$. A finite algebra Eilenberg-Moore for this monad, i.e.~a finite $\monad_\Sigma$-algebra, consists of a universe, call it $Q$, and a multiplication operation, which can be seen as a function
\begin{align*}
	\delta : Q \cdot \Sigma^* \to Q.
\end{align*}
The associativity of multiplication says that 
\begin{align*}
	\delta(\delta(q \cdot w) \cdot v ) = \delta (q \cdot w \cdot v),
\end{align*}
and therefore $\delta$ is uniquely defined by its values on $Q \cdot \Sigma$. Stated differently, a $\monad_\Sigma$ algebra is the same thing as a deterministic finite word automaton with input alphabet $\Sigma$, without designated initial and accepting states.
\end{myexample}

\paragraph*{Connections with $\Sigma$-algebras.} Recall that in universal algebra, a $\Sigma$-algebra consists of a universe $A$ together, together with an operation $f : A^n \to A$ for each $f \in \Sigma$ of rank $n$.  To go from a $\monad_\Sigma$-algebra   $\alg$ in the sense of Eilenberg-Moore to a $\Sigma$-algebra in the sense of universal algebra, one defines the universe to be $A$, and the operation corresponding to a $n$-ary letter $f \in \Sigma$ to be
\begin{align*}
	(a_1,\ldots,a_n) \in A^n \mapsto \mult_\alg(f(a_1,\ldots,a_n)).
\end{align*}
In the terminology of Section~\ref{sec:representing-an-algebra}, this is the $\Sigma$-reduct of $\alg$, where we view $\Sigma$ as the (finitely complete and effective) subfunctor  
\begin{align*}
	\Sigma A = \set{f(a_1,\ldots,a_n) : \mbox{$f$ is an $n$-ary symbol in $\Sigma$}} \subseteq \monad_\Sigma A
\end{align*}
Every $\Sigma$-algebra is obtained this way, and therefore the two notions are essentially the same. This sameness extends to morphisms\footnote{Actually, this sameness works for a more general notion of ranked set used in category theory, i.e.~when $\Sigma$ is an arbitrary functor. This more general setting can be used to describe unranked trees, when $\Sigma$ is the functor 
\begin{align*}
	X \mapsto X^*
\end{align*}
or unranked trees without sibling order, when $\Sigma$ is the functor which takes $X$ to finite multisets over $X$.  A problem with these  more general settings is that  their Eilenberg-Moore algebras model automata that are too strong, in the sense that the transition function need not be describable in a finite way.}.

\paragraph*{Connection with tree automata.}  As shown in Example~\ref{ex:dfa}, if the alpahbet $\Sigma$ contains letters of rank at most one, then $\monad_\Sigma$-algebras are essentially the same thing as deterministic word automata. For other alphabets,  the correspondence is with deterministic bottom-up tree automata.
For a $\monad_\Sigma$ algebra $\alg$, there is a  unique $\monad_\Sigma$-morphism
\begin{align*}
	h : \monad_\Sigma \emptyset \to \alg.
\end{align*}
When interpreting an element of $\monad_\Sigma \emptyset$ as a tree over the ranked alphabet, the algebra $\alg$ maps every tree to an element of its universe. When the algebra is finite, this is the same thing as a deterministic bottom-up tree automaton, with the only difference being that an automaton also has an accepting subset of states, which indicates when a tree belongs to the language. Therefore, $\monad_\Sigma$-recognisable languages are the same thing as the classical notion of regular languages of finite trees over the ranked alphabet $\Sigma$.

\begin{myexample}\label{ex:tree-fo-pseudo}
Consider the following variant of first-order logic on trees over a ranked alphabet $\Sigma$. To a tree $t \in \monad_\Sigma \Gamma$, one assigns a logical structure, where the universe is the nodes of the tree, and there are the following predicates: a unary predicate that is true in nodes with label $a$, a binary predicate for the descendant relation, and binary predicates for the $i$-th child relation for every $i$. A subset of $\monad_\Sigma \Gamma$ is called definable in first-order logic if there is a formula of first-order logic that is true in the logical structures corresponding to trees in the subset, and false in logical structures corresponding to tree outside the subset. A well-known open problem stated in~\cite{DBLP:conf/caap/Thomas84} is: can one decide if a recognisable language of trees is definable in first-order logic?  

Here we show the, already known, result that tree languages definable in first-order logic form a language pseudovariety. Since the assumptions of the Pseudovariety Theorem apply to the monad $\monad_\Sigma$, this will imply  that first-order definability of a tree language  depends only on its syntactic algebra (whether or not this dependence is computable is the open problem).

Recall that a language pseudovariety is a class of languages that is closed under Boolean combinations, polynomial derivatives, and inverse images under morphisms. (The other conditions in the definition are vacouous when there is only one sort, as is the case here.) Boolean combinations are for free in first-order logic. Closure under polynomial derivatives is shown that same way as closure under  inverse morphisms, so we only show the latter closure. We need to show that for every $\monad_\Sigma$-morphism 
\begin{align}\label{eq:term-monad-morphism}
	h : \monad_\Sigma \Gamma \to \monad_\Sigma \Delta,
\end{align}
inverse images under $h$ of first-order languages are also first-order definable. One way of proving this statement is to show that $h$ is a special case of a more general notion of  \emph{copying first-order interpretation}, and first-order definable languages are closed under inverse images of such interpretations. Another way, which we use here, is to use  Ehrenfeucht-Fra\"iss\'e games. Let us write $s \sim_n t$ if player Duplicator has a winning strategy in the $n$-round Ehrenfeucht-Fra\"iss\'e game for the logical structures corresponding to $s$ and $t$.  It is easy to see that closure under inverse morphisms of  first-order definable languages is implied by the following observation:
\begin{align}\label{eq:ef-stealing}
	s \sim_n t \quad \mbox{implies} \quad h(s) \sim_n h(t) \qquad \mbox{for every $n \in \Nat$}.
\end{align}
The above observation is proved using a straightforward strategy copying argument, which we describe in more detail below. The key to the strategy copying argument is that  every node  in an image tree $h(t)$ is uniquely identified by  two pieces of information, which we call the \emph{origin} and \emph{offset}, whose definition is explained by example in Figure~\ref{fig:origin-offset}.

\begin{figure}[htbp]
	\centering
		\includegraphics[scale=0.7]{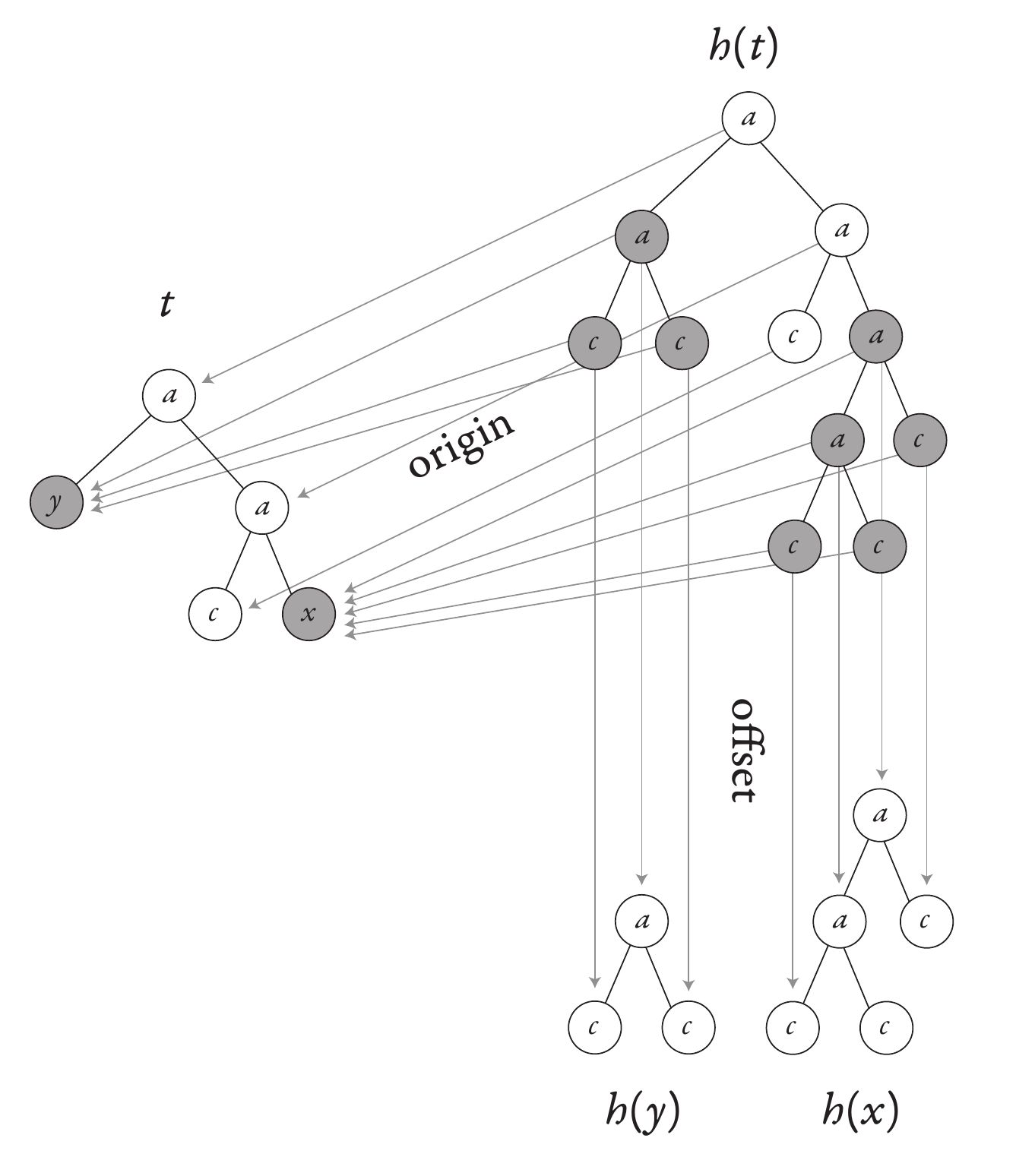}
	\caption{The origin and offset functions. In this example, $\Sigma$ has one symbol $a$ of rank two, and one symbol $c$ or rank zero. The origin function is from nodes of $h(t)$ to nodes of $t$. The offset function is defined on the ``new'' nodes in $h(t)$, i.e.~those nodes in $h(t)$ that have  labels in  $\Gamma$, and which have darker colour in the picture. The offset function  maps such a node to the corresponding node $h(\sigma)$, where $\sigma \in \Gamma$ is the label of the origin.}
	\label{fig:origin-offset}
\end{figure}
To prove~\eqref{eq:ef-stealing}, in the game corresponding to $h(s)$ and $h(t)$, Duplicator preserves the following invariant. 
\begin{itemize}
	\item [(*)] Suppose that $i$ rounds have been played so far, and that the nodes selected in those rounds were $x_1,\ldots,x_i$ in the tree $h(s)$ and $y_1,\ldots,y_i$ in the tree $h(t)$.
Then the offsets, if defined, are the same for each $x_j$ and $y_j$, and   Duplicator has a winning strategy for the remaining rounds in the game  for $s$ and $t$, assuming that the selected nodes $x_1,\ldots,x_i$ and $y_1,\ldots,y_i$ are replaced by their origins.
\end{itemize}
Assuming that $s \sim_n t$ holds, it is not  difficult to show that Duplicator can preserve the invariant for $n$ rounds in the game between $h(s)$ and $h(t)$.  Although simple, the strategy copying argument is a bit delicate --  as we  will see in Example~\ref{ex:clone-non-pseudo}, closure under inverse morphisms  will fail in a different monad for modelling trees, where morphisms can duplicate subtrees.
\end{myexample}
\paragraph*{Dependence on $\Sigma$.} In the monad $\monad_\Sigma$, there is a different monad for every $\Sigma$. In the following two sections, we  present two approaches where the monad is independent of the alphabet. The price we will pay is using categories of ranked sets.

\subsection{Clones}
\stepcounter{monadcounter}
In this section,  we consider a monad which is used to describe clones.  We begin by recalling the definition of a clone from universal algebra: a clone over a universe $A$ is a set of functions of the form $A^n \to A$, of possibly different arities $n \in \Nat$,
which includes all projections, i.e.~functions of the form $(a_1,\ldots,a_n) \mapsto a_i$, and which is closed under composition in the sense that if the clone contains an $n$-ary operation $f$ and $k$-ary operations $f_1,\ldots,f_n$, then it also contains the $k$-ary operation
\begin{align*}
\bar a \in A^k \qquad \mapsto \qquad f(f_1(\bar a),\ldots,f_n(\bar a))\in A.
\end{align*}

\paragraph*{The category of ranked sets.} To model clones by a monad, we use a different category than sets. The category is ranked sets, i.e.~sorted sets where the sort names are natural numbers.  Recall that the notions of language theory are parametrised by notions of finite object and finite algebra. We make the following design decisions for the clone monad: a finite ranked set is one with finitely many elements, in particular only finitely many ranks can be achieved in a finite ranked set.  We come back to the notion of finite algebra later on.%

\begin{figure}[htb]
	\centering
	\begin{tabular}{cc}
		\includegraphics[scale=0.5]{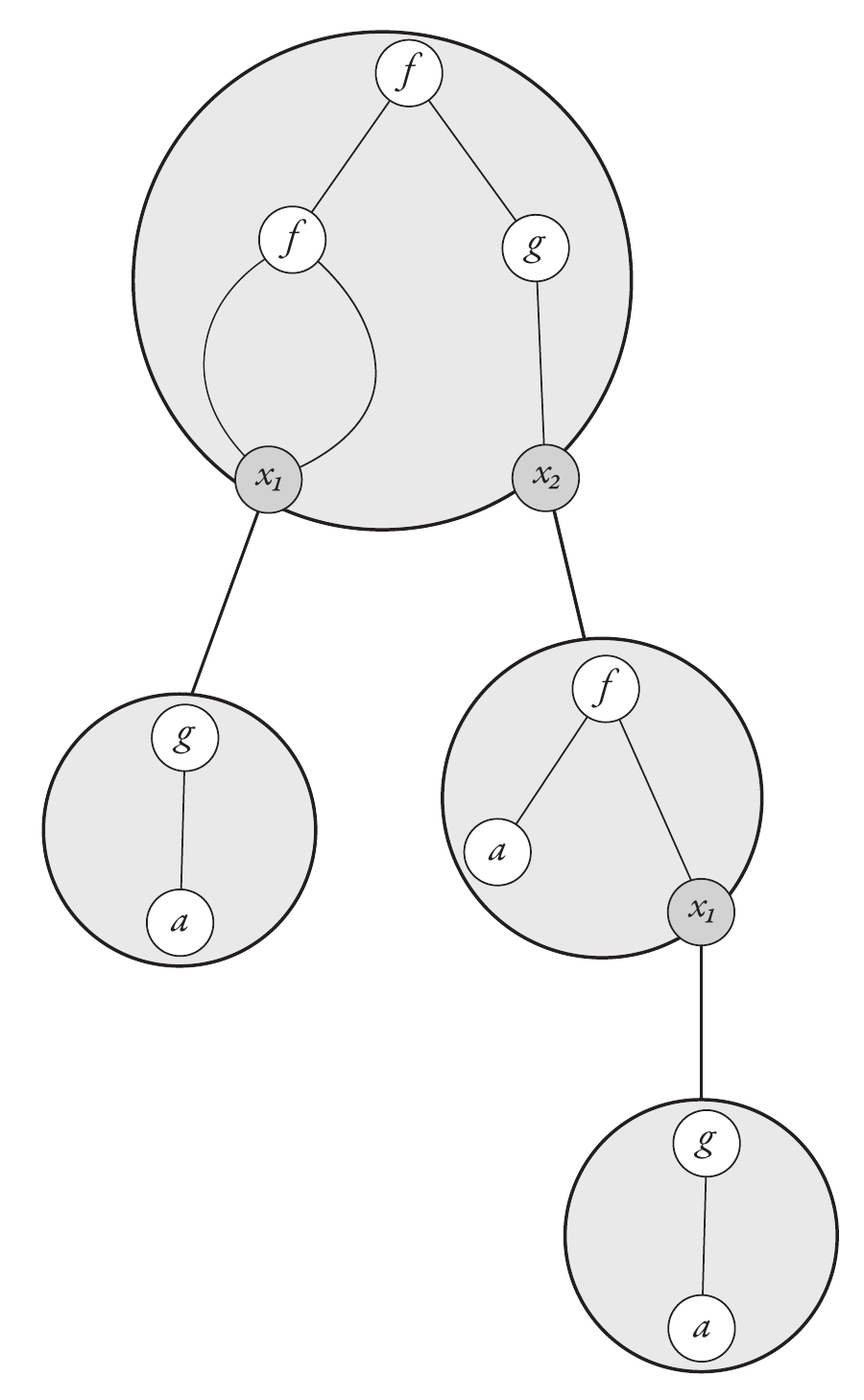} &  				\includegraphics[scale=0.5]{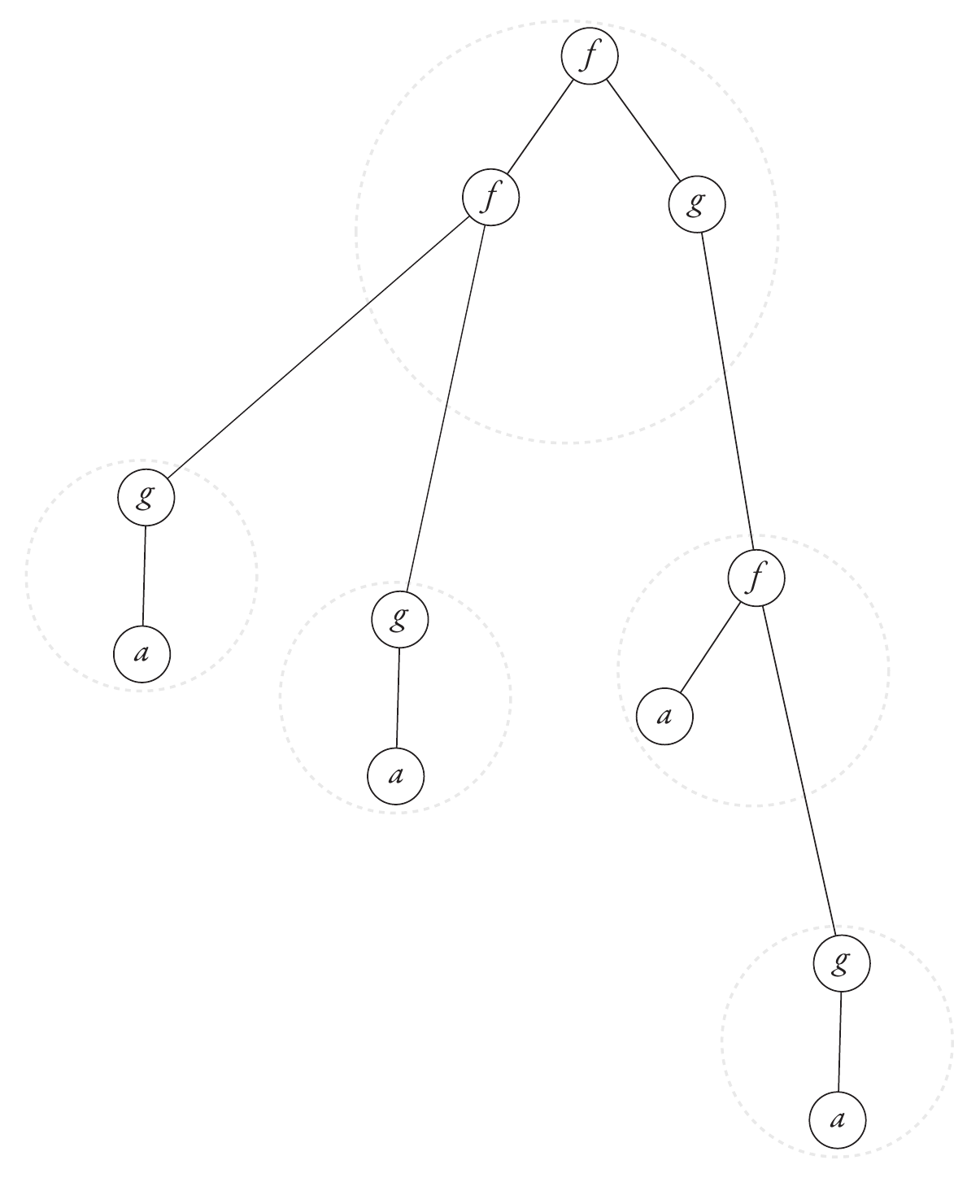}\\
				$t \in \clo\ \! \clo\ \!  \Sigma$ & $\monmul \Sigma (t) \in \clo\ \! \Sigma$\\
				$t$ has 4 nodes and rank 0 & $\monmul \Sigma (t)$ has 11 nodes and rank 0.
	\end{tabular}
	\caption{Example of multiplication in a $\clo$-algebra. The ranked alphabet $\Sigma$ has elements $a,g,f$ of arities $0,1,2$ respectively. The left picture represents a tree $t \in \clo\ \!\clo\ \! \Sigma$, where variable $x_1$ is used twice in the label of the root, which is drawn using parallel edges. This double use  results in duplication after multiplication is applied. The light grey dotted circles on the right are not part of $\mult_{\clo \Sigma} (t)$, they just highlight how $\monmul \Sigma (t)$ is obtained from $t$. }
	\label{fig:clo}
\end{figure}

\paragraph*{The clone monad.} The \emph{clone monad} maps a ranked set $\Sigma$ to the ranked set $\clo \Sigma$, where elements of rank $n$ are terms over $\Sigma$ that use $n$ variables $x_1,\ldots,x_n$ (the sequence of variables $x_1,x_2,\ldots$ is chosen so that they are fresh with respect to $\Sigma$). The terms need not use all variables, and variables may appear with repetitions.  The monad multiplication operation
\begin{align*}
\monmul  \Sigma :	 \clo \clo \Sigma \to \clo \Sigma
\end{align*}
 is  substitution, as illustrated in Figure~\ref{fig:clo}. 

\paragraph*{Comparison with clones.} We use the name $\clo$-algebra for an Eilenberg-Moore algebra in the monad of clones. A $\clo$-algebra is almost the same thing as a clone in the sense of universal algebra, with the following differences. 
\begin{itemize}
	\item Clones are more general than $\clo$-algebras in the sense that clones admit  a distinction between the universe and the operations of rank zero (constants). In other words, it is not necessarily the case that every element of a clone's universe is a constant. (If this is the case, then a clone is called a \emph{polynomial clone}.)
	\item Clones are less general than $\clo$-algebras in the sense that in a clone, unlike in a $\clo$-algebra, there is an extensionality property with respect to the universe: elements of the clone are uniquely determined by the transformations that they induce on the universe. This is similar to the finite observability condition used in the Pseudovariety Theorem from Section~\ref{sec:pseudovariety}.
\end{itemize}
Therefore, a polynomial clone is the same thing as a $\clo$-algebra that is zero-extensional in the sense every element is determined by its transformation on rank zero elements. 
\paragraph*{Finitary clones.}  There is no sense in considering $\clo$-algebras that have a finite universe, because the requirement on projections means that the universe is nonempty on every rank. In $\clo$-algebras, we call a $\clo$-algebra \emph{finite} if  it has finitely many elements for every rank, and is finitely generated.  The finite generation axiom is natural in the context of recognising languages (every recognisable $\clo$-language over a finite alphabet is recognised by a finitely generated $\clo$-algebra), but it  is not superfluous -- there exist clones over a three element universe  that are not fintiely generated, as shown by Yanov and Muchnik in~\cite{yanovmuchnik}, and this is even the case for polynomial clones~\cite{agoston1983number}.

\begin{myexample}\label{ex:clone-non-pseudo}
	This is a non-example of a pseudovariety. Let us revisit first-order logic on trees as defined in Example~\ref{ex:tree-fo-pseudo}.	A language of ranked trees can be seen as a special case of a $\clo$-language, which happens to contain only elements of rank zero.  Such languages are not closed under inverse images of $\clo$-morphisms, which is witnessed by the following example, essentially due to Potthoff~\cite{potthoff}. (Recall that first-order definable language were closed under inverse morphisms for the monad of ranked trees, as shown in Example~\ref{ex:tree-fo-pseudo}. What worked in Example~\ref{ex:tree-fo-pseudo} and no longer works in this example is that for $\clo$-morphisms,  a node is not uniquely determined by its offset and origin.)
	Consider  letters $a_0,a_1,a_2$ with ranks $0,1,2$ respectively, and consider the $\clo$-morphism
	\begin{align*}
	h : 	\clo \set{a_0,a_1} \to \clo \set{a_0,a_2}
	\end{align*}
	which maps $a_0$ to $a_0$, and which maps $a_1$ to the term $a_2(x_1,x_1)$. This morphism sends trees that look like words to complete binary trees, as shown below:
	\begin{center}
		\begin{tabular}{cc}
			\includegraphics[scale=0.3]{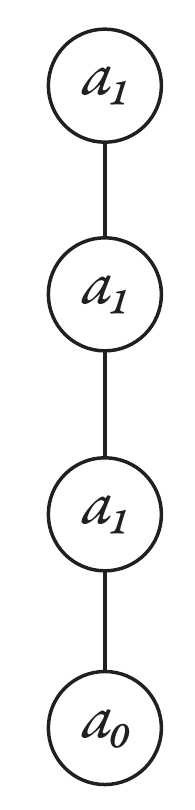} &
						\includegraphics[scale=0.3]{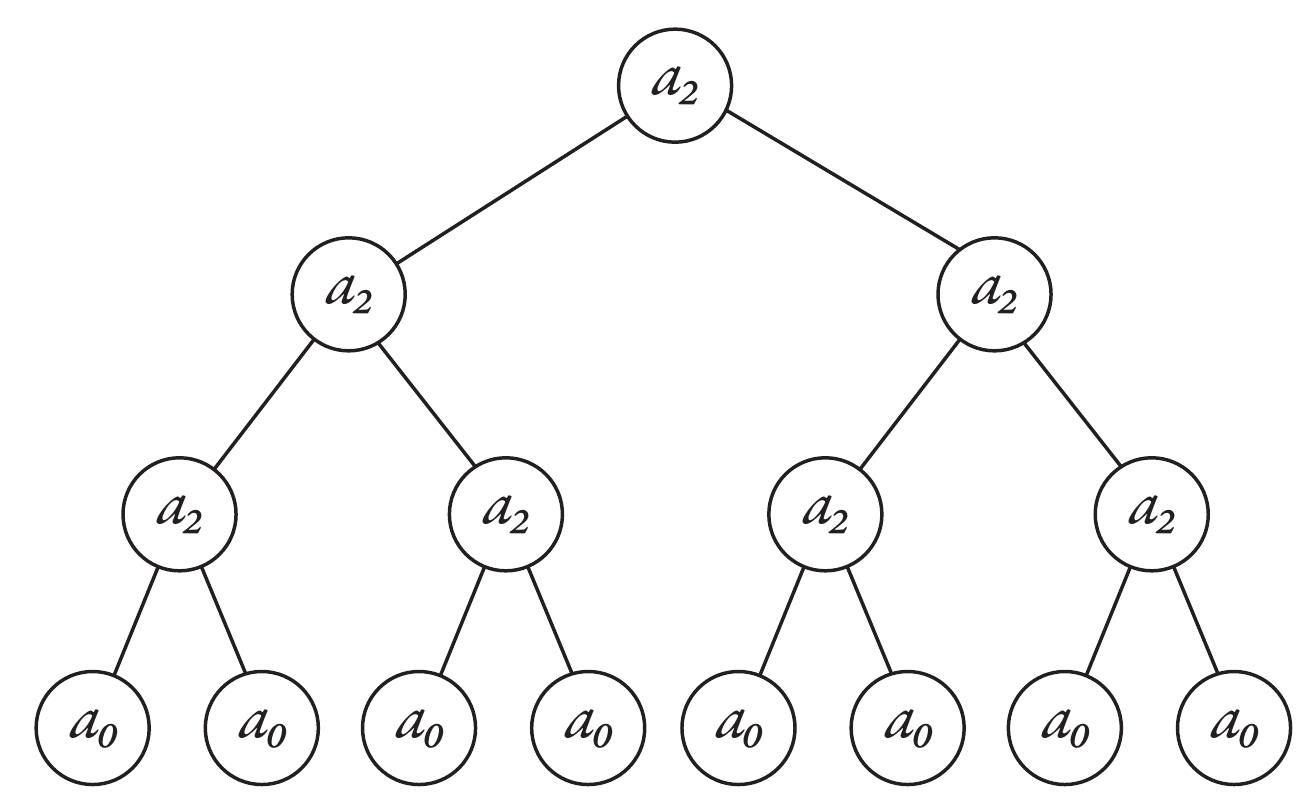} \\
						$t \in \clo\set{a_0,a_1}$ & $h(t) \in \clo\set{a_0,a_2}$
		\end{tabular}
	\end{center}
	There is a first-order formula $\varphi$ that is true in complete binary trees of even depth, and false in complete binary trees of odd depth. The formula says that if one follows the unique path that begins in the root, and then turns left, right, left, right, etc., then one ends up in a leaf that is a left child. The inverse image, under the $\clo$-morphism $h$, of the language defined by $\varphi$ is the set of trees over alphabet $\set{a_0,a_1}$ which have even depth. This inverse image is not definable in first-order logic, and therefore first-order definable tree languages are not closed under inverse images of $\clo$-morphisms.
	
	In particular, first-order logic does not form a pseudovariety of $\clo$-languages. Therefore clones, or at least syntactic clones, are not the right tool to study first-order logic on trees.    As shown in~\cite{esik2003logically}, his problem can be solved by using preclones, which are a  variant of nonduplicating clones where every variable is used only once.   The inadequacy of clones in this context is  a bit of a shame, because clones have a better developed theory than preclones, e.g.~Rosenberg classifies  clones with a minimal set of operations that contains something other than projections~\cite{rosenberg1986minimal} or clones with a maximal set of operations that does not contain all operations~\cite{rosenberg1970funktionale}, while Hobby and McKenzie classify congruences in a finite clone~\cite{hobby1988structure}.
\end{myexample}

\subsection{Forests of unranked trees}
\stepcounter{monadcounter}
We present a  monad for modelling trees, which corresponds to forest algebra~\cite{DBLP:conf/birthday/BojanczykW08}. As in the  previous two monads, the  trees are finite (finitely many nodes) and labelled (each node comes with a label). Unlike for the two previous monads trees are  unranked, i.e.~the number of children of a node is not determined by its label, and can be arbitrarily large. We also assume that  trees are sibling-ordered, i.e.~the children of a node come with a total order. Finally, instead of trees it will be more convenient to talk about \emph{forests}, which we define to be ordered sequences of trees, i.e.~ordered sequences of trees that are unranked, labelled and sibling-ordered. Here is a picture of a forest:
\begin{align}\label{eq:forest}
	\vcenter{\hbox{\includegraphics[scale=0.5]{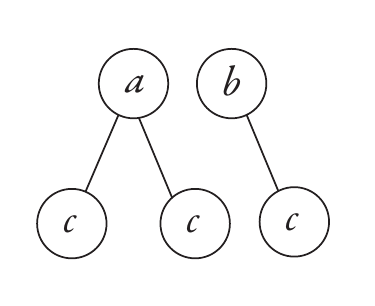}			}}
\end{align}

\paragraph*{Forests and contexts.} The monad in this section will correspond to forest algebra. The principal idea behind forest algebra is to use two kinds of objects, namely forests and contexts.  Forests have already been described above. 
A \emph{context}  is defined to be a forest with exactly one distinguished leaf, which is called the \emph{port} of the context\footnote{One could consider a variant of this monad without the requirement that the port  appears in exactly one leaf, we keep this requirement so that the monad ends up describing forest algebra introduced  in~\cite{DBLP:conf/birthday/BojanczykW08}. Furthermore, allowing ports in many leaves would break the argument in Example~\ref{ex:forest-fo-pseudo}, actually first-order logic would no longer be a pseudovariety.}. Here is a picture of a context, with the port being labelled by $x_1$:
\begin{align}\label{eq:context}
\begin{aligned}
\vcenter{\hbox{	{\includegraphics[scale=0.5]{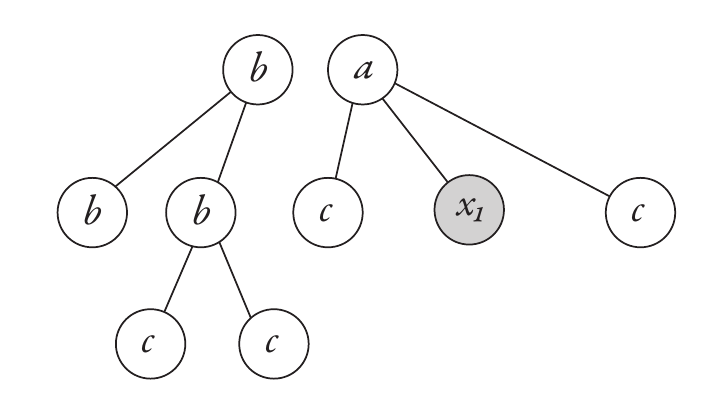}}}}
\end{aligned}
\end{align}
The idea behind the port is that it  can be replaced  with a forest (or another context). One needs to be careful with the notion of replacement, because a port is a single node, while the forest that will replace it might have multiple roots, e.g.~the forest  in~\eqref{eq:forest}. The result of the replacement is that the all the roots of the inserted forest become children of the parent of the port, e.g.~the result of replacing the port of~\eqref{eq:context} by the forest~\eqref{eq:forest} is illustrated below, with the grey background indicating what used to be the port:
\begin{align}\label{eq:forest-in-context}
\vcenter{\hbox{	\includegraphics[scale=0.5]{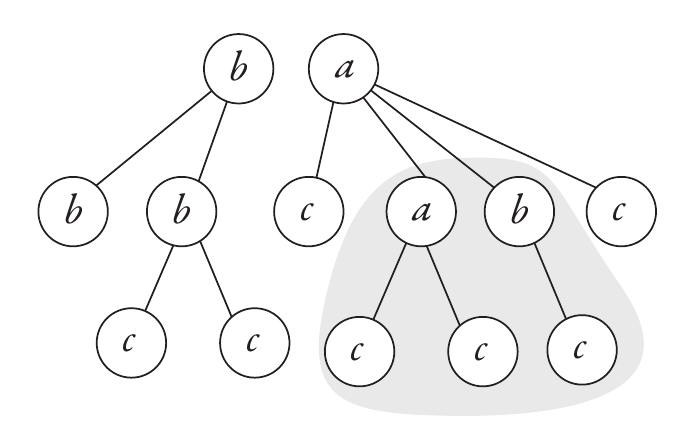}}}
\end{align}

\paragraph*{The forest monad.} We now define the monad for forests and contexts, which is called the \emph{forest monad}. Line in all previously considered monads, the main idea is that one can replace any node with another element of the monad. In the forest monad, we will use the following discipline: leaves in a forest or context are be replaced by forests, while non-leaves are be replaced by contexts. This leads to a two-sorted alphabet: there are \emph{forest labels}, which are found in leaves, and there are \emph{context labels}, which are found on non-leaves.  

\begin{figure}[htbp]
	\centering
		\begin{tabular}{cc}
			\includegraphics[scale=0.5]{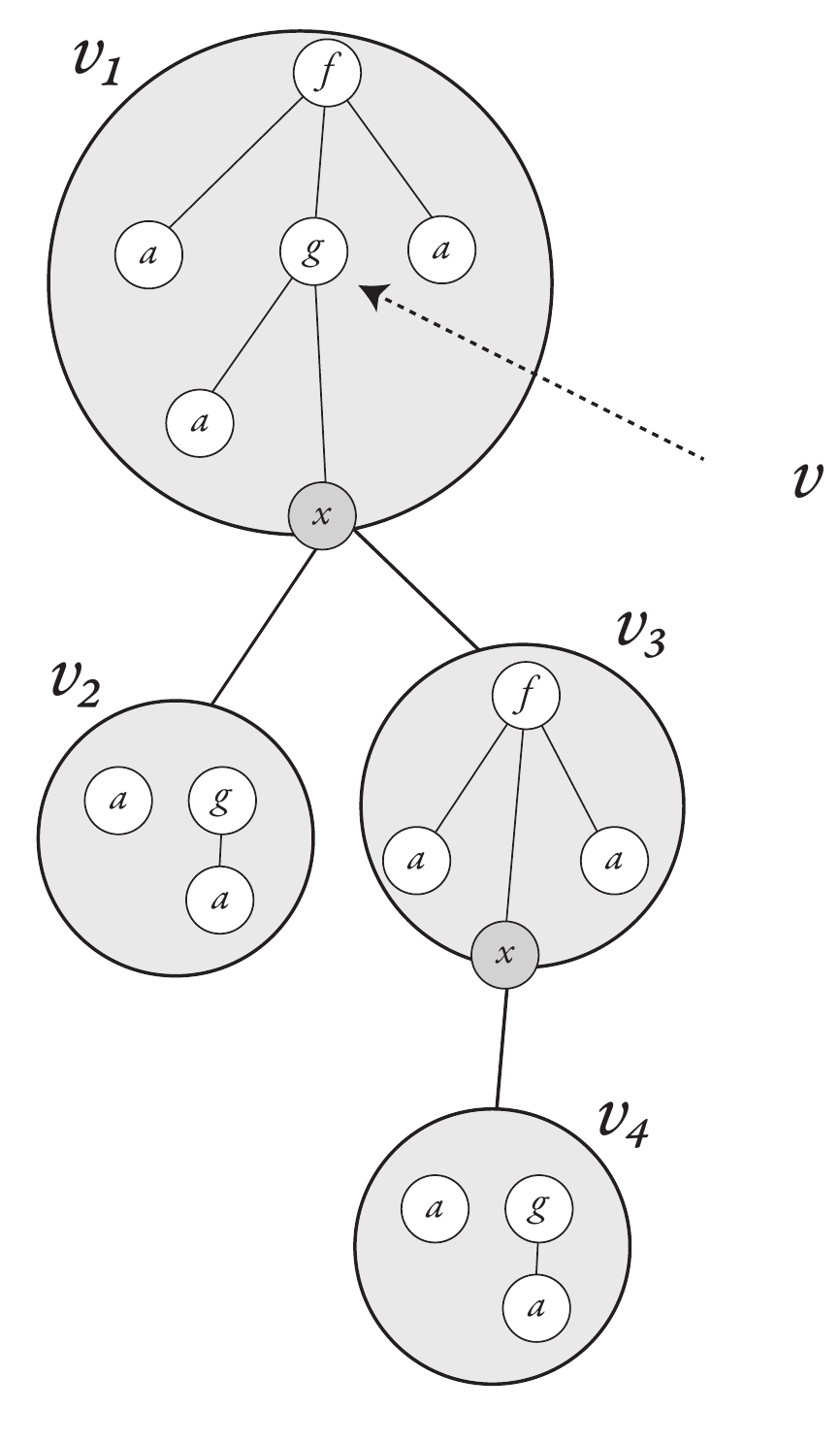} & 
			\includegraphics[scale=0.5]{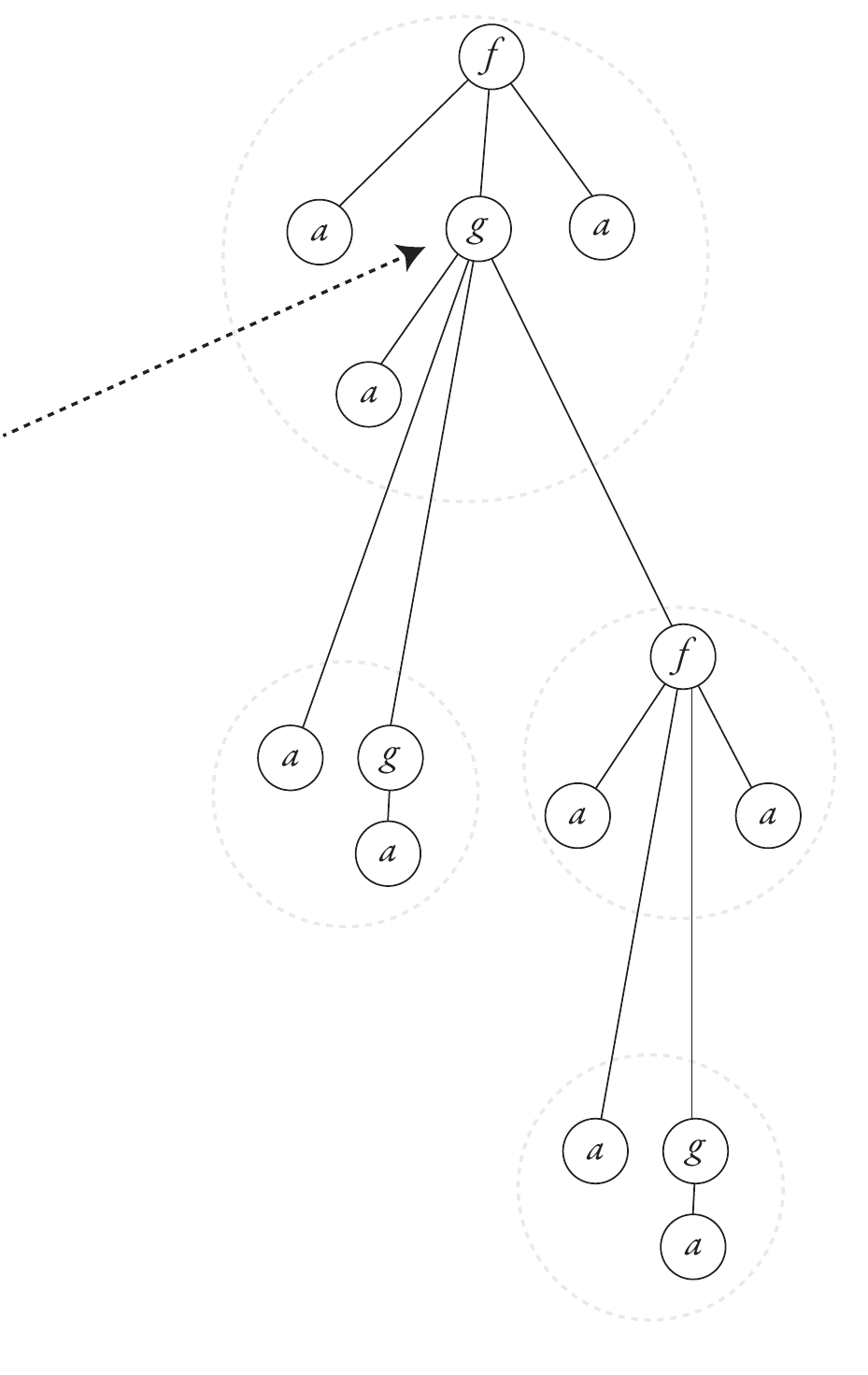} \\
			$t \in \forest \forest \Sigma$ & $\mult_{\forest \Sigma} (t) \in \forest \Sigma$
		\end{tabular}
	\caption{Example of multiplication in the forest monad. Before multiplication, $t$ has two context nodes $v_1$ and $v_3$ and two forest nodes $v_2$ and $v_4$.   After multiplication, $t$ has fourteen nodes, which correspond to the non-variable nodes in the labels of $v_1,\ldots,v_4$. Note how the $x$ in the label of node $v_1$ is replaced by three nodes, namely the two roots of $v_2$ and the one root of $v_3$, resulting in a change of the number of children for node $v$.}
	\label{fig:forest}
\end{figure}

More formally, the forest monad, denoted by $\forest$, is in the category of two-sorted sets, where the sort names are ``forest'' and ``context''.  When applied to a  sorted set $\Sigma$, the forest monad $\forest$ yields the following sorted set $\forest \Sigma$
\begin{itemize}
	\item on the forest sort, $\forest \Sigma$ contains nonempty forests labelled by $\Sigma$ such that leaves are labelled by letters of ``forest'' sort, while non-leaves are labelled by letters of ``context'' sort;
		\item on the context sort, $\forest \Sigma$ contains contexts labelled by $\Sigma$ in the same way as in the previous item.
\end{itemize}
The unit operation in the monad $\forest$ maps a forest  element $a$ to 
\emph{unit forest} that looks like this
\begin{center}
	\includegraphics[scale=0.5]{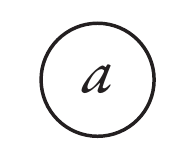}	
\end{center}
and maps a context element $a$ to a \emph{unit context} that looks like this
\begin{center}
	\includegraphics[scale=0.5]{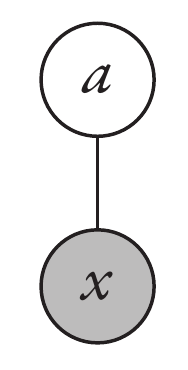}
\end{center}
  The multiplication   operation in the monad is based on the intuitions of replacement depicted in pictures~\eqref{eq:forest},~\eqref{eq:context} and~\eqref{eq:forest-in-context}. The operation is   illustrated in Figure~\ref{fig:forest}.  

A \emph{finite alphabet} is  a two-sorted set finite that is finite on both sorts, and  a \emph{ finite $\forest$-algebra} is one whose universe is finite. 
\begin{lemma}\label{lem:forest-algebra}
	Every $\forest$-algebra is spanned  by the subfunctor $\forest_0$ which maps $\Sigma$ to
	\begin{align*}
		\includegraphics[scale=0.4]{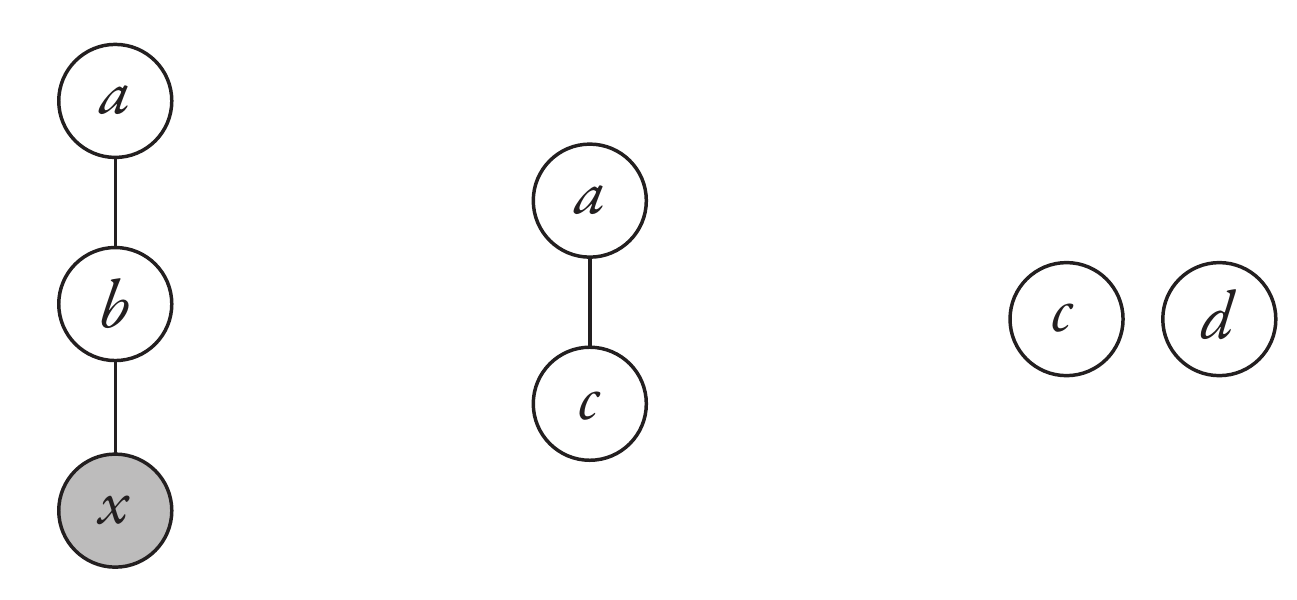}
	\end{align*}
	where    $a,b$ are context elements of $\Sigma$, and $c,d$  are forest element of $\Sigma$.
\end{lemma}
\begin{pr}
	The lemma boils down to  the following easy fact. Every forest or context can be built out of units forests and unit contexts by the following operations: replacing the port of a context by another context or a forest, and concatenating two forests.
\end{pr}

By the above lemma, every $\forest$-algebra is uniquely determined by its  $\forest_0$-reduct. This reduct is exactly the same thing as a forest algebra from~\cite{DBLP:conf/birthday/BojanczykW08}, in the variant of forest algebra where there is no empty forest or context. One advantage of seeing forest algebras as a special case of monads is the we can apply the general theorems from the first part to see that forest algebra has a syntactic morphism theorem (already known) or a pseudovariety theorem (not present in the literature).

\begin{myexample}\label{ex:forest-fo-pseudo}
	Let us revisit first-order login on trees, as considered in Examples~\ref{ex:tree-fo-pseudo} and~\ref{ex:clone-non-pseudo}. To a forest one can assign a logical structure, where the universe is the nodes of the forest, there are unary predicates for the labels, and two binary predicates for the descendant and document orders (document order is the order in which nodes are visited in depth first search, which takes into account the order on siblings). For a context, the structure is defined the same way, except there is constant which denotes the port. A language $L \subseteq \forest \Sigma$ is called first-order definable if it can be defined by a formula of first-order logic in terms of the logical structure defined above.  One can show that first-order definable language form a pseudovariety. The interesting case is to show that if 
	\begin{align*}
		h : \forest \Sigma \to \forest \Gamma
	\end{align*}
	is a $\forest$-morphism, then 
	\begin{align*}
		t \sim_n s \quad \mbox{implies} \quad h(t) \sim_n h(s) \qquad \mbox{for every $n \in \Nat$,}
	\end{align*}
	where $\sim_n$ says that the corresponding logical structures have the same first-order theory of rank $n$.  This is proved using the same origin and offset argument as in Example~\ref{ex:tree-fo-pseudo}. 	
\end{myexample}

\newcommand{\omegaforest}{\omega\forest}
\subsection{A monad for infinite unranked forests.} \stepcounter{monadcounter} The $\omega$-forest monad, denoted by $X \mapsto \omegaforest X$, is defined like the monad $\forest$, with the difference that  infinite forests  and  infinite contexts are also allowed (assume finite branching, though). The  problem with this monad is that it is unclear what finite algebra should be in this case. Clearly, the algebra needs to be finite on both sorts, but this is not sufficient, as the following example shows.

\newcommand{\dense}{\mathsf{dense}}
\begin{myexample}
	Consider an alphabet $\Sigma$ in the sense of the monad $\omegaforest$, i.e.~an alphabet with elements of sorts ``forest'' and ``context''. Let $L$ be an arbitrary set of trees over the alphabet $L$, not necessarily \mso definable. Define $\dense L$ to be those forests where every node has a some descendant with a subtree in $L$. We claim that $\dense L$ is recognised by an $\omegaforest$-algebra $\alg$ with a four element universe. There forest sort has elements ``forests in $L$'' and ``forests not in $L$''. The context sort has elements:  ``every node outside the port path has some descendant with a subtree in $L$'' and ``some node outside the port path has  no descendants with a subtree in $L$''; where the port path is defined to be the ancestors of the port. The dependence on $L$ in the algebra $\alg$ is seen in the multiplication operation
	\begin{align*}
		\mult_{\alg} : \omegaforest A \to A
	\end{align*}
	which maps infinite objects to elements of the universe $A$.
	 In particular, there are uncountably many $\omegaforest$-algebras with finite universes, and there is no hope of representing them in a finite way.
	\end{myexample}

As witnessed by the above example, the notion of finite algebra should have some additional requirements. Let us make the design decision that
 languages recognised by finite algebras should be exactly those that can be defined in \mso.  The question of finding an adequate notion of finite $\omegaforest$-algebra is a monad formulation of an open problem in the community of algebraic language theory, namely the problem of a finding an algebraic model for \mso on infinite tees. The fact that we use monads, or that the trees are unranked, does not seem to be important. 

A simpleminded solution is to define an $\omegaforest$-algebra $\alg$ to be finite if its universe is finite on both sorts, and the multiplication operation is \mso definable, in the sense that every language
\begin{align*}
	\mult_{\alg}^{-1}(a) \subseteq \omegaforest A \qquad \mbox{with $a \in A$}
\end{align*}
is \mso definable. Adjusting for a different terminology, this is the solution proposed in~\cite{idziaszek_ef}, where it is  shown that syntactic algebras can be computed, one can check if an algebra satisfies given equalities, and the algebras can be used to decide questions such as ``is a given language of infinite trees definable in the temporal logic $\mathsf{EF}$?''.  This definition of finite algebra is compatible with the results from Part I, in particular with the Syntactic Morphism Theorem and the Pseudovariety Theorem. Examples of language classes that are pseudovarieties include: languages defined in weak \mso, i.e.~only using existential quantification over finite sets; languages recognised by nondeterministic (respectively, alternating) tree automata that use parity ranks from a given subset $\Omega \subseteq \Nat$.

\section{Future work}
\label{sec:more}
This section sketches some potential monads to study in the future, with reasons for studying them.

\begin{itemize}
	\item Unranked trees with possibly infintie branching (or graphs, which should not make a difference) modulo bisimulation. The hope would be that recognisable languages, under a suitably chosen notion of finite algebra, would be the same thing as definable in $\mu$-calculus.
	\item Edge labelled hypergraphs. This looks like a monad, because a hypergraph with $n$ distinguished port vertices can be substituted for a hyperedge of rank $n$, in the same spirit as Figure~\ref{fig:clo}. The hope would be to describe tree width or clique width as submonads generated by finite subfunctors (as defined in Section~\ref{sec:representing-an-algebra}).
	\item Typed terms of $\lambda$-calculus with fixpoints, modulo equivalence. The hope would be to describe the work of Salvati and Walukiewicz.
	\item Relations on words with origin information, as a generalisation of transducers with origin information from~\cite{DBLP:conf/icalp/Bojanczyk14}. The hope would be to give an algebraic framework for asynchronous relations on words with origin. The origin information would cure problems like no syntactic object, or undecidability of universality, which plague asynchronous relations without origin.
\end{itemize}
%

\mypart{Profinite Monads}{In this part, we show that for every monad $\monad$, at least in the category of sets, there is a profinite version $\promonad$. This gives immediately definitions, and basic theorems about, things like profinite words, profinite countable chains, profinite trees, etc.
We also study the special case  of profinite words, and show how the generic notion of recognisable language instantiates to an interesting class of languages of profinite words.}
\section{Stone duals and topology on an algebra}
\label{sec:topology}
Profinite constructions are common in mathematics. For recognisable languages, the best known profinite construction is the semigroup of profinite words. In this section, we show how profinite are defined on the abstract level of monads. The main results of this section are:
\begin{itemize}
	\item Lattices of languages are exactly those families of languages that can be defined by profinite implications. This generalises to monads a result that was proved for semigroups in~\cite{gehrke2008duality}.	As a corollary, we get a monad generalisation of the Reiterman theorem~\cite{reiterman1982birkhoff}, which says that pseudovarieties are exactly those families of languages that can be defined by profinite identitites, which are a stronger form of profinite implications. 
	\item Every class of languages, e.g.~context-free or decidable, can be used to yield get some kind of profinite object, but only recognisable languages can be used if we want algebraic operations to be uniformly continuous. These results are monad generalisations of results that were proved for semigroups in~\cite{gehrke2010topological}.
\end{itemize}

 \subsection{Stone duals of Boolean algebras}
 \label{sec:stone-boolean}
 In this section, we recall the definition of the Stone dual of a Boolean algebra, and how Stone duals can be used to characterise lattices. Section~\ref{sec:stone-boolean} does not talk about monads.
 
  Consider  a Boolean algebra
 \begin{align*}
 	(A, \cap, \cup, \neg).
 \end{align*}
 Define an \emph{ultrafilter} in $A$ to be a proper subset $U \subset A$ which  is  closed under intersections, and which contains every element of $A$ or its complement but not both. The \emph{Stone dual} of $A$, denoted by $\stone A$, is defined to be the following topological space. The points in the space are ultrafilters in the Boolean algebra. The topology is  generated by base open sets which are of the form
 \begin{align*}
 	\bar a \eqdef \set{U : \mbox{$U$ is an ultrafilter containing $a$}} \qquad \mbox{for }a \in A.
 \end{align*}
 The topology on the  Stone dual is known to compact and Hausdorff.
 One of the advantages of the Stone dual is that it can be used to describe lattices of languages, including the special case of pseudovarieties of languages.

 \paragraph*{Profinite implications.} We begin by repeating a result from~\cite{gehrke2008duality}, which says that for an arbitrary Boolean algebra,  lattices are exactly those sets which are defined by profinite implications. (This terminology is different than~\cite{gehrke2008duality}, which uses the name ``equation'' for what we call an implication.)
 
 Consider a Boolean algebra $A$. (In the context of this paper, it is convenient to think of $A$ as being all recognisable subsets of $\monad \Sigma$. In this case, elements of the Boolean algebra are themselves sets.)
 A \emph{profinite implication} over a  Boolean algebra $A$ is  an expression of the form $w \to v$, where $w,v \in \stone A$. The arrow is just part of the syntax, so formally a profinite implication is simply a pair of elements from $\stone A$. An element  $a \in A$  (e.g.~a recognisable language, when the Boolean algebra consists of recognisable languages) is said to satisfy the implication if  $a \in w$  implies  $a \in v$. A subset $B \subseteq A$  (e.g.~a family of recognisable languages when the Boolean algebra consists of recognisable languages) is said to be defined by a set of profinite implications if it contains exactly the elements that satisfy all profinite implications from the set. In the following lemma, $B$  is called a \emph{lattice} if it contains the $0$ and $1$ in a Boolean algebra, and  is closed under finite unions and finite intersections.
The following theorem can be found implicitly in~\cite{gehrke2008duality}, and maybe earlier as well.
 \begin{theorem}\label{thm:implications}
 	Let $A$ be a Boolean algebra and  let   $B \subseteq A$. Then $B$ is a lattice if and only if it is definable by profinite implications.
 \end{theorem}
 \begin{pr}
 	It is easy to see that if a subset of $A$ is definable by profinite implications, then it is a lattice. To prove the other implication, consider  a lattice  $B \subseteq A$.  We  claim that $B$ is defined by the set of profinite implications:
 	\begin{align}\label{eq:set-of-implications}
 		\set{x \to y : \mbox{for every $a \in B$, if $a \in x$ then $a \in y$}}.
 	\end{align}
 By definition, every element of $B$ satisfies the profinite implications above. Let then $a \in A$ be such that $a$ satisfies all the profinite implications above. To prove the theorem, we will to show $a \in B$. 

Recall the definition of $\bar a$ in the definition of the Stone dual, which is that $\bar a$ is the set defined by $a \in x$ iff $x \in \bar a$. We first claim that every $x \in \bar a$ satisfies
 	\begin{align}\label{eq:claim-covers}
 		x \in \bar b \subseteq \bar a \qquad \mbox{for some $b \in B$}
 	\end{align}
 	For $x \in \bar a$, define $[x] \subseteq \stone A$ to be the intersection
 		\begin{align*}
  	\bigcap_{b \in B \cap x} \bar b.
 		\end{align*}
 		It is easy to see that $[x]$ is the set of all $y \in \bar A$ such that the profinite implication $x \to y$ belongs to the set~\eqref{eq:set-of-implications}.  By assumption that $a$ satisfies all these  profinite implications, it follows that $[x] \subseteq \bar a$. Note that $[x]$ is  an intersection of sets that are closed. By compactness of the Stone dual,  $[x]$ is equal to  an intersection of finitely many $\bar b$ with $x \in b \in B$. Furthermore, the intersection is nonempty, because $B$ contains the greatest element of the Boolean algebra, being a lattice. Because  $B$ is closed under finite intersections, it follows that $[x]=\bar b$ for some $b \in B$ with $x \in \bar b$. Together with $[x] \subseteq \bar a$, this proves~\eqref{eq:claim-covers}.

 From~\eqref{eq:claim-covers} it follows that $\bar a$ is the union of all $\bar b$ ranging  over $b \in B$ such that $\bar b \subseteq \bar a$.  By compactness, the union can be made finite, and by closure of $B$ under finite union, it follows that there is some $b \in B$ such that $\bar a = \bar b$. Finally, since $a,b$ are in the Boolean algebra, it follows that $a=b$.
 \end{pr}

\subsection{Stone duals of $\monad$-algebras}
Section~\ref{sec:stone-boolean} did not use monads and recognisability. In this section, we consider the special case of Stone duals of recognisable languages in a $\monad$-algebra. Using this Stone dual, we prove a monad version of the Reiterman theorem, which says that a class of recognisable languages is a language pseudovariety if and only if it can be defined by profinite identities.

\paragraph*{The Stone dual of a $\monad$-algebra.} Fix a monad $\monad$ in the category of sets. We assume that finite alphabets are finite sets, and finite algebras are algebras with finite universes. The results can be easily generalised to sorted sets. Let $\alg$ be a $\monad$-algebra, not necessarily finite. Define $\reco \alg$ to be the subsets of the universe of $\alg$ that are recognised  by $\monad$-morphisms from $\alg$ into finite $\monad$-algebras.   Since $\reco \alg$ is a Boolean algebra,  it has a Stone dual, which we denote by  $\stone \alg$.  

\begin{myexample}\label{ex:profinite-words} 
	Consider the monad of finite nonempty words, where $+$-algebras are semigroups, and $+$-morphisms are semigroup morphisms. Consider the semigroup $\Sigma^+$ where $\Sigma$ is a  a finite alphabet. An element of $\stone \Sigma^+$ is an ultrafilter in the Boolean algebra of recognisable languages over $\Sigma$. Recalling the definition of an ultrafilter, an element of $\stone \Sigma^+$ is a family of  recognisable languages over $\Sigma$, which is closed under intersection, and which contains every recognisable language or its complement.

  A simple example of such an ultrafilter is  one that is induced by a word $w \in \Sigma^+$, namely the ultrafilter of recognisable languages which contain $w$.  Stated differently, $\stone \Sigma^+$ can be seen as a generalisation of $\Sigma^+$.
 
Here is  a more exciting ultrafilter, which corresponds to taking the idempotent power of a finite word. Recall the well known fact that in every finite semigroup $\alg$ of size $n$, the function $a \mapsto a^{n!}$ maps every element of $\alg$ to an idempotent, i.e.~
 \begin{align*}
 	a^{n!} \cdot a^{n!} = a^{n!},
 \end{align*}
 and this element is the unique idempotent power of $a$, i.e.~if
 \begin{align*}
 	a^k \cdot a^k = a^k \qquad \mbox{implies} \qquad a^k = a^{n!}.
 \end{align*}
We write $a^{\#}$ for this idempotent power. A common notation would be $w^\omega$, but choose $\#$ to avoid conflict with the $\omega$ power in infinite words.  For every semigroup morphism $h : \Sigma^+ \to \alg$, and every $w \in \Sigma^+$, we have
\begin{align*}
	h(w^{n!})  = h(w)^\# \qquad \mbox{for all but finitely many $n$}.
\end{align*}
This implies that for every recognisable language $L \subseteq \Sigma^+$ and every word $w \in \Sigma^+$, either $L$ contains $w^{n!}$ for all but finitely many $L$; or $L$ does not contain $w^{n!}$ for all but finite many $L$. This in turn implies  that  the set of languages
 \begin{align*}
 	\set{L \subseteq \Sigma^+ : \mbox{$L$ is recognisable and $w^{n!} \in L$ for all but finitely many $n$}}
 \end{align*}
 is an ultrafilter, which we denote by $w^{\#}$. 
\end{myexample}

\begin{running}\label{ex:infty-profinite-idempotent}
	Consider the monad of $\infty$-words used in the running example. As for semigroups, for every $\infty$-algebra $\alg$ and every $a \in \alg$ there is a unique idempotent power $a^\#$. Let $w \in \Sigma^+$ is a finite nonempty word. As in Example~\ref{ex:profinite-words}, one can also define profinite $\infty$-word $w^{\#}$, namely the ultrafilter
    \begin{align}\label{eq:run-w-stab}
    	\set{L \subseteq \Sigma^\infty : \mbox{$L$ is recognisable and $w^{n!} \in L$ for all but finitely many $n$}}.
    \end{align}
Note that in the monad for $\infty$-words, the notation $w^\omega$ stands for an actual infinite word, which can then be treated as a profinite word, i.e.~
\begin{align}\label{eq:run-w-omega}
	\set{L \subseteq \Sigma^\infty : w^\omega \in L} 
\end{align}
\end{running}

Theorem~\ref{thm:implications} can be applied to $\stone \alg$; for instance if the monad is the monad of finite words, and $\alg$ is $\Sigma^+$, then Theorem~\ref{thm:implications} says that a family of recognisable languages over $\Sigma$ is a lattice if and only if it is definable by a set of profinite identities.

\begin{running}
	Consider a language $L \subseteq \Sigma^\infty$.  Define the \emph{first difference distance} between two $\infty$-words to be zero if they are equal, and  otherwise to be $1/n$ where $n$ is the first position where the words have a different label. Define a \emph{safety} language to be a set of $\infty$-words which is closed under limits with respect to first difference distance. In other words, safety says that if $w$ is an infinite word such that every finite prefix of $w$ can be extended to some word from the language, then $w$ itself belongs to the language. 
	
	It is easy to see that recognisable safety $\infty$-languages form a lattice, and therefore by Theorem~\ref{thm:implications} they must be characterised by a set of profinite implications. One can show that the set of profinite implications is 
	\begin{align}\label{eq:run-implication-infty}
		vw^\#u \to vw^\omega,
	\end{align}
	where $v,w,u \in \Sigma^+$ and the powers $^{\#}$ and $^\omega$ are understood as profinite words in the same sense as in~\eqref{eq:run-w-stab} and~\eqref{eq:run-w-omega}, i.e.~the two sides of the above profinite implication are the following ultrafilters, respectively.
	    \begin{align*}
&	 \set{L \subseteq \Sigma^\infty : \mbox{$L$ is recognisable and $vw^{n!}u \in L$ for all but finitely many $n$}}\\
&		\set{L \subseteq \Sigma^\infty : vw^\omega \in L} 
	\end{align*}
Indeed, suppose that $L$ is a safety language, and it satisfies the left side of the profinite implication for some $v,w,u$, which means that it contains $vw^{n!}u$ for almost all $n$. By safety, the language $L$ must also contain  the limit of the sequence $vw^{n!}u$, which is $vw^\omega$, and therefore $L$ satisfies the right side of the profinite implication. The more interesting case is the  converse, i.e.~showing that if $L$ satisfies all profinite implications of the form~\eqref{eq:run-implication-infty}, then it is a safety language. To prove that $L$ is a safety language, assume that it contains all words
\begin{align*}
	w_1,w_2,\ldots
\end{align*}
which tend, under first difference distance, to some word $w$. We need to show that $L$ also contains $w$. If $w$ is finite, then all but finitely many of the words $w_i$ are equal to $w$, and therefore $w \in L$. Assume therefore that $w$ is infinite. By the Ramsey Theorem, $w$ can be factorised as
\begin{align*}
	w= v_0 v_1 v_2 \cdots
\end{align*}
such that all word $v_1,v_2,\ldots$ have the same image under some 
$\infty$-morphism
\begin{align*}
	h: \Sigma^\infty \to \alg
\end{align*}
which recognises $L$. This means that for every $i$, all but finitely many of the words $w_j$ have a prefix of the form $v_0 v_1 \cdots v_i$. Without loss of generality, we may assume that 
\begin{align*}
	w_i = v_0 v_1 \cdots v_i u_i,
\end{align*}
and also without loss  of generality we may assume that all words $u_i$ have the same image under the $\infty$-morphism $h$. Since $L$ is recognised by $h$, it follows that $L$ contains all words of the form $v_0 (v_1)^n u_1$. By~\eqref{eq:run-implication-infty}, $L$ also contains $v_0 (v_1)^\omega$, which has the same image under $h$ as $w$, and therefore $L$ also contains $w$.
\end{running}



 \paragraph*{Defining pseudovarieties by identities} As mentioned in its proof, Theorem~\ref{thm:implications} does not use any properties of recognisability over a monad.  We now present a corollary of the theorem, which is more specific to monads, and which says that pseudovarieties can be defined by identities.

To define profinite identities, we observe that both $\monad$-morphisms and unary polynomials can be naturally lifted to profinite objects, as described below. Suppose that 
\begin{align*}
	f : \alg \to \balg
\end{align*}
is a function, not necessarily a $\monad$-morphism, which has the property that recognisable languages are preserved under inverse images of $f$, i.e.
\begin{align}\label{eq:inverse-f-preserves-reco}
	L \in \reco \balg \qquad \mbox{implies} \qquad f^{-1}(L) \in \reco \alg.
\end{align}
Then for every ultrafilter $U$ of recognisable languages subsets of $\alg$, the family
\begin{align*}
	(\stone f) (U) \eqdef \set{L \in \reco \balg : f^{-1}(L) \in U }
\end{align*}
is an ultrafilter of recognisable subsets of $\balg$. In other words, $f$ lifts to a function
\begin{align*}
	\stone f : \stone \alg \to \stone \balg.
\end{align*}
One can show that the mapping $\stone$ defined this way is a functor, whose domain is the category of $\monad$-algebras with functions that satisfy~\eqref{eq:inverse-f-preserves-reco}.
We will be interested in two special cases of functions $f$ with property~\eqref{eq:inverse-f-preserves-reco}, i.e.~when $f$ is a $\monad$-morphism and when $f$ is a unary polynomial.  The following fact is an immediate consequence of the definitions.

\begin{fact}\label{fact:stone-pull}
	If $f : \alg \to \balg$ satisfies~\eqref{eq:inverse-f-preserves-reco}, and $L \subseteq \balg$, then
	\begin{align*}
		f^{-1}(L) \mbox{ satisfies } w\to v \qquad \mbox{iff} \qquad L \mbox{ satisfies } (\stone f)(w) \to (\stone f)(v)
	\end{align*}
\end{fact}

 Define a \emph{profinite identity} to be an expression of the form $	w = v$ where  $w,v \in \stone \monad X$ for some finite set   $X$ of variables. As in profinite implications, the equality sign is just part of the syntax, and formally a profinite identity is simply the pair $(w,v)$. If $\alg$ is a $\monad$-algebra, then we  say that  $L \in \reco  \alg$ satisfies a profinite identity $w=v$ if it satisfies
 \begin{eqnarray*}
 		 (\stone (p \circ h))  (w)   \leftrightarrow 	 (\stone (p \circ h)) (v)
 \end{eqnarray*}
 for every unary polynomial $p \in \upol 1 \alg$ and every $\monad$-morphism $h : \monad X \to \alg$, where $\leftrightarrow$ means that the profinite implication is satisfied both ways. As mentioned above, the mapping $\stone$ is a functor, and therefore $\stone (p \circ h)$ is the same as $(\stone p) \circ (\stone h)$. Intuitively speaking, for every substitution of the variables, i.e.~every morphism $h$, and in every environment, i.e.~for every unary polynomial $p$, the two sides of the profinite identity are equivalent.
 
 \begin{myexample}
 	Consider the monad of finite words, and the profinite identity 
	\begin{align*}
		xy=yx.
	\end{align*}  Formally speaking, the profinite identity uses   profinite words, call them ``$xy$'' and ``$yx$'', which correspond to the finite words $xy$ and $yx$, as described in the second paragraph of  Example~\ref{ex:profinite-words}. A recognisable language $L \subseteq \Sigma^+$ satisfies this profinite identity if 
	\begin{align*}
		p^{-1}L \in \mbox{``$wv$''} \qquad \mbox{iff} \qquad p^{-1}L \in \mbox{``$vw$''}
	\end{align*}
		holds for every unary polynomial $p$ over $\Sigma^+$ and every $w,v \in \Sigma^+$. 
	Unraveling the definitions of the profinite words ``$xy$'' and ``$yx$'', this means that
	\begin{align*}
		wv  \in p^{-1}L \qquad \mbox{iff} \qquad vw \in p^{-1}L.
	\end{align*}
This means that the language must be commutative.
 \end{myexample}
 
  \begin{myexample}
  	Consider again the monad of finite words. Recall the  profinite word $w^{\#}$ that was described in Example~\ref{ex:profinite-words}. In a similar way, we can define a profinite word $w^{\#+1}$ to be the ultrafilter
	\begin{align*}
		\set{L \subseteq \Sigma^+ : \mbox{$L$ is recognisable and $w^{n!+1} \in L$ for all but finitely many $n$}}.
	\end{align*}
	Consider the following  profinite identity over a single variable $x$:
	\begin{align*}
		x^{\#}=x^{\#+1},
	\end{align*}
	which the reader might recognise as the identity defining aperiodic semigroups. We now check that this is the case under the definitions of this section.
A recognisable language $L \subseteq \Sigma^+$ satisfies this profinite identity if for every unary polynomial $p$ over the semigroup $\Sigma^+$ and every $w \in \Sigma^+$, the following conditions are equivalent
	 \begin{itemize}
	 	\item $w^{n!}$ belongs to $p^{-1}L$ for all but finitely many $n$;
		\item $w^{n!+1}$ belongs to $p^{-1}L$ for all but finitely many $n$.
	 \end{itemize}
	 This implies that for every word $w$ and unary polynomial $p$, the language $p^{-1}L$ contains either finitely many, or all but finitely many, of the powers $w^n$.	 
This means that the syntactic semigroup of the language is aperiodic, which means that the language is definable in first-order logic, by Sch\"utzenberger's theorem.
  \end{myexample}
 %

The above two examples showed that, in the monad of finite words, some classes of recognisable languages can be characterised via profinite identities. 
The following theorem, which is a monad version of the Reiterman Theorem~\cite{reiterman1982birkhoff}, says that this is the case for all pseudovarieties, although infinite sets of identities might need to be used. Note that although profinite identities can be evaluated in a recognisable subset of an arbitrary $\monad$-algebra, in the following theorem we talk only about $\monad$-languages, i.e.~recognisable subsets of algebras of the form $\monad \Sigma$ where $\Sigma$ is a finite alphabet.

The following theorem uses the polynomial variant of language pseudovarieties that is mentioned in Section~\ref{sec:poly-pseudovariety}, i.e.~this is a class of recognisable languages that is closed under polynomial derivatives, inverse morphisms, and Boolean combinations.
 \begin{theorem}\label{thm:reiterman}
 	Let $\langvar$ be class of recognisable $\monad$-languages. Then $\langvar$ is a pseudovariety if and only if it is defined by a set of profinite identities.
 \end{theorem}
\begin{pr}	
	The right-to-left implication  is essentially checking the definitions, while the left-to-right implication is a corollary of Theorem~\ref{thm:implications}.
	\paragraph*{Right-to-left implication.} 
	Suppose that $I$ is a set of profinite identities, and let $\langvar$ be the set of  recognisable $\monad$-languages which satisfy all of these identities.  We need to show that $\langvar$ is a pseudovariety.   As mentioned    in the proof of Theorem~\ref{thm:implications}, satisfying profinite implications is closed under unions and intersections. It is easy to see that satisfying a profinite identity is invariant under complementation. Therefore, $\langvar$ is closed under Boolean combinations. It remains to show that $\langvar$ is closed under inverse images of morphisms and under polynomial derivatives. Let then $L \subseteq  \monad \Gamma$ be a language that satisfies all identities from $I$, and suppose that 
	\begin{align*}
		h : \monad \Sigma \to \monad \Gamma
	\end{align*}
 is a $\monad$-morphism. We will show that  $h^{-1}(L) \subseteq  \monad \Sigma$ also satisfies all identities in $I$. (The proof for polynomial derivatives is the same and is ommitted.) By definition, we need to show that for every profinite identity $w = v$ in $I$ which is over variables $X$, and every 
	\begin{align*}
		f : \monad X \to \monad \Sigma \qquad q \in \upol 1 \monad \Sigma
	\end{align*}
	which are a $\monad$-morphism and unary polynomial, respectively, we have 
	\begin{align*}
		h^{-1}(L) \quad \mbox{ satisfies } \quad (\stone (q \circ f))(w) \leftrightarrow (\stone (q \circ f))(v)
	\end{align*}
By Fact~\ref{fact:stone-pull} and functoriality of $\stone$, the above is equivalent to saying that 
	\begin{align*}
		L \quad \mbox{ satisfies } \quad (\stone (h \circ q \circ f))(w) \leftrightarrow (\stone (h \circ q \circ f))(v),
	\end{align*}
	which  is the same as saying that
	\begin{align*}
		L \quad \mbox{ satisfies } \quad (\stone (r \circ h \circ f))(w) \leftrightarrow (\stone (r \circ h \circ f))(v),
	\end{align*}
	where $r \in \upol 1 \monad \Gamma$ is the image of $q$ under $h$, see~\eqref{eq:commute-polynomials}. Since $h \circ f$ is itself a $\monad$-morphism, the above holds by assumption that $L$ satisfies all profintie identities from~$L$.
	
\paragraph*{Left-to-right implication.}	 Let $\langvar$ be a class of recognisable $\monad$-languages which is a language pseudovariety. We need to show that $\langvar$ is definable by profinite identitites. For a finite alphabet $\Sigma$, define $\langvar_\Sigma$ to be all languages from $\langvar$ over alphabet $\Sigma$, and let  $I_\Sigma$  be the set of profinite implications that are satisfied by all languages in $\langvar_\Sigma$. Define $I$ to be the set of profinite identities $w = v$ such that some $I_\Sigma$ contains the  profinite implication $w \to v$. 
We show below that $\langvar$ is defined by $I$, i.e.~a language belongs to $\langvar$ if and only if it satisfies all identities in $I$. 

\begin{itemize}
	\item Suppose that $L$ satisfies all profinite identities from $I$. In particular, this means that $L$ satisfies all profinite implications from $I_\Sigma$. Since $\langvar$ is a pseudovariety, it follows that $\langvar_\Sigma$ is a lattice, and therefore by Theorem~\ref{thm:implications}, $\langvar_\Sigma$ is defined by the profinite implications from $I_\Sigma$, which means that $L$ belongs to~$\langvar_\Sigma$.
	\item Suppose that $L$ belongs to $\langvar$. We need to show that $L$  satisfies 
	all profinite identities from $I$.  In other words, we need to show that if $w \to v$ is a profinite implication from $I_\Gamma$, then 
	\begin{align*}
		L \quad \mbox{satisfies} \quad (\stone (p \circ h)) (w) \leftrightarrow (\stone (p \circ h)) (v)
	\end{align*}
	for every unary polynomial $p \in \upol 1 \monad \Sigma$ and every $\monad$-morphism $h : \monad \Gamma \to \monad \Sigma$. By Fact~\ref{fact:stone-pull}, and closure propeties of a pseudovariety, this boils down to the profinite implications $w \leftrightarrow v$ being satisfied by a language from $\langvar_\Gamma$, which holds by definition of $I_\Gamma$.
\end{itemize}
\end{pr}

\subsection{Uniform continuity}
The results in Section~\ref{sec:stone-boolean} and~\ref{sec:stone-algebras} did not really use assumptions on recognisability. Actually, Theorem~\ref{thm:implications} would also be true for non-recognisable languages, as shown in the following example.

\begin{myexample}
	Consider the monad of finite words.  For the purpose of this example, defin $\stone \Sigma^+$  to be the Stone dual of the Boolean algebra of decidable languages over the alphabet $\Sigma$, as opposed to the recognisable languages considered in the previous secition. Also for the purpose of this example, define a pseudovariety to be a class of decidable languages that is closed under Boolean combinations, inverse morphisms and polynomial derivatives, e.g.~the polynomial time complexity class {\sc p} is such a pseudovariety. Inspection of the proofs in Section~\ref{sec:stone-boolean} shows that Theorem~\ref{thm:reiterman} would also work in this setup, in particular {\sc p} is definable by profinite identities.
\end{myexample}

In this section, we show that recognisable languages are special in some sense. The result in this section is a  generalisation of  Theorem 4.1 in~\cite{gehrke2010topological} from semigroups to a certain class of monads over sets.

%
%
%
%
%

\paragraph*{Uniformly continuous operations.}  We begin by defining the notion of a uniformly continuous operation in a $\monad$-algebra, with respect to a chosen class of languages. Let
\begin{align*}
	\Ll = \set{L_1,L_2,\ldots}
\end{align*}
be a countable family of subsets of a set $A$, along with some enumeration.  We say that $L \in \Ll$ separates two elements of $A$ if it contains exactly one of them.    Define the $\Ll$-distance on $A$ to be
\begin{align*}
	\distance_{\Ll} (a,b) = \frac 1 {2^{n}} \qquad \mbox{where $n$ is minimal such that $L_n$ separates $a,b$. }
\end{align*}
It is easy to see that this is a distance, assuming that every two elements of~$A$ are separated by some element of $\Ll$. Note how countability is used in the definition.  Unravelling the classical  definition of uniform continuity,  a  function
\begin{align*}
	f : A^n \to A
\end{align*}
 is 
uniformly continuous with respect to  $\Ll$-distance if for every finite set  $\Kk \subseteq \Ll$ there is some finite set $\Mm \subseteq \Ll$ such that
\begin{align*}
\bigwedge_{i} v_i \equiv_\Mm w_i  \quad \mbox{implies} \quad f(v_1,\ldots,v_n) \equiv_\Kk f(w_1,\ldots,w_n)
\end{align*}
where $\equiv_\Kk$ says that elements cannot be separated by languages from $\Kk$, and $\equiv_\Mm$ is the analogous equivalence but lifted pointwise to functions. It follows that although the definition of  $\Ll$-distance depends on the enumeration of $\Ll$, the notion of uniformly continuous function does not. 

The goal of this section is to investigate conditions on $\Ll$ which guarantee that all polynomials of finite arity define uniformly continuous functions. The answer will be that $\Ll$ needs to contain only recognisable languages. We begin by two examples, which show the result for the special case of semigroups.


\begin{myexample}
	\label{example:uniformly-continuous}
	Let $\Sigma$ be a finite alphabet, let $X$ be a set of finite semigroups (e.g. all finite semigroups, or all aperiodic semigroups), and consider the  $\Ll$-distance on  $\Sigma^+$ where $\Ll$ is all subsets of $\Sigma^+$ recognised by semigroups in $X$. We claim that concatenation, which can be seen as a binary polynomial
	\begin{align*}
		(w,v) \mapsto wv
	\end{align*}
	 is uniformly continuous with respect to $\Ll$-distance. We need to show that for every finite $\Kk \subseteq \Ll$ there is some finite $\Mm \subseteq \Ll$ such that 
\begin{align*}
	w_1 \equiv_\Mm w_2 \mbox{ and } v_1 \equiv_\Mm v_2 \quad \mbox{implies} \quad w_1 v_1 \equiv_\Ll w_2 v_2.
\end{align*}
holds for every words $w_1,w_2,v_1,v_2 \in \Sigma^+$.
The languages $\Mm$ can be taken to be all languages recognised by those semigroups that are used to recognise the languages from $\Kk$. The family $\Mm$ is finite  because there are finitely many possible semigroup morphisms from $\Sigma^+$ to a finite set of finite semigroups.  The same solution works for other operations  in  $\Sigma^+$ that can be built using concatenation.
\end{myexample}

\begin{myexample}
	As in the previous example, consider  the $\Ll$-distance on $\Sigma^+$ where $\Ll$ contains some language that is not recognisable. We show that with respect to $\Ll$-distance, concatenation might continuous, but not uniformly continuous.
	
To show that concatenation might be continuous, suppose that $\Ll$ contains all singleton languages, e.g.~$\Ll$ is the decidable languages. This implies that  the topology generated by $\Ll$-distance is discrete, because every singleton set  is open. Therefore, the topology on finite powers of $\Sigma^+$ is also discrete, and thus concatenation is continuous with respect to $\Ll$-distance, like any other operation on this semigroup.

Let us now show that  concatenation is not uniformly continuous. 	Consider some non-recognisable language $L \in \Ll$.  We will show that there is no finite set $\Mm$ of decidable languages such that
\begin{align*}
	w_1 \equiv_\Mm w_2 \mbox{ and } v_1 \equiv_\Mm v_2 \quad \mbox{implies} \quad w_1 v_1 \equiv_{\set L} w_2 v_2.
\end{align*}
 for every words $w_1,w_2,v_1,v_2 \in \Sigma^+$. Let then $\Mm$ be a finite set of languages from $\Ll$, or any languages for that matter. Because  $L$ is not recognisable, there are infinitely many left derivatives, i.e.~languages of the form $x^{-1}L$. Since there are finitely many equivalence classes of $\equiv_\Mm$, there must exist some two words $w_1,w_2$ such that
	\begin{align*}
		w_1 \equiv_\Mm v_2 \qquad \mbox{and} \qquad w_1^{-1}L \neq w_2^{-1}L.
	\end{align*}
The inequality of derivatives means that there is some $v$ such that
\begin{align*}
	w_1 v \not \equiv_{\set L} w_2 v,
\end{align*}
which proves that concatenation is not uniformly continuous.
\end{myexample}

The two examples above are essentially Theorem 4.1 of~\cite{gehrke2010topological}.
The goal of this section is to generalise that result to  algebras over  abstract monads. The role of concatenation will be played by polynomials of finite arity. In our generalisation we assume that the monad is finitary, and that it is over the category of (unsorted) sets. The proof can be easily generalised to finitely sorted sets. There is one additional assumption in our generalisation, which will require some more definitions. 

\newcommand{\nondupol}{\mathsf{nondupol}}

\paragraph*{Observationally complete polynomials.} The idea behind observational completeness is that sometimes, instead of using all unary polynomials in the definition of the syntactic congruence, one can use a smaller subset, e.g.~unary polynomials that use the variable only once (whatever that may mean in an abstract monad).

	Let $\alg$ be a $\monad$-algebra. We write $\upol n \alg$ for   polynomials with $n$ argumetns in the algebra $\alg$; we do not need to indicate the sorts of these arguments  because we use unsorted sets. We use the convention that the variables in a polynomial from $\upol n \alg$ are called $x_1,\ldots,x_n$. Therefore, formally 
	\begin{align*}
		\upol n \alg  = \monad (A \sqcup \set{x_1,\ldots,x_n}).
	\end{align*}
	  A set $P \subseteq \upol 1 \alg$ is called \emph{observationally complete for $\alg$} if the following conditions are equivalent for every $a,b$ in the universe of  $\alg$ and every  subset $L$ of  the universe of $\alg$:
	\begin{eqnarray}
				w \in p^{-1}L \qquad \mbox{iff} \qquad w' \in p^{-1}L &\qquad \mbox{for every $p \in \upol 1 \alg$}
				\label{eq:observationally-complete-sm}\\
				w \in p^{-1}L \qquad \mbox{iff} \qquad w' \in p^{-1}L &\qquad \mbox{for every $p \in \upol 1 \alg \cap P$}.\label{eq:observationally-complete}
	\end{eqnarray}
	Recall that the condition in~\eqref{eq:observationally-complete-sm} is the equivalence relation defined in the proof of the Syntactic Morphism Theorem.

	\begin{myexample}\label{ex:observationally-complete}
		Consider the monad of finite words where algebras are semigroups. Call a unary polynomial nonduplicating if it uses its  variable exactly once. Such a polynomial is of the form  $			w x_1 v$
		where $w,v$ are   possibly empty words over the universe of the  semigroup. Without loss of generality one could also assume that $w,v$ have length zero or one. It is not difficult to show that the unary nonduplicating polynomials are observationally complete in every semigroups.  
	\end{myexample}
	
	\begin{running}\label{ex:run-observationally-complete}
	   		Consider the monad of ultimately periodic $\infty$-words. It is not difficult to show that in every algebra $\alg$ for this monad, an observationally complete set of unary polynomials is 
			\begin{align*}
				\set{wx_1v,w(x_1v)^\omega: \mbox{ where $w,v \in A^*$}}.
			\end{align*}
			These unary polynomials correspond to the \emph{Arnold congruence} from~\cite{DBLP:journals/tcs/Arnold85}.			
	\end{running}

\paragraph*{Finite covers.} Define an  \emph{$n$-ary term} to be an element of $\monad \set{x_1,\ldots,x_n}$, where $x_1,\ldots,x_n$ are the variables used for polynomials. In other words, a term is the special case of a polynomial that does not use any constants, and therefore an $n$-ary term is an $n$-ary polymomial in every algebra.
  We say that a unary polynomial $p \in \upol 1 \alg$ is \emph{covered} by an $n$-ary term $q$ 
if there exist $a_2,\ldots,a_n$ in the universe of $\alg$ such that
\begin{align*}
	p  = q (x_1,a_2,\ldots,a_n).
\end{align*}
A set $P \subseteq \upol 1 \alg$ is  said to have a finite cover, if there is a finite set $Q$  of terms, of possibly different arities, such that every polynomial in $P$ is covered by some term in $Q$. 

\begin{myexample}\label{ex:finite-covers}
	The  nonduplicationg polynomials mentioned in Example~\ref{ex:observationally-complete} are covered by the 3-ary term    $x_2 x_1 x_3$. The nonduplicationg polynomials mentioned in Running Example~\ref{ex:run-observationally-complete} are  covered by the two 3-ary terms $x_2x_1x_3$ and $x_2(x_1x_3)^\omega$.
Summing up, in both these  monads, every algebra has a finite cover for the set of nonduplicating unary polynomials.
\end{myexample}

\paragraph*{Characterisation of uniformly continuous term operations.} We are now ready to state the theorem that characterises recognisability as a necessary and sufficient condition for uniform continuity of term operations. In the theorem, we write $\deriv \Ll$ for the set of all polynomial derivatives of languages from~$\Ll$.

\begin{theorem}\label{thm:consistent-topology}
	Consider a finitary monad $\monad$ in the setting of  sets. Let $\alg$ be a finitely generated $\monad$-algebra  which has an observationally complete set of unary polynomials that has a finite cover. Let   $\Ll $ be a  countable family of   subsets of the universe of $\alg$. Then
	 \begin{enumerate}
	 	\item  if $\Ll$ contains only $\monad$-recognisable languages, then every term operation is uniformly continuous for $\deriv \Ll$-distance;
		\item if $\Ll$ contains at least one language that is not $\monad$-recognisable, then some term operation is not uniformly continuous for $\Ll$-distance.
	 \end{enumerate}
\end{theorem}

Before proving the theorem, note that by the discussion in Example~\ref{ex:finite-covers}, the assumptions of the theorem are satisfied by every algebra in the monad of finite words, and by every algebra in the monad of ultimately periodic words.

\medskip

\begin{pr}
	We skip the proof of the first item, which  is proved as in Example~\ref{example:uniformly-continuous}, and does not use the assumption on the observationally complete set of unary polynomials, but uses the assumption on finite generation.
	
	Let us consider the second item.	
	Let $P$ be a set of observationally complete polynomials with a finite cover $Q$, as in the assumptions of the theorem.  Let $L \in \Ll$ be some language that is not recognisable. Since $\alg$ is finitary, we can use the Syntactic Morphism Theorem. It follows that the equivalence in~\eqref{eq:observationally-complete-sm} has infinite index,  and therefore the equivalence relation defined in~\eqref{eq:observationally-complete} as applied to $P$ has infinite index. For a unary polynomial $q$ in the finite cover $Q$, define $\sim_q$ to be the relation as in~\eqref{eq:observationally-complete}  but with polynomials restricted to those that are covered by $q$, i.e.~$\sim_q$ identifies $a,b \in A$ if
	\begin{align*}
		a \in p^{-1}L \qquad \mbox{iff} \qquad b \in p^{-1}L &\qquad \mbox{for every $p \in \upol 1 \alg$ covered by $q$}.
	\end{align*}
	Since the relation~\eqref{eq:observationally-complete} has infinite index and  is the intersection of the finitely many relations $\sim_q$, there must be some $n$-ary term  $q \in Q$ such that $\sim_q$ has infinite index.  We claim that $q$, when seen as a polynomial in $\upol n \alg$, is not uniformly continuous. To prove this, consider some finite set $\Kk \subseteq \Ll$. Because the index of $\sim_q$ is infinite and the index of $\equiv_\Kk$ is finite, there must be some $a,b \in A$ such that
	 \begin{align*}
	 	a \not \sim_q b \qquad \mbox{and} \qquad a \equiv_\Kk b.
	 \end{align*}
	 Unraveling the definition of $\sim_q$, this means that there are some 
	 $a_2,\ldots,a_n$ in the universe of $\alg$ such that
	 \begin{align*}
	 	q(a,a_2,\ldots,a_n) \quad \not \equiv_{\set L} \quad  q(b,a_2,\ldots,a_n),
	 \end{align*}
	 which proves that $q$ is  not uniformly continuous.
\end{pr}

 %
%

\section{Profinite monads}
\label{sec:profinite-monads}
In this section, we prove that the Stone dual considered in the previous section has sufficient structure to make it into a monad, i.e.~for every monad $\monad$ the
mappings
\begin{eqnarray*}
	\Sigma &\qquad \mapsto \qquad&   \stone  (\monad \Sigma) \\
	f : \Sigma \to \Gamma & \qquad \mapsto \qquad & \stone (\monad f)
\end{eqnarray*}
can be equipped with unit and multiplication to make it a  monad, which we will denote by   $\promonad$. Because $\promonad$ is a monad, it has its own notion of recognisability, which is related to but richer than the notion of recognisability of the original $\monad$. This richer notion of recgnisability is studied in Section~\ref{sec:profinite-words}, on the example of profinite words.

\subsection{Definition of the profinite monad}
\label{sec:definition-profinite-monad}
Fix for the rest of this section  a monad $\monad$, in the category of sets, the generalisation to sorted sets being straightforward. 
We  explain how to convert $\monad$ into a monad, which we denote by $\promonad$, that describes profinite objects over $\monad$.

\newcommand{\recset}{\mathsf{rec}_\monad}

\paragraph*{Types.}  Instead of Stone duals as studied in the previous section, we will use in this section an alternative definition, which has a more algebraic flavour.
	For a $\monad$-algebra $\alg$, not necessarily finite, 
	%
	 define a \emph{$\monad$-morphism type over $\alg$} to be a function $\tau$ which maps every surjective $\monad$-morphism
	\begin{align}\label{eq:a-surjective-morphism}
		h : \alg \to \balg \qquad \mbox{with $\balg$ finite} 
	\end{align}
	to  an  element $h^{\tau} \in \balg$, subject to the condition that 
	\begin{align}\label{eq:type-condition}
		(g \circ h)^\tau = g(h^\tau) \qquad \mbox{for every }h : \alg \to \balg\mbox{ and }g : \balg \to \calg
	\end{align}
	where $g$ is a surjective $\monad$-morphism between finite $\monad$-algebras.
The set of of $\monad$-morphism types over a $\monad$-algebra $\alg$ is called its \emph{compactification}, and is denoted by $\bar \alg$. As a topological space, the set of $\monad$-morphism types is an equivalent definition of the Stone dual defined in the previous section, as stated in the following fact.

\begin{fact}\label{fact:same-as-stone-dual}
	If $\alg$ is a $\monad$-algebra, then $\stone \alg$ is homeomorphic to $\bar \alg$, assuming that the base open sets are of the form
	\begin{align*}
		\set{ \tau  \in \bar \alg : h^\tau = b} 
	\end{align*}
for $h : \alg \to \balg$ a surjective $\monad$-morphism into a finite $\monad$-algebra and $b \in \balg$.
\end{fact}

Nevertheless,  we use $\monad$-morphism types instead of the Stone dual because they  will be more convenient to study the algebraic structure.
 Define the  \emph{profinite extension}  of a surjective $\monad$-morphism
 \begin{align*}
 	h: \alg \to \balg 
 \end{align*}
into a finite $\monad$-algebra $\balg$ to be the function 
\begin{align*}
\bar h :\bar \alg \to \balg	
\end{align*}
defined by $\bar h (\tau ) = h^\tau$.
In terms of profinite extensions~\eqref{eq:type-condition}, says that the following diagram commutes.
\begin{align}\label{eq:morphism-type}
	\xymatrix{
	\bar \alg \ar[r]^{\bar h} \ar[dr]_{\overline{g \circ h}} & \balg \ar[d]^{\bar g} \\ & \calg
	}
\end{align}

\begin{myexample}
	Consider the monad  $+$ of finite words, where $+$-algebras are semigroups, and $+$-morphisms are semigroup morphisms. 	Consider the semigroup $\Sigma^+$ where $\Sigma$ is a  a finite alphabet. As stated in Fact~\ref{fact:same-as-stone-dual}, $+$-morphism types, or semigroup morphism types, over $\Sigma^+$ are the same thing  as elements of the Stone dual $\stone \Sigma^+$, which are profinite words as described in Example~\ref{ex:profinite-words}.  We now revisit the element $w^{\#}$ of the Stone dual that was described in Example~\ref{ex:profinite-words}, and describe its corresponding semigroup morphism types.
	
By definition,  a  semigroup morphism type over $\Sigma^+$ is a function which maps every semigroup morphism
	\begin{align*}
		h : \Sigma^+ \to \alg \qquad \mbox{with $\alg$ a finite semigroup}
	\end{align*}
 to an element of $\alg$, in a way that is consistent with composition. 
  The  idempotent power $w^{\#}$ described in Example~\ref{ex:profinite-words} is the morphism type which maps a morphism $h$ to $h(w)^{\#}$. Let us check that $w^{\#}$ defined this way  is indeed a semigroup morphism type, we need to show that for every semigroup morphisms 
\begin{align*}
	h : \Sigma^* \to \alg \qquad g : \alg \to \balg
\end{align*}
with $\alg$ and $\balg$ being finite semigroups, and $g$ being surjective,  we have
\begin{align*}
	g(h(w)^\#) = (g \circ h(w))^{\#}.
\end{align*}
This is checked below, assuming that $n$ and $m$ are the sizes of $\alg$ and $\balg$.
\begin{eqnarray*}
	g(h(w)^\#)  &=& \eqexplain{by definition}\\
	g(h(w)^{n!}) & =& \eqexplain{because $g$ is a semigroup morphism} \\
	 g(h(w))^{n!} &=& \eqexplain{because $m \le n$ and $m!$ is an idempotent powe}\\
	    (g(h(w))^{m!} &=& \eqexplain{by definition}\\
	    (g \circ h(w))^{\#}.
\end{eqnarray*}
\end{myexample}



\paragraph*{The functor of $\promonad$.}  We now define the profinite monad $\promonad$. We assume that the unit and multiplication in the original monad $\monad$ are denoted by $\monun \Sigma$ and $\monmul \Sigma$. An object $\Sigma$ is mapped by $\promonad$   the compactification  of the $\monad$-algebra $\monad \Sigma$, i.e.
\begin{align*}
	\promonad \Sigma \eqdef \overline {\monad \Sigma}.
\end{align*}
The remaining components of the monad, i.e.~how $\promonad$ acts on functions, as well as the unit and multiplication operations, are defined and proved correct in the following theorem.

\begin{theorem}\label{thm:definition-of-promonad}
	There are unique operations
	\begin{eqnarray*}
		\promonad f &:& \promonad \Gamma \to \promonad \Sigma  \qquad \mbox{for $f : \Gamma \to \Sigma$}\\
		\promonun {\Sigma} &:& \Sigma \to \promonad \Sigma \\
		\promonmul {\Sigma} &:& \promonad \promonad \Sigma \to \promonad \Sigma
	\end{eqnarray*}
	such that for every finite $\monad$-algebra $\alg$ and every surjective  $\monad$-morphism
	\begin{align*}
		h : \monad \Sigma \to \alg,
	\end{align*}
into a finite $\monad$-algebra,	the following diagrams commute
	\begin{align*}
		\xymatrix{
		\promonad \Gamma \ar[r]^{\promonad f} \ar[dr]_{\overline{h \circ \monad f}} & \promonad \Sigma \ar[d]^{\bar h} \\ & \alg
		} \qquad 
		\xymatrix @C=3pc{
		\Sigma \ar[r]^{\promonun {\Sigma}} \ar[d]_{\monun {\Sigma}} & \promonad \Sigma\ar[d]^{\bar h} \\ 
		\monad \Sigma \ar[r]^{ h} & \alg
		} \qquad 
	    \xymatrix @R=2pc
	   				{ \promonad \promonad \Sigma \ar[d]_{\promonad \bar h}  \ar[rr]^{\promonmul {\Sigma}} && \promonad \Sigma \ar[d]^{\bar h}\\
	   				\promonad A \ar[rr]^{\overline{\mult_\alg}}&  & \alg
	   				}.
	\end{align*}
	Furthermore, equipped with the above operations, $\promonad$ is a monad.
\end{theorem}

The rest of Section~\ref{sec:definition-profinite-monad} is devoted to proving the above theorem.
First observe that the operations from the statement of the theorem, if they exist, are uniquely  specified by the diagrams in the statement of the theorem,  because an element of $\promonad \Sigma$ is uniquely specified by its values under all possible profinite extensions $\bar h$. We need to check that the operations actually produce  morphism types, i.e.~the values that they produce satisfy~\eqref{eq:type-condition}.
Let us first check that $\promonad f$ produces types, i.e.~that
\begin{align*}
	\overline{g \circ h} \circ \promonad  = g \circ    \bar h \circ \promonad f  
\end{align*}
holds for every finite $\monad$-morphisms
\begin{align*}
	h : \monad \Sigma \to \alg \qquad g : \alg \to \balg
\end{align*}
where $\alg,\balg$ are finite. This is checked below:
\begin{eqnarray*}
 \overline{g \circ h} \circ \promonad f  &=& \eqexplain{by definition of $\promonad f$} \\
  \overline{g \circ h \circ \monad f}  &=& \eqexplain{by~\eqref{eq:morphism-type}} \\
g \circ    \overline{ h \circ \monad f}  &=&   \eqexplain{by definition of $\promonad f$} \\
g \circ    \bar h \circ \promonad f  
\end{eqnarray*}
For the unit operation, the check is even simpler:
\begin{eqnarray*}
 \overline{g \circ h} \circ \promonun {\Sigma}  &=& \eqexplain{by definition of $\promonun {\Sigma}$} \\
 {g \circ h} \circ \monun {\Sigma}  &=& \eqexplain{by definition of $\promonun {\Sigma}$} \\
  g \circ \bar h \circ \promonun {\Sigma} 
\end{eqnarray*}
Before checking that the  multiplication operation defined in the theorem produces types, we check that $\promonad$ is a functor, i.e.
\begin{align*}
	\promonad (f \circ g) = \promonad f \circ \promonad g  \qquad \mbox{for every $f : \Delta \to \Sigma$ and $g : \Gamma \to \Delta$}.
\end{align*}
To prove the above equality, we show that the  two sides of the  equality are equal after being composed with functions of the form $\bar h$ with 
\begin{align*}
	h : \monad \Sigma \to \alg
\end{align*}
a $\monad$-morphism into a finite $\monad$-algebra.
 This is checked below and illustrated in Figure~\ref{fig:promonad-is-a-functor}.
\begin{eqnarray*}
 \bar h \circ \promonad (f \circ g)  &=& \eqexplain{by definition of $\promonad (f \circ g)$} \\
 \overline{ h \circ \monad (f \circ g)}  &=& \eqexplain{because $\monad$ is a functor} \\
	   \overline {h \circ \monad f \circ \monad g}  &=& \eqexplain{by definition of $\promonad g$} \\
	     \overline {h \circ \monad f }\circ  \promonad g  &=& \eqexplain{by definition of $\promonad f$} \\
		 \bar h \circ \promonad f \circ \promonad g
\end{eqnarray*}
%

\begin{figure}[htbp]
	\begin{align*}
		\xymatrix @C=3pc{
		\promonad \Sigma \ar@/_4pc/[dd]_{\promonad (f \circ g)} \ar[d]_{\promonad g} \ar[dr]^{\ \ \ \overline{h \circ \monad f \circ \monad g}=\overline{h \circ \monad (f \circ  g)}} \\ 
		\promonad \Gamma \ar[d]_{\promonad f} \ar[r]^{\overline {h \circ \monad f}}& \alg \\
		\promonad \Delta \ar[ur]_{\bar h}
		}
	\end{align*}	
	\caption{$\promonad$ is a functor.}
	\label{fig:promonad-is-a-functor}
\end{figure}

\paragraph*{Multiplication in $\promonad$}
To prove that the multiplication operation of the monad $\promonad$ is produces types, one uses the following lemma in the special case of $\alg = \monad \Sigma$.
\begin{lemma}\label{lem:define-mult}
	If $\alg$ is a $\monad$-algebra, then there is a unique operation
\begin{align*}
	\mult_{\bar \alg} : \promonad \bar \alg \to \bar \alg,
\end{align*}
which makes the following diagram commute 
\begin{align}\label{eq:promonad-mult}
 \xymatrix @R=2pc
				{ \promonad \bar \alg \ar[d]_{\promonad \bar h}  \ar[rr]^{\mult_{\bar \alg}} && \bar \alg \ar[d]^{\bar h}\\
				\promonad \balg \ar[rr]^{\overline{\mult_\alg}}&  & \balg
				}
\end{align}
for every $\monad$-morphism $h : \alg \to \balg$ into a finite $\monad$-algebra $\balg$.
\end{lemma}
\begin{pr}
The diagram~\eqref{eq:promonad-mult} leaves no choice in the definition of $\mult_{\bar \alg}$, 
since an element of $\bar \alg$ is uniquely defined by its images under all possible $\bar h$. 
We check below that the multiplication operation is well-defined, i.e.~it produces $\monad$-morphism types. Let then
\begin{align*}
	h : \alg \to \balg \qquad g : \balg \to \calg
\end{align*}
be $\monad$-morphisms with $\balg$ and $\calg$ being finite $\monad$-algebras. We need to show that 
\begin{align*}
	 \overline{g \circ h} \circ \promonmul {\bar \alg}=  	 g \circ  \bar h \circ \promonmul {\bar \alg}.
\end{align*}
\begin{figure}[htbp]
\begin{align*}
	\xymatrix@R=4pc @C=4pc{
	\promonad \bar \alg \ar@/_4pc/[dd]_{\promonad {(g \circ h)}} \ar[d]_{\promonad \bar h}  \ar[r]^{\mult_{\bar \alg}} & \bar \alg \ar[d]^{\bar h} \ar@/^4pc/[dd]^{\overline{g \circ h}}\\	
	\promonad \balg  \ar[dr]^{\overline{g \circ \mult_\balg}}_{\overline{\mult_\calg \circ \monad g}}
	\ar[d]_{\promonad g}\ar[r]^{\overline{\mult_\balg}} & \balg \ar[d]^g \\
	\promonad \calg \ar[r]_{\overline{\mult_\calg}} & \calg
	}
\end{align*}
	\caption{Multiplication in $\promonad$ is well-defined.}
	\label{fig:promonad-multiplication-well-defined}
\end{figure}
 This is done below and illustrated in Figure~\ref{fig:promonad-multiplication-well-defined}.
\begin{eqnarray*}
 \overline{g \circ h} \circ \promonmul {\bar \alg}  &=& \eqexplain{by~\eqref{eq:promonad-mult}} \\
\overline{\mult_\calg} \circ \promonad\ \overline{g \circ h}  &=& \eqexplain{by~\eqref{eq:morphism-type}} \\
\overline{\mult_\calg} \circ \promonad(g \circ \bar h)  &=& \eqexplain{because $\promonad$ is a functor} \\
\overline{\mult_\calg} \circ \promonad g \circ \promonad \bar h  &=& \eqexplain{by definition of $\promonad g$} \\
\overline{\mult_\calg \circ \monad g} \circ \promonad \bar h  &=& \eqexplain{because $g$ is a $\monad$-morphism} \\
\overline{g \circ \mult_\balg} \circ \promonad \bar h  &=& \eqexplain{by~\eqref{eq:morphism-type}} \\
g \circ \overline{ \mult_\balg} \circ \promonad \bar h  &=& \eqexplain{by~\eqref{eq:promonad-mult}} \\
  g \circ \bar h \circ \circ \mult_{\bar \alg}
\end{eqnarray*}
\end{pr}

So far we have proved that the operations in the statement of Theorem~\ref{thm:definition-of-promonad} are well defined, i.e.~they produce $\monad$-morphism types, and that $\promonad$ is a functor. We now check the remaining axioms of a monad. We skip proving that multiplication and unit are natural, i.e.~the upper two diagrams in Figure~\ref{fig:monad-axioms}. We only show that multiplication is associative and consistent with the unit, i.e.~the lower two diagrams in Figure~\ref{fig:monad-axioms}.

To prove that the multiplication operation in the monad is associative, we apply the following lemma to the special case of $\alg = \monad \Sigma$.

\begin{lemma}\label{lem:compactify-associative}
	Let $\alg$ be a $\monad$-algebra, and let  $\mult_{\bar \alg}$ be as in Lemma~\ref{lem:define-mult}. Then the following diagram commutes:
		\begin{align*}
			\xymatrix @R=2pc @C=3pc {\promonad \promonad \bar \alg  \ar[r]^{\promonmul {\bar \alg}} \ar[d]_{\promonad{\mult_{\bar \alg}}} & \promonad \bar \alg \ar[d]^{\mult_{\bar \alg}}  \\
					\promonad \bar \alg \ar[r]_{\mult_{\bar \alg}}& \bar \alg
					}
		\end{align*}
\end{lemma}
\begin{pr}
	Because an element of $\bar \alg$ is uniquely determined by its values under $\bar h$, with $h$ ranging over $\monad$-morphisms from $\alg$ into finite $\monad$-algebras, it suffices to show that the diagram commutes when extended with such a $\bar h$, i.e.
	\begin{align*}
				\bar h \circ \mult_{\bar \alg} \circ \promonad \mult_{\bar \alg}= 
		\bar h \circ \mult_{\bar \alg}  \circ {\promonmul {\bar \alg}}.
	\end{align*}
Let us then fix $h : \alg \to \balg$ and prove the above equality.
The calculation is performed below and also illustrated in Figure~\ref{fig:promonad-is-associative}.
	\begin{figure}[htbp]
		\centering
			\begin{align*}
				\xymatrix @R=2pc @C=3pc
				{
				&&& \promonad \bar \alg \ar[d]^{\mult_{\bar \alg}} \ar[ddl]^{\promonad \bar h}\\
				\promonad \promonad \bar \alg
				\ar[d]_{\promonmul {\bar \alg}} \ar@/_/[drr]_{\promonad \overline{\mult_\balg  \circ \monad \bar h}}
				  \ar[urrr]^{\promonad \mult_{\bar \alg}} \ar[rr]^{\promonad \promonad \bar h} && \promonad \promonad B 
				 \ar@{}[dl]_{} 
				 \ar[d]_{\promonad\ \overline{\mult_\balg}} & \bar \alg \ar[ddd]^{\bar h}\\
				\promonad \bar \alg
				 \ar[dddrr]_{\mult_{\bar \alg}}\ar[ddrr]^{\promonad \bar h} \ar@/^/[ddrrr]^{\overline{\mult_\balg  \circ \monad \bar h}}&  &
				 \promonad B   \ar[ddr]^{\overline{\mult_\balg}} & \\ \\
				&& \promonad B   \ar[r]^{\overline{\mult_\balg}} & B\\
				& &\bar \alg \ar[ur]_{\bar h}
				}
			\end{align*}
		\caption{The four-sided faces in the diagram commute by~\eqref{eq:promonad-mult}, or by $\promonad$ applied to~\eqref{eq:promonad-mult}. The three-sided faces in the diagram commute by the definition of $\promonad$ on functions, or by  $\promonad$ applied to the definition of $\promonad$ on functions.
		}
		\label{fig:promonad-is-associative}
	\end{figure}
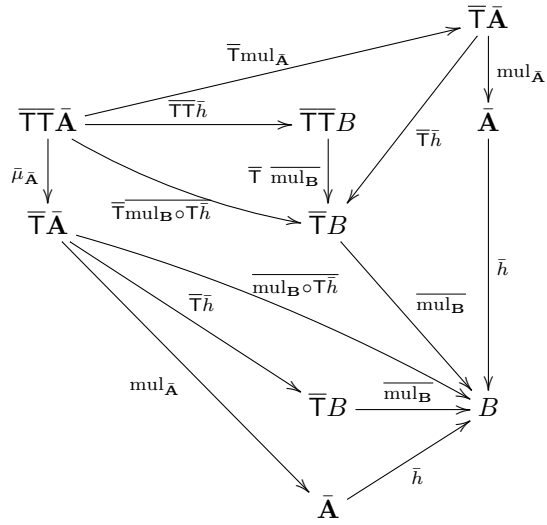
		\begin{eqnarray*}
		\bar h \circ \mult_{\bar \alg} \circ \promonad \mult_{\bar \alg} & = &  \eqexplain{by~\eqref{eq:promonad-mult}} \\
		\overline{\mult_\balg} \circ \promonad \bar h \circ \promonad \mult_{\bar \alg} & = &  \eqexplain{by $\promonad$ applied to~\eqref{eq:promonad-mult}} \\
		\overline{\mult_\balg} \circ  \promonad \ \overline{\mult_\balg} \circ \promonad \promonad \bar h & = &  \eqexplain{because $\promonad$ is a functor} \\
		\overline{\mult_\balg} \circ  \promonad \big( \overline{\mult_\balg} \circ  \promonad \bar h\big) & = &  \eqexplain{by definition of $\promonad \bar h $} \\
\overline{\mult_\balg} \circ \promonad \overline{\mult_\balg  \circ \monad \bar h}& = &  \eqexplain{by~\eqref{eq:promonad-mult} with the $\monad$-morphism being $\mult_\balg  \circ \monad \bar h$} \\
\overline{\mult_\balg  \circ \monad \bar h} \circ {\promonmul {\bar \alg}}& = &  \eqexplain{by definition of $\promonad \bar h$} \\
  \overline{\mult_\balg} \circ \promonad \bar h  \circ {\promonmul {\bar \alg}}& = &  \eqexplain{by~\eqref{eq:promonad-mult}} \\
\bar h \circ \mult_{\bar \alg}  \circ {\promonmul {\bar \alg}}
	\end{eqnarray*}
	
\end{pr}	

We now check the last axiom of a monad, namely that the following diagram commutes:
\begin{align*}
	 \xymatrix @R=2pc { \promonad \Sigma \ar[dr]^{\mathrm{id}_\Sigma}  \ar[r]^{\promonun {\promonad \Sigma}} \ar[d]_{\promonad \promonun {\Sigma}} & \promonad \promonad \Sigma \ar[d]^{\promonmul {\Sigma}}  \\
	\promonad \promonad \Sigma \ar[r]_{\promonmul {\Sigma} }& \promonad \Sigma}
\end{align*}
Let us first check the upper triangular face of the diagram:
\begin{eqnarray*}
\bar h \circ \promonmul {\Sigma} \circ	\promonun {\promonad \Sigma} & = & \eqexplain{by~\eqref{eq:promonad-mult}}\\
\overline{\mult_\alg} \circ \promonad \bar h \circ	\promonun {\promonad \Sigma} & = & \eqexplain{by definition of $\promonad \bar h$}\\
\overline{\mult_\alg \circ \monad \bar h}  \circ	\promonun {\promonad \Sigma} & = & \eqexplain{by definition of $\promonun {\Sigma}$}\\
\mult_\alg \circ \monad \bar h  \circ	\monun {\promonad \Sigma} & = & \eqexplain{because $\monad$ is a monad}\\
\mult_\alg \circ \monun { A} \circ  \bar h   & = & \eqexplain{because $\mult_\alg$ is the identity on units} \\
  \bar h   
\end{eqnarray*}
Let us now check the lower triangular face of the diagram:
\begin{eqnarray*}
\bar h \circ \promonmul {\Sigma} \circ\promonad	\unit_{ \promonad \Sigma} & = & \eqexplain{by~\eqref{eq:promonad-mult}}\\
\overline{\mult_\alg} \circ \promonad \bar h \circ	\promonad	\unit_{ \promonad \Sigma} & = & \eqexplain{because $\promonad$ is a functor}\\	
\overline{\mult_\alg} \circ \promonad (\bar h \circ		\unit_{ \promonad \Sigma}) & = & \eqexplain{by definition of $\promonun {\Sigma}$}\\	
\overline{\mult_\alg} \circ \promonad ( h \circ		\unit_{ \monad \Sigma}) & = & \eqexplain{because $\promonad$ is a functor}\\	
\overline{\mult_\alg} \circ \promonad  h \circ \promonad		\unit_{ \monad \Sigma} & = & \eqexplain{by definition of $\promonad  h$}\\	
\overline{\mult_\alg \circ \monad  h} \circ \promonad		\unit_{ \monad \Sigma} & = & \eqexplain{by definition of $\promonad \monun {\Sigma}$}\\	
\overline{\mult_\alg \circ \monad  h \circ \monad		\unit_{ \monad \Sigma}} & = & \eqexplain{because $h$ is a $\monad$-morphism}\\	
\overline{h \circ \monmul {\Sigma} \circ \monad		\unit_{ \monad \Sigma}} & = & \eqexplain{because $\monad$ is a monad}\\	
\bar{h} 
\end{eqnarray*}
This completes the proof that $\promonad$ is a monad.

\subsection{From a $\monad$-algebra to a $\overline {\mathsf T}$-algebra.}
\label{sec:from-monad-to-promonad}
Having defined the monad $\promonad$, it is natural to ask what are finite $\promonad$-algebras, and what are the languages recognised by them.  In this section, we discuss how every $\monad$-algebra can be transformed into a $\promonad$-algebra. Since this transformation preserves finiteness, it gives a source of examples of finite $\promonad$-algebras. However, the algebras produced by this transformation are not very interesting, because they are essentially decorations of $\monad$-algebras.  More interesting examples will  be given in Section~\ref{sec:profinite-words}.

\paragraph*{From $\alg$ to $\bar \alg$.} An element of $a \in \alg$ can  can be interpreted as an element of $\bar \alg$, namely as the   $\monad$-morphism type which maps a $\monad$-morphism $h$ to $h(a)$. We denote this interpretation by $\iota_\alg$, by definition it makes the following diagram commute:
\begin{align}\label{eq:iota-definition}
\xymatrix{
\alg \ar[dr]_h	\ar[r]^{\iota_\alg}  & \bar \alg \ar[d]^{\bar h}\\ & \balg}.
\end{align}
for every $\monad$-morphism $h$ into a finite $\monad$-algebra $\balg$.
It is tempting to think of $\iota_\alg$ as an embedding. However, for $\iota_\alg$ to be an embedding, one would require that every distinct elements of $ \alg$ can be distinguished by some $\monad$-morphism into a finite $\monad$-algebra.  This additional assumption is true in all monads studied in this paper, at least for 
finitely generated $\monad$-algebras,
 but it can be false, e.g.~with a very restrictive notion of finite $\monad$-algebra.
 
\paragraph*{An algebraic structure on $\bar \alg$.}
From  Lemmas~\ref{lem:define-mult} and~\ref{lem:compactify-associative} it follows that if $\alg$ is a $\monad$-algebra, then there is a multiplication operation
\begin{align*}
	\mult_{\bar \alg} : \promonad \bar \alg \to \bar \alg
\end{align*}
which turns the compactification $\bar \alg$ into a $\promonad$-algebra. The following lemma implies that compactification preserves finiteness of algebras.

\begin{lemma}\label{lem:compactification-preserves-fintieness}
	If $\alg$ is a finite $\monad$-algebra, then $\bar \alg$ is isomorphic to the $\promonad$-algebra where the universe is the universe of $\alg$, and  multiplication is defined to be the profinite extension of $\mult_\alg$, i.e.~by
	\begin{align*}
		\overline{ \mult_\alg} : \promonad A \to A.
	\end{align*}
\end{lemma}
\begin{pr}
	We claim that the isomorphism is  the
		profinite extension
		\begin{align*}
			\overline {\ident \alg} : \bar \alg \to \alg
		\end{align*}
		of the identity on $\alg$.
		We claim that the above is a bijection, because its inverse is $\iota_{\alg}$. To prove bijectivity, we need to show that 		\begin{align*}
		\iota_\alg \circ	\overline{\ident \alg} \qquad  			\overline{\ident \alg}  \circ \iota_\alg 
		\end{align*}
		are the identity functions on $\bar \alg$ and $\alg$ respectively. For the latter, we invoke~\eqref{eq:iota-definition}. The former is explained in the following diagram
		\begin{align*}
			\xymatrix{
			\bar \alg \ar[d]_{\bar h} \ar[r]^{\overline{\ident \alg}} & \alg \ar[d]^{\iota_\alg} \ar[dl]^h \\
			 \balg & \bar \alg \ar[l]^{\bar h}
			}
		\end{align*}

\end{pr}

\begin{lemma}\label{lem:define-hat}
	If $h : \alg \to \balg$ is a $\monad$-morphism, then there is  a unique function
	\begin{align*}
		\bar  h : \bar \alg \to \bar \balg
	\end{align*}
	which makes the following diagram commute
	\begin{align*}
		\xymatrix
		{
		\bar \alg \ar[r]^{\bar h} \ar[dr]_{\overline{g \circ h}} & \bar \balg \ar[d]^g \\ & \calg
		}
	\end{align*}
	for every $\monad$-morphism $g : \balg \to \calg$ with $\calg$ finite.
\end{lemma}
\begin{pr}
	Note that the definition of $\promonad f$  is actually a special case of this lemma, because $\promonad f$ is makes the diagram in the lemma commute for $\monad f$, i.e.~$\promonad f = \overline {\monad f}$.  The lemma is proved the same way as  we proved that $\promonad f$ is well defined.
\end{pr}

The above lemma introduces a little clash of notation.  If 
\begin{align*}
	h : \alg \to \balg
\end{align*}
is a $\monad$-morphism such that $\balg$ if finite, then $\bar h$ has two definitions:  namely the profinite extension of $\alg$, which is of  type $\bar \alg \to \balg$, and the definition from the above lemma, which is of the type $\bar \alg \to \bar \balg$.  However, the two definitions are essentially the same mapping, because they are equal up to the isomorphism from Lemma~\ref{lem:compactification-preserves-fintieness}.

\begin{lemma}\label{lem:bar-is-promonad-morphism}
	If $h : \alg \to \balg$ is a $\monad$-morphism, then $\bar h$ defined in Lemma~\ref{lem:define-hat} is a $\promonad$-morphism and makes the following diagram commute:
	\begin{align*}
		\xymatrix
		{
		\alg \ar[r]^h \ar[d]_{\iota_\alg}  & \balg \ar[d]^{\iota_\balg} \\
		\bar \alg \ar[r]_{\bar h} & \bar \balg
		}
	\end{align*}
\end{lemma}
\begin{pr}
	We first check the diagram in the statement of the lemma. It suffices to show that the diagram commutes after the lower right corner is extended with the profinite extension of a $\monad$-morphism  $f : \balg \to \calg$  into a finite $\monad$-algebra. This is shown in the following diagram:
	\begin{align*}
		\xymatrix@C=3pc{
		\alg \ar[rr]^{h} \ar[dd]_{\iota_{\alg}}&& \balg \ar[dl]^f \ar[dd]^{\iota_{\balg}} \\  & \calg \\
		\bar \alg \ar[ur]^{\overline{f \circ h}} \ar[rr]_{\bar h} && \bar \balg \ar[ul]^{\bar f}
		}
	\end{align*}
	The proof that $\bar h$ is a $\promonad$-morphism is in the following diagram.
	\begin{align*}
		\xymatrix {
		\promonad \bar \alg \ar[dr]^{\promonad \ \overline{f \circ h}} \ar[dd]_{\promonad \bar h} \ar[rrr]^{\mult_{\bar \alg}} & &  & \bar \alg\ar[dd]^{\bar h} \ar[dl]_{\overline{f \circ h}} \\
		& \promonad C \ar[r]^{\overline{\mult_\calg}}  & \calg  \\
		\promonad \bar \balg \ar[ur]^{\promonad \bar f} \ar[rrr]_{\mult_{\bar \balg}} &  &  & \bar \balg \ar[ul]^{\promonad \bar f}
		}
	\end{align*}
	The upper and lower faces commute by Lemma~\ref{lem:define-mult},  the right face commutes by Lemma~\ref{lem:define-hat}, and the left face commutes by applying the functor $\promonad$ to Lemma~\ref{lem:define-hat}.	
\end{pr}

\subsection{From a $\overline {\mathsf T}$-algebra to a $\monad$-algebra.}
In the previous section, we showed how to convert a $\monad$-algebra into a $\promonad$-algebra. We now discuss the opposite direction.
To go from a  $\promonad$-algebra $\alg$ to a $\monad$-algebra, call it $\restrictalg \alg$, one keeps the same universe and defines the multiplication operation by
 \begin{align*}
 	\xymatrix{
	\monad A \ar[dr]_{\mult_{\restrictalg \alg}} \ar[r]^{\iota_A} & \promonad A \ar[d]^{\mult_\alg} \\ & A
	}
 \end{align*}
The following lemma shows that this construction is correct. In the specific case of the monad of finite words, the lemma says that a profinite semigroup is actually a semigroup (it has other structure as well).

\begin{lemma}\label{lem:promonad-is-a-monad-algebra} 
	If $\alg$ is a $\promonad$-algebra, then $\restrictalg \alg$ is a $\monad$-algebra.
	If $h : \alg \to \balg$ is a $\promonad$-morphism, then the function underlying  $h$ is a $\monad$-morphism from $\restrictalg \alg$ to $\restrictalg \balg$.
\end{lemma}
%
%
%
 		\newcommand{\doubleiota}{\iota\iota}
%
%

\begin{pr}
		By Lemma~\ref{lem:bar-is-promonad-morphism}  applied to $\bar h$ being
	\begin{align*}
		\monad \iota_{\monad \Sigma} : \monad \monad \Sigma \to \monad \promonad \Sigma
	\end{align*}
	we see that the following diagram commutes
			\begin{align*}
				\xymatrix{
				\monad \monad \Sigma \ar[d]_{\monad \iota_\Sigma} \ar[r]^{\iota_{\monad \Sigma}} & \promonad \monad \Sigma \ar[d]^{\promonad \iota_\Sigma}\\
				\monad \promonad \Sigma \ar[r]_{\iota_{\promonad \Sigma}} & \promonad \promonad \Sigma
				}
			\end{align*}
			Let us write $\doubleiota_\Sigma$ for  the diagonal of the above diagram.
	
	To prove that $\restrictalg \alg$ is a $\monad$-algebra, we will show that the following diagram commutes (the outer perimeter of the diagaram says that ${\mult_{\restrictalg \alg}}$ is associative as required in a $\monad$-algebra):
	\begin{align*}
		\xymatrix @R=1pc @C=3pc
		{
		\monad \monad A \ar[dr]^{\doubleiota_A} \ar[rrr]^{\monmul {A}} \ar[ddd]_{\monad {\mult_{\restrictalg \alg}}} &&& \monad A  \ar[dl]^{\iota_A}\ar[ddd]^{\mult_{\restrictalg \alg}}\\ 
		& \promonad \promonad A \ar[r]^{\promonmul {A}} \ar[d]_{\promonad \mult_\alg} & \promonad A \ar[ddr]^{\mult \alg}\\
		& \promonad A \ar[rrd]_{\mult_\alg} \\
		\monad A \ar[ur]^{\iota_A} \ar[rrr]_{\mult_{\restrictalg \alg}} &&& \alg
		}
	\end{align*}
	The middle face of the diagram is the assumption that $\alg$ is a $\promonad$-algebra. The right and bottom faces are the definition of ${\mult_{\restrictalg \alg}}$. The top face can be shown using the definition of multiplication in $\promonad A$, and does not use the algebraic structure on $A$. Finally, for the left face, we use the following diagram:
	\begin{align*}
			\xymatrix @R=1pc @C=7pc
			{
			\monad \monad A \ar[ddr]_{\monad \iota_A} \ar[dddr]_{\monad {\mult_{\restrictalg \alg}}} \ar[rrdd]^{\doubleiota_\Sigma} \\ \\
			& \monad \promonad A  \ar[d]^{\monad \mult_\alg}\ar[r]_{\iota_{\promonad A}} & \promonad \promonad A \ar[d]^{\promonad \mult_\alg}\\
			& \monad A \ar[r]_{\iota_A} & \promonad A
			}
	\end{align*}
	The rectangular face commutes by Lemma~\ref{lem:bar-is-promonad-morphism}. The lower triangular face commutes by applying the functor $\monad$ to the definition of ${\mult_{\restrictalg \alg}}$. The upper triangular face commutes because it is the definition of $\doubleiota_A$.
	
	This completes the proof that $\restrictalg \alg$ is a $\monad$-algebra. To prove that $h$ as in the statement of the lemma is a $\monad$-morphism, we consider the  following diagram:
\begin{align*}
	\xymatrix {
	\monad A \ar[dr]^{\iota_A} \ar[dd]_{\mult_{\restrictalg \alg}} \ar[rrr]^{\monad h} & &  & \monad B\ar[dd]^{\mult_{\restrictalg \balg}} \ar[dl]_{\iota_B} \\
	& \promonad A \ar[r]^{\promonad h} \ar[dl]^{\mult_\alg} & \promonad B \ar[dr]_{\mult_\balg} \\
	A\ar[rrr]_h &  &  & B
	}
\end{align*}
The left and right faces commute by definitions of multiplication in $\restrictalg \alg$ and $\restrictalg \balg$. The bottom face commutes by assumption that $h$ is a $\promonad$-morphism. The top face commutes by Lemma~\ref{lem:bar-is-promonad-morphism}.
\end{pr}

%
%
%
%
%
%

\subsection{Clopen languages and Stone algebras}
\label{sec:stone-algebras} A well known result for profinite words is that there is a one-to-one correspondence between  clopen subsets of the profinite monoid over $\Sigma$, and recognisable subsets of  $\Sigma^+$. In this section we prove Theorem~\ref{thm:clopen-languages}, which generalises this observation to monads.

Recall that if $\alg$ is a $\monad$-algebra, then by Fact~\ref{fact:same-as-stone-dual} there is a topological structure on $\bar \alg$, which is homeomorphic to the Stone dual  $\stone \alg$. Recall also the mapping $\iota_\alg : \alg \to \bar \alg$. Using these two notions,  for   $L \subseteq \alg$, we define $\bar L \subseteq \bar \alg$ to be the  closure, in the topology of $\bar \alg$,  of the image of $L$ under the $\iota_\alg$. In Theorem~\ref{thm:clopen-languages}, we will show that the clopen subsets of $\bar \alg$ are exactly the closures, in the sense just defined, of $\monad$-recognisable subsets of $\alg$.

\begin{theorem}\label{thm:clopen-languages}
Let $\alg$ be a $\monad$-algebra. A subset of $\bar \alg$ is clopen if and only if it is of equal to $\bar L$ for some recognisable $L \subseteq \alg$.
\end{theorem}

Before proving the theorem, we present a lemma.

	\begin{lemma}\label{lem:closure-characterised}
If   $h : \alg \to \balg$   is a $\monad$-morphism into a finite $\monad$-algebra, then
		\begin{align*}
			\overline{h^{-1}(F)} = (\bar h)^{-1}(F) \qquad \mbox{for every $F \subseteq \balg$.}
		\end{align*}
	\end{lemma}
	\begin{pr}
		Recall that a base open set in $\bar \alg$ is a set of the form
			\begin{align*}
				\bar h^{-1}(b) \qquad \mbox{for some $\monad$-morphism $h :\alg \to \balg $ and $b \in B$},
			\end{align*} 
			and such sets are also closed. Therefore, the set on the right side  of the equality in the statement of the lemma is closed, as a finite union of base  sets. To complete the proof, we show that  the  image of $h^{-1}(F)$ under $\iota_\alg$ is dense in the set on the right side, i.e.~every open subset of the right side  contains $\iota_\alg(a)$ for some $a \in \alg$ with $h(a) \in F$. Every open set contains a base open set, and therefore it suffices to show that if 
			\begin{align*}
				g : \alg \to \calg
			\end{align*}
			is a $\monad$-morphism into a finite $\monad$-algebra, and the base open set $\bar g^{-1}(c)$ is included in the right side of the equality, then $\bar g^{-1}(c)$ contains $\iota_\alg(a)$ for some $ a \in \alg$ with $h(a) \in F$.  For $a$ it suffices to choose any element of $g^{-1}(c)$, which is easily shown to belong to $h^{-1}(F)$.
	\end{pr}

\begin{pr}[of Theorem~\ref{thm:clopen-languages}]
Let us begin with the left-to-right implication. Consider a clopen subset of $\bar \alg$. Like any clopen set in a compact space, this is a finite union of base open sets.  By definition of closed base open sets in $\bar \alg$ and  Lemma~\ref{lem:closure-characterised}, we see that every clopen subset of $\bar \alg$ can be represented as a finite union
\begin{align*}
	\bigcup_i \overline{h_i^{-1}(F_i)} = 	\overline {\bigcup_i {h_i^{-1}(F_i)}}.
\end{align*}
The right side is as required in the statement of the theorem.  The right-to-left implication is done by reversing the above reasoning.
\end{pr}

\paragraph*{Stones} In Section~\ref{sec:from-monad-to-promonad}, we showed how to convert $\monad$-algebras into $\promonad$-algebras, and how to convert $\monad$-morphisms into $\promonad$-morphisms. We now explain that the $\monad$-algebras and $\monad$-morphisms produced this way have special topological properties.

A $\promonad$-algebra $\alg$ is called \emph{Stone} if its universe is finite, and the multiplication operation is continuous assuming the discrete topology on the universe. The reason for this name is that the discrete topology is the only one which makes the finite universe a Stone space, i.e.~a compact totally disconnected Hausdorff topological space\footnote{Actually, already the Hausdorff requirement implies discreteness, but Stone spaces are closely connected to profiniteness.}. As shown in the following lemma, Stone $\promonad$-algebras are essentially the same thing as finite $\monad$-algebras. In Section~\ref{sec:profinite-words} we will show examples of finite $\promonad$-algebras that are not Stone.

\begin{theorem}\label{thm:back-and-forth-stone}  
Up to isomorphism,	the mappings 
			\begin{eqnarray*}
				\alg  \qquad &\mapsto& \qquad \closealg \alg\\
						h : \alg \to \balg \qquad &\mapsto& \qquad \bar h : \bar \alg \to \bar \balg
			\end{eqnarray*}
			are one-to-one correspondences  between, respectively:
	\begin{itemize}
		\item 	
finite $\monad$-algebras and finite $\promonad$-algebras that are Stone; and
\item  $\monad$-morphisms into finite $\monad$-algebras and continuous $\promonad$-morphisms into finite  $\promonad$-algebras that are Stone.
	\end{itemize}
\end{theorem}
\begin{pr}
	We only consider the first mapping, the second is proved in a similar way.	We will show that if $\alg$ is a finite $\monad$-algebra then $\bar \alg$ is Stone, and if $\alg$ is a Stone, then $\alg_\monad$ is a finite $\monad$-algebra.
	In Lemma~\ref{lem:compactification-preserves-fintieness}, we have shown that if $\alg$ is a finite $\monad$-algebra, then $\bar \alg$ is isomorphic to the algebra whose universe is the universe of $\alg$, and whose multiplication operation is 
	\begin{align*}
		\overline{\mult_\alg} : \promonad A \to A,
	\end{align*}
	i.e.~the profinite extension of the original multiplication. Every profinite extension is continuous (assuming the discrete topology on the image) by definition of the topology in $\bar \alg$, and therefore $\bar \alg$ is Stone. To prove that the correspondence is one-to-one up to isomorphism, we need to show that if $\alg$  is a Stone $\promonad$-algebra then
	  it is isomorphic to $ \closealg{\restrictalg \alg}$, and if $\alg$ is a finite $\monad$-algebra then it is isomorphic to $\restrictalg{(\closealg \alg)}$. We only prove the former isomorphism.  Let then $\alg$ be a Stone algebra.  By Lemma~\ref{lem:compactification-preserves-fintieness}, it suffices to show that  the two multiplication operations
	  \begin{eqnarray*}
	  	\mult_\alg &:& \promonad A \to A \\
       \overline{\mult_{{\restrictalg \alg}}} &:& \promonad A \to A 
	  \end{eqnarray*}
	  are equal. By definition, the two operations agree on elements in the image of $\monad A$ under
	  \begin{align*}
	  	\iota_{\monad A} : \monad A \to \promonad A.
	  \end{align*}
	This image is a dense subset of $\promonad A$. Because $A$ is finite,  $\promonad A$ is a metric space (we implicitly assume that there are countably many finite $\monad$-algebras up to isomorphism), and therefore both multiplication operations are uniformly continuous functions that agree on a dense subset. Such functions must be equal.
\end{pr}

\section{Profinite words}
\label{sec:profinite-words}
\newcommand{\profwords}{{\bar +}}
In this section, we illustrate the profinite monad construction from Section~\ref{sec:profinite-monads} in the special case of words.
Consider the monad $\Sigma \mapsto \Sigma^+$ of finite words. Let us denote by $\Sigma \mapsto \Sigma^{\profwords}$ the profinite version of this monad, as defined in Section~\ref{sec:profinite-monads}.  In particular, $\Sigma^{\profwords}$ is a semigroup thanks to Lemma~\ref{lem:promonad-is-a-monad-algebra}, and $\Sigma^{\profwords}$ is has a topology which makes it a Stone space by Fact~\ref{fact:same-as-stone-dual}. An element of $\Sigma^\profwords$ is called a \emph{profinite word} over the alphabet $\Sigma$. One of the results of this section is Theorem~\ref{thm:msoinf-undecidable}, which implies that \mso is undecidable over profinite words, already with two predicates.

As shown in Section~\ref{sec:from-monad-to-promonad}, every finite  semigroup  $\salg$, can be extended to a $\profwords$-algebra $\bar \salg$ with the same universe, where the multiplication operation
\begin{align*}
	\mult_{\bar \salg} : S^{\profwords} \to S
\end{align*}
 is continuous assuming the profinite topology on the domain and the  discrete topology on the image.  
In this section we give an example of a finite $\profwords$-algebra that is not obtained this way, because the multiplication operation is not going to be continuous assuming the discrete topology on the image.

\subsection{The unboundedness language}
We say that a profinite word $w \in \Sigma^\profwords$  has \emph{at least $n$ letters} in a subset $\Gamma\subseteq \Sigma$ if it has value $n$ under $\bar h$ where   
	\begin{align*}
		h : \Sigma^+ \to \set{0,1,\ldots,n}
	\end{align*}
	is the semigroup morphism which counts the number of letters in $\Gamma$ up to threshold $n$.  A profinite word is said to have  exactly $n$ letters from a set if it has at least $n$ letters from the set but not at least $n+1$. If a profinite word has at least $n$ letters in $\Gamma$ for every $n$, then we say that it has \emph{an unbounded number} of letters in $\Gamma$.

\begin{lemma}\label{lem:}
	The set of profinite words in $\set{0,1}^\profwords$ which have unboundedly many ones is  $\profwords$-recognisable.
\end{lemma}
\begin{pr}
We show that the set in the statement of the lemma  is recognised by a $\profwords$-morphism 
	\begin{align*}
		h : \set{0,1}^\profwords \to \alg
	\end{align*}
	where $\alg$ is the finite $\profwords$-algebra defined as follows. 
	The universe of  $\alg$ has three elements, call them $0,1$ and $\infty$, which represent profinite words that have zero ones,  a bounded number of ones, and an unboundedly number of ones respectively. The  multiplication operation 
	\begin{align*}
		\mult_\alg : A^\profwords \to A
	\end{align*}
	is defined as follows.  If the argument has only zeros, the value is zero. If the argument has at least  one letter $\infty$, or unboundedly many ones, then the value is $\infty$. Otherwise the value is one. Note that the multiplication operation is \emph{not}  continuous, at least assuming a discrete topology on the universe, because the inverse image of $1$ is not closed. We now prove that this multiplication is associative, i.e.~that the following diagram commutes:
	\begin{align*}
		\vcenter{\xymatrix  { (A^\profwords)^\profwords  \ar[rr]^{\promonmul A} \ar[d]_{(\mult_\alg)^\profwords} && A^\profwords \ar[d]^{\mult_\alg}  \\
		A^\profwords \ar[rr]_{\mult_\alg}& & A
		}}
	\end{align*}
	where $\promonmul A$ denotes the multiplication operation of the profinite monad.
%
%

%
To prove that the above diagram commutes, we need to show that
\begin{align}\label{eq:goal-unboundedness-algebra}
{\mult_\alg}((\mult_\alg)^\profwords(w))= 	{\mult_\alg}(\promonmul A(w)).
\end{align}
holds for every profinite word of profinite words  $w \in (A^\profwords)^\profwords$. 
We consider two cases, depending on whether $w$ has an unbounded number of letters in the set $A^\profwords - 0^\profwords$.
\begin{itemize}
	\item The word $w$ has an unbounded number of letters outside $0^\profwords$. We will show that~\eqref{eq:goal-unboundedness-algebra} holds, because both sides are equal to $\infty$. Consider first
	the left side.
	Let 
	\begin{align*}
		h_n : A^+ \to \set{0,\ldots,n}
	\end{align*}
	be the semigroup morphism that counts the number of nonzero letters. Our assumption on $w$ says that $(\overline h_1)^\profwords(w)$ has unboundedly many ones.  Since the image of $h_1$ is a subset of $A$, it makes sense to compare values of $\overline h_1$ with values of ${\mult_\alg}$, in particular the following observation is easy to get:
		\begin{align*}
		\overline h_1 (v)\ \le\  {\mult_\alg}(v) \qquad \mbox{for every $v \in A^\profwords$}.
	\end{align*}
	As in the proof of Lemma~\ref{lem:mso-recognisable}, a binary relation $R \subseteq X \times Y$ lifts to a relation $R^\monad \subseteq \monad X \times \monad Y$. Apply this construction to the natural ordering on $A$, and call $\le$ the resulting relation on $A^\profwords$. As we have observed,
	\begin{align*}
		(\overline h_1)^\profwords(w)\ \le^\profwords\ (\mult_\alg)^\profwords(w).
	\end{align*}
	The profinite word on the left of the above inequality has an unbounded number of ones by our assumption, and therefore it is mapped by ${\mult_\alg}$ to $\infty$.
	It is not difficult to see that the mapping ${\mult_\alg}$ is monotone with respect to $\le$, and therefore the left side of the equality in~\eqref{eq:goal-unboundedness-algebra} is $\infty$. 
	
	To prove that the right side of the equality in~\eqref{eq:goal-unboundedness-algebra} is also $\infty$, by definition of ${\mult_\alg}$ we need to show that every $n$ satisfies
	\begin{align*}
		\overline {h_n} (\promonmul A(w)) = n.
	\end{align*}
	Let $\mult_n$ be the multiplication operation in the semigroup $\set{0,\ldots,n}$.
	Theorem~\ref{thm:definition-of-promonad} says that 
		\begin{align*}
			\vcenter{\xymatrix @R=2pc
				{  A^{\profwords \profwords} \ar[d]_{ \overline{h_n}^\profwords}  \ar[rr]^{\promonmul A} && \promonad A \ar[d]^{\overline {h_n}}\\
				\promonad  \set{0,\ldots,n} \ar[rr]^{\overline{\mult_n}}&  & \set{0,\ldots,n}
				}}
	\end{align*}
	To prove that the right side of the equality in~\eqref{eq:goal-unboundedness-algebra} is also $\infty$, from the definition of $\mult_\alg$ we need to show that for every 
	$n$, if start with $w$ and  consider the right-down path in the above diagram, then we get $n$. Because the diagram commutes, we can also consider the down-right path. Our assumption on $w$ says that $(\overline {h_n})^\profwords (w)$ is a profinite word which has an unbounded number of nonzero letters. On such words, $\overline {h_n}$ gives result $n$.

	\item The other case is when  $w$ has a bounded number  of letters outside $0^\profwords$. 
	We begin with a straightforward lemma, which uses the semigroup structure of profinite words that was described in Lemma~\ref{lem:promonad-is-a-monad-algebra}.
Let us denote the unit of the profinite monad by $\promonun \Sigma$, i.e.~if $a \in \Sigma$ then $\unitt a \Sigma \in \Sigma^\profwords$ is the corresponding profinite word.
		\begin{lemma}\label{lem:finite-decomposition-of-profinite-word}
			If  $w \in \Sigma^\profwords$ has a bounded number of letters in  $\Gamma \subseteq \Sigma$ then it admits  a finite decomposition
			\begin{align*}
	w=			w_0 \cdot \unitt {a_1} \Sigma \cdot  w_1 \cdots w_{n-1} \cdot \unitt{a_n}\Sigma \cdot w_n
			\end{align*}
			where $w_0,\ldots,w_n$ are profinite words over the alphabet $\Sigma-\Gamma$, $a_1,\ldots,a_n$ are letters in $\Gamma$, and the dot stands for concatenation in the profinite semigroup.
		\end{lemma}
	
	 By applying Lemma~\ref{lem:finite-decomposition-of-profinite-word}, there is a decomposition
	\begin{align*}
		w =  w_0  \cdot \unitt{a_1}{A^\profwords} \cdot  w_1 \cdots w_{n-1}  \cdot \unitt{a_n}{A^\profwords} \cdot  w_n
	\end{align*}
	where $w_i \in (0^\profwords)^\profwords$,  $a_i \in A^\profwords - 0^\profwords$, and the dot is concatenation in the profinite semigroup over alphabet $A^\profwords$.  Lemma~\ref{lem:promonad-is-a-monad-algebra} implies that
	\begin{align*}
\promonmul A(w) = \promonmul A(w_0) \cdot a_1 \cdot \promonmul A(w_1)\cdots \promonmul A(w_{n-1}) \cdot a_n \cdot \promonmul A (w_n) 
	\end{align*}
	where the dot  is concatenation in the profinite  semigroup over alphabet $A$. Since ${\mult_\alg}$ is a semigroup morphism, and it maps words in $0^\profwords$ to the identity in $A$, it follows that 
	\begin{align*}
		{\mult_\alg}(\promonmul A(w))= {\mult_\alg}(a_1) \cdots {\mult_\alg}(a_n).
	\end{align*}
	Let us now consider ${\mult_\alg}((\mult_\alg)^\profwords(w))$.  Lemma~\ref{lem:promonad-is-a-monad-algebra} says that $(\mult_\alg)^\profwords$ is a semigroup morphism, and therefore $		{\mult_\alg}^\profwords(w)$ is equal to
	\begin{align*}
 {\mult_\alg}^\profwords(w_0) \cdot {\mult_\alg}^\profwords(\unitt{a_1}{A^\profwords}) \cdot {\mult_\alg}^\profwords(w_1) \cdots {\mult_\alg}^\profwords(w_{n-1}) \cdot {\mult_\alg}^\profwords(\unitt{a_n}{A^\profwords}) \cdot {\mult_\alg}^\profwords(w_n).
	\end{align*}
	By the axioms of a monad, we have 
	\begin{align*}
		{\mult_\alg}^{\profwords}(\unitt{a_i}{A^\profwords}) = \unitt{{\mult_\alg}(a_i)}{A}.
	\end{align*}
	Each word ${\mult_\alg}^\profwords(w_i)$ in the decomposition of ${\mult_\alg}^{\profwords}(w)$ belongs to $0^\profwords$. Therefore, because ${\mult_\alg}$ is a semigroup morphism that maps $0^\profwords$ to the identity, we get
	\begin{align*}
		{\mult_\alg}({\mult_\alg}^\profwords(w)) =  {\mult_\alg}(\unitt{{\mult_\alg}(a_i)}{A}) \cdots{\mult_\alg}(\unitt{{\mult_\alg}(a_n)}{A}).
	\end{align*}
	The result follows because ${\mult_\alg}(\unitt a {A})=a$  holds for every $a \in A$.
\end{itemize}
	This completes the proof that ${\mult_\alg} : A^\profwords \to A$ is a $\profwords$-morphism.
\end{pr}

Let us define \msoinf to by applying the abstract notion of \mso defined in Section~\ref{sec:abstract-mso}, with the base predicates being the language of unboundedly many ones from the previous lemma, and the profinite closure of the language ``some $a$ comes before some $b$''. This class of languages of profinite words was considered in~\cite{torunczyk_phd} and~\cite{DBLP:conf/icalp/Torunczyk12}, adjusting for the monad terminology. From Lemma~\ref{lem:mso-recognisable} it follows that \msoinf contains only  $\profwords$-recognisable languages. It is not clear if it contains all $\profwords$-recognisable languages.  

\begin{theorem}\label{thm:msoinf-undecidable}
	The satisfiability problem for \msoinf is undecidable.
\end{theorem}
\begin{pr}
Consider 	\mso{\sc+u} on infinite words, which is an extension of \mso. This logic is shown undecidable in~\cite{DBLP:journals/corr/msou}. 
 Corollary 2 of~\cite{DBLP:conf/icalp/Torunczyk12} shows that decidability  of \mso{\sc+u} on infinite words reduces to decidability of \msoinf on profinite words.
\end{pr}

\bibliographystyle{alpha}
\bibliography{bib}

\begin{thebibliography}{{\'A}DH83}

\bibitem[{\'A}DH83]{agoston1983number}
I~{\'A}goston, J~Demetrovics, and L~Hann{\'a}k.
\newblock The number of clones containing all constants (a problem of {R.
  McKenzie}).
\newblock In {\em Colloquia mathematica ocietatis Janos Bolyai}, volume~43,
  pages 21--25, 1983.

\bibitem[Arn85]{DBLP:journals/tcs/Arnold85}
Andr{\'{e}} Arnold.
\newblock A syntactic congruence for rational omega-language.
\newblock {\em Theor. Comput. Sci.}, 39:333--335, 1985.

\bibitem[BI09]{idziaszek_ef}
Miko{\l}aj Boja{\'n}czyk and Tomasz Idziaszek.
\newblock Algebra for infinite forests with an application to the temporal
  logic {EF}.
\newblock In {\em CONCUR}, pages 131--145, 2009.

\bibitem[Boj13]{DBLP:journals/mst/Bojanczyk13}
Mikolaj Bojanczyk.
\newblock Nominal monoids.
\newblock {\em Theory Comput. Syst.}, 53(2):194--222, 2013.

\bibitem[Boj14]{DBLP:conf/icalp/Bojanczyk14}
Mikolaj Bojanczyk.
\newblock Transducers with origin information.
\newblock In Javier Esparza, Pierre Fraigniaud, Thore Husfeldt, and Elias
  Koutsoupias, editors, {\em Automata, Languages, and Programming - 41st
  International Colloquium, {ICALP} 2014, Copenhagen, Denmark, July 8-11, 2014,
  Proceedings, Part {II}}, volume 8573 of {\em Lecture Notes in Computer
  Science}, pages 26--37. Springer, 2014.

\bibitem[BR12]{DBLP:journals/jcss/BedonR12}
Nicolas Bedon and Chlo{\'{e}} Rispal.
\newblock Sch{\"{u}}tzenberger and {Eilenberg} theorems for words on linear
  orderings.
\newblock {\em J. Comput. Syst. Sci.}, 78(2):517--536, 2012.

\bibitem[B{\"u}c62]{buchi_decision}
Julius~Richard B{\"u}chi.
\newblock On a decision method in restricted second-order arithmetic.
\newblock In {\em Proc. 1960 Int. Congr. for Logic, Methodology and Philosophy
  of Science}, pages 1--11, 1962.

\bibitem[BW08]{DBLP:conf/birthday/BojanczykW08}
Mikolaj Bojanczyk and Igor Walukiewicz.
\newblock Forest algebras.
\newblock In J{\"{o}}rg Flum, Erich Gr{\"{a}}del, and Thomas Wilke, editors,
  {\em Logic and Automata: History and Perspectives [in Honor of Wolfgang
  Thomas].}, volume~2 of {\em Texts in Logic and Games}, pages 107--132.
  Amsterdam University Press, 2008.

\bibitem[CCP11]{DBLP:conf/icalp/CartonCP11}
Olivier Carton, Thomas Colcombet, and Gabriele Puppis.
\newblock Regular languages of words over countable linear orderings.
\newblock In Luca Aceto, Monika Henzinger, and Jiri Sgall, editors, {\em
  Automata, Languages and Programming - 38th International Colloquium, {ICALP}
  2011, Zurich, Switzerland, July 4-8, 2011, Proceedings, Part {II}}, volume
  6756 of {\em Lecture Notes in Computer Science}, pages 125--136. Springer,
  2011.

\bibitem[Eil74]{0317.94045}
S.~Eilenberg.
\newblock {\em Automata, languages, and machines. Vol. A.}
\newblock 1974.

\bibitem[{\'E}W03]{esik2003logically}
Zolt{\'a}n {\'E}sik and Pascal Weil.
\newblock On logically defined recognizable tree languages.
\newblock In {\em FST TCS 2003: Foundations of Software Technology and
  Theoretical Computer Science}, pages 195--207. Springer, 2003.

\bibitem[GGP08]{gehrke2008duality}
Mai Gehrke, Serge Grigorieff, and Jean-{\'E}ric Pin.
\newblock Duality and equational theory of regular languages.
\newblock In {\em Automata, languages and programming}, pages 246--257.
  Springer, 2008.

\bibitem[GGP10]{gehrke2010topological}
Mai Gehrke, Serge Grigorieff, and Jean-{\'E}ric Pin.
\newblock A topological approach to recognition.
\newblock In {\em Automata, Languages and Programming}, pages 151--162.
  Springer, 2010.

\bibitem[HM88]{hobby1988structure}
David~Charles Hobby and Ralph McKenzie.
\newblock {\em The structure of finite algebras}, volume~76.
\newblock American Mathematical Society Providence, 1988.

\bibitem[MB15]{DBLP:journals/corr/msou}
Szymon~Toru\'nczyk Miko{\l}aj~Boja\'nczyk, Pawe{\l}~Parys.
\newblock The {\sc mso+u} theory of {$(\Nat,<)$} is undecidable.
\newblock {\em CoRR}, arXiv:1502.04578, 2015.

\bibitem[Pot95]{potthoff}
A.~Potthoff.
\newblock First-order logic on finite trees.
\newblock {\em Lecture Notes in Computer Science}, 915:125--139, 1995.

\bibitem[PP04]{perrin_pin_words}
Dominique Perrin and {Jean-\'{E}ric} Pin.
\newblock {\em Infinite Words: Automata, Semigroups, Logic and Games}.
\newblock Elsevier, 2004.

\bibitem[Rei82]{reiterman1982birkhoff}
Jan Reiterman.
\newblock The {Birkhoff} theorem for finite algebras.
\newblock {\em Algebra Universalis}, 14(1):1--10, 1982.

\bibitem[Ros70]{rosenberg1970funktionale}
Ivo Rosenberg.
\newblock {\em {\"U}ber die funktionale Vollst{\"a}ndigkeit in den mehrwertigen
  Logiken: Struktur der Funktionen von mehreren Ver{\"a}nderlichen auf
  endlichen Mengen}.
\newblock Academia, 1970.

\bibitem[Ros86]{rosenberg1986minimal}
Ivo Rosenberg.
\newblock Minimal clones i: the five types.
\newblock In {\em Lectures in Universal Algebra (Proc. Conf. Szeged 1983)},
  volume~43, pages 405--427. North-Holland Amsterdam, 1986.

\bibitem[She75]{shelah_composition}
Saharon Shelah.
\newblock The monadic theory of order.
\newblock {\em The Annals of Mathematics}, 102(3):379--419, 1975.

\bibitem[Ste92]{DBLP:books/el/treeauto1992/Steinby92}
Magnus Steinby.
\newblock A theory of tree language varieties.
\newblock In {\em Tree Automata and Languages}, pages 57--82. 1992.

\bibitem[Tho84]{DBLP:conf/caap/Thomas84}
Wolfgang Thomas.
\newblock Logical aspects in the study of tree languages.
\newblock In {\em {CAAP}}, pages 31--50, 1984.

\bibitem[Tho96]{thomas_languages}
Wolfgang Thomas.
\newblock Languages, automata and logics.
\newblock Technical Report 9607, Institut f{\"u}r Informatik und Praktische
  Mathematik, Christian-Albsechts-Universit{\"a}t, Kiel, Germany, 1996.

\bibitem[Tor11]{torunczyk_phd}
Szymon Toru{\'n}czyk.
\newblock {\em Languages of profinite words and the limitedness problem}.
\newblock PhD thesis, University of Warsaw, 2011.

\bibitem[Tor12]{DBLP:conf/icalp/Torunczyk12}
Szymon Toru\'nczyk.
\newblock Languages of profinite words and the limitedness problem.
\newblock In {\em Automata, Languages, and Programming - 39th International
  Colloquium, {ICALP} 2012, Warwick, UK, July 9-13, 2012, Proceedings, Part
  {II}}, pages 377--389, 2012.

\bibitem[TW98]{DBLP:conf/stoc/TherienW98}
Denis Th{\'{e}}rien and Thomas Wilke.
\newblock Over words, two variables are as powerful as one quantifier
  alternation.
\newblock In Jeffrey~Scott Vitter, editor, {\em Proceedings of the Thirtieth
  Annual {ACM} Symposium on the Theory of Computing, Dallas, Texas, USA, May
  23-26, 1998}, pages 234--240. {ACM}, 1998.

\bibitem[Wil91]{DBLP:conf/icalp/Wilke91}
Thomas Wilke.
\newblock An {Eilenberg} theorem for infinity-languages.
\newblock In Javier~Leach Albert, Burkhard Monien, and Mario
  Rodr{\'{\i}}guez{-}Artalejo, editors, {\em Automata, Languages and
  Programming, 18th International Colloquium, ICALP91, Madrid, Spain, July
  8-12, 1991, Proceedings}, volume 510 of {\em Lecture Notes in Computer
  Science}, pages 588--599. Springer, 1991.

\bibitem[Wil93]{wilke_algebraic}
Thomas Wilke.
\newblock An algebraic theory for regular languages of finite and infinite
  words.
\newblock {\em Int. J. Alg. Comput.}, 3:447--489, 1993.

\bibitem[YM59]{yanovmuchnik}
Y.~I. Yanov and A.~A. Muchnik.
\newblock Existence of k-valued closed classes without a finite basis.
\newblock {\em Dokl. Akad. Nauk.}, 127:44--46, 1959.

\end{thebibliography}
\pagebreak
\appendix

\refstepcounter{monadcounter}\label{counter:monadcounter}

\end{document}